\RequirePackage{fix-cm}

\documentclass[smallcondensed]{svjour3}     
\smartqed  
\usepackage{newtxtext,newtxmath}      
\usepackage{amsmath}
\usepackage{amsfonts}
\usepackage{array}
\usepackage{graphicx}
\usepackage{color}
\usepackage{fullpage}
\usepackage{natbib,url}
\usepackage{graphics}
\usepackage[colorlinks=true,breaklinks=true,linkcolor={blue}]{hyperref}
\numberwithin{equation}{section}
\usepackage{etoolbox}

\newcommand*{\affaddr}[1]{#1} 
\newcommand*{\affmark}[1][*]{\textsuperscript{#1}}
\journalname{Journal of Elasticity}
\begin{document}

\title{Stable Spatially Localized Configurations in a Simple Structure---A Global Symmetry-Breaking Approach}



\author{Shrinidhi S. Pandurangi\affmark[1]
\and Ryan S. Elliott\affmark[2,*]
\and Timothy J. Healey\affmark[3,1]
\and Nicolas Triantafyllidis\affmark[4,5,6]}


\institute{\affaddr{\affmark[1]Field of Theoretical and Applied Mechanics, Cornell University, Ithaca, NY, USA}\\
\affaddr{\affmark[2]Aerospace Engineering and Mechanics, University of Minnesota, Minneapolis, MN, 55455 USA}\\
\affaddr{\affmark[3]Department of Mathematics, Cornell University, Ithaca. NY, USA}\\
\affaddr{\affmark[4]LMS, \'{E}cole Polytechnique, CNRS UMR7649, Institut Polytechnique de Paris, 91128 Palaiseau, France}\\
\affaddr{\affmark[5]D\'{e}partement de M\'{e}canique, \'{E}cole Polytechnique, Palaiseau 91128, France}\\
\affaddr{\affmark[6]Aerospace Engineering Department \& Mechanical Engineering Department (emeritus),
The University of Michigan, Ann Arbor, MI, USA}\\
\affaddr{\affmark[*]Corresponding Author: \url{relliott@umn.edu}}}

\date{Received: date / Accepted: date}

\maketitle

\begin{abstract}

We revisit the classic stability problem of the buckling of an inextensible, axially compressed beam on a nonlinear elastic foundation with a semi-analytical approach to understand how spatially localized deformation solutions emerge in many applications in mechanics.  Instead of a numerical search for such solutions using arbitrary imperfections, we propose a systematic search using branch-following and bifurcation techniques along with group-theoretic methods to find all the bifurcated solution orbits (primary, secondary, etc.) of the system and to examine their stability and hence their observability.  Unlike previously proposed methods that use multi-scale perturbation techniques near the critical load, we show that to obtain a spatially localized deformation equilibrium path for the perfect structure, one has to consider the secondary bifurcating path with the longest wavelength and follow it far away from the critical load.  The novel use of group-theoretic methods here illustrates a general methodology for the systematic analysis of structures with a high degree of symmetry.

\end{abstract}

\tableofcontents

\newpage
\section{Introduction}
\label{sec:Section_1}

Surface instabilities in soft elastic materials under compression, leading to highly localized deformation regions known as creases, have been reported in the recent experiments by \cite{Gent_Cho}.  Since then, a number of studies have discussed the computations of such deformations \citep[e.g.][]{Hohlfeld_Mahadevan_1, Hohlfeld_Mahadevan_2, Cao_Hutchinson, Ciarletta, Chen_et_al_1, Li_Cao_et_al, Silling, Chen_et_al_2, Hong_Zhao_Suo, Auguste_et_al, Wang_Zhao, Chen_Yang_Wheeler, Diab_Kim, Zhao_et_al, Jin_et_al_1, Jin_et_al_2}.  It is well-known that a linearized stability analysis of the flat surface predicts wrinkles \citep[see][]{Biot} and hence precludes a perfect half-space model from \emph{directly} bifurcating from a flat state to a creased one.  While the methods proposed in the existing literature overcome this difficulty by introducing an {\it{a priori}} imperfection which biases the system to a desired configuration, the choice of imperfection introduces an arbitrariness in the procedure.  In this work, we present an approach which is based on applying a systematic numerical continuation coupled with analytical group-theoretic methods (equivariant bifurcation theory).  We demonstrate our methodology on the classical problem of stability of an infinite linear beam\footnote{The reader should note that we employ a linear beam model for its simplicity, both analytically and numerically, in this work.  Thus, undue weight should not be given to the physical interpretation of fine-scale details associated with the highly-deformed solutions obtained here.  Instead, the primary focus here is an analysis of the simplest model leading to the existence of such spatially localized solutions.  As we show, all of the important phenomena (primary bifurcation, cascades of secondary bifurcations, etc.) occur well within the small-displacement regime.  Thus, the linear beam model poses no limitation on our ability to determine the correct mechanics of the nonlinear beam--foundation system of interest.} on a nonlinear elastic foundation.  The nonlinearity in the foundation allows a diverse set of stable and unstable deformations of the beam.  Of particular interest in this study are the stable spatially localized configurations of the beam which are the equilibrium solutions to systems having a softening foundation in the small deformation regime.  Although the higher regularity required by the beam--foundation system precludes the existence of a true ``crease,'' its fully localized ``single-hump'' solutions\footnote{Such as those illustrated in Fig.~\ref{Fig:soft-none-mode}(a).} will serve as reasonable surrogates.  Thus, the model is well-suited for the purpose at hand.  Indeed, the behavior of the beam--foundation system is representative of a much larger class of problems which exhibit spatially localized patterns \citep[e.g.][]{makrides:sandstede:2019, Audoly, Diamant_Witten, Rivetti, Pocivavsek}.

Owing to its rich post-critical behavior, several aspects of this simple system and its variants are much studied.  For example, the existence of global bifurcations corresponding to spatially localized large deformation solutions for fourth-order ordinary differential equations with quadratic nonlinearity was established by \cite{Champneys_Toland:1993}.  Similarly, the existence of such periodic solutions for a system with cubic-quintic nonlinearity was explored by \cite{Beardmore_et_al:2005}.  Asymptotic approaches employing an ansatz having two different length scales lead to the so-called amplitude equations that predict spatially localized solutions. These methods have been applied for linear beams on softening foundations \citep[e.g.][]{Hunt_Bolt_Thompson, Hunt_Wadee, Potier-Ferry_1, Potier-Ferry_2} and geometrically nonlinear beams on linear elastic foundations \citep[e.g.][]{Hunt_Wadee_Shiacolas}.  While these approaches give an amplitude modulated solution, their validity is restricted to axial loads in a small neighborhood of the critical load.  Moreover, it can be shown that all such solutions are unstable and hence not observable.  Highly spatially localized stable solutions, as will be shown through this work, exist only far from the trivial flat state and well below the buckling load where uniformly wrinkled deformations first bifurcate---a behavior common to the creasing problem.  To overcome the restriction to axial loads in a neighborhood of the critical load, a localized Rayleigh-Ritz method was proposed by \cite{Wadee_Hunt_Whiting:1997}.  Deformations of the beam are represented as a sum of terms involving sinusoidal and hyperbolic functions.  This approach was successful in obtaining approximate localized deformation solutions far from the initial buckling load \citep[e.g.][]{Wadee_Hunt_Whiting:1997, Wadee_Bassom:2000}.  Once such localized deformation solutions were found, it was noticed that beams with foundations that soften for small displacements and reharden as the displacement increases display an oscillatory behavior, termed ``snaking,'' in their axial load--displacement response \citep[see e.g.][]{Woods_Champneys:1999, Budd_Hunt_Kuske:2001, Peletier:2001}.  Here the bifurcating branch initially develops a highly localized ``single-hump'' configuration as the load sharply decreases (with increasing average axial compression) from the critical value.  Eventually the load reaches a minimum at a turning point and the branch stabilizes with increasing load as the average axial compression increases.  As the path continues, it oscillates (snakes) between turning points with alternating stable and unstable segments.  With each oscillation an additional hump is added to the localized deformation pattern and eventually a wave-packet-like shape is achieved.  This phenomenon is also referred to as cellular buckling \citep[][]{Hunt_et_al}.  The existence of ``multi-hump'' solutions, unrelated to snaking behavior and associated with beams on softening foundations, was shown by \cite{Champneys_Toland:1993}.  These solutions are characterized by packets of large amplitude-modulated deformation separated by sections of nearly flat, undeformed, beam.  In this regard, perturbation approaches for ``double-hump'' solutions associated with softening foundations have been discussed by \cite[]{Wadee_Bassom:1999}.  For rehardening foundations, similar solutions have been obtained by \cite{Wadee_Coman_Bassom:2002}.  Finally, similar to Biot's instability of an elastic half-space, the deformations on the primary bifurcation path for a perfect beam--foundation system represent uniformly wrinkled solutions.  However, it has been shown that the deformations localize on the primary bifurcation path if an inhomogeneity (imperfection) is introduced in the stiffness of the foundation or in the bending modulus of the beam \citep[][]{Amazigo_Budiansky_Carrier:1970, Luongo:1991, Wadee:2000, Coman:2006, Coman:2010}.

In contrast to the above localized behavior for softening foundations, beams on foundations having a purely hardening response buckle to form stable small-amplitude wrinkles.  As the axial load is increased the wrinkles can change periodicity through an unstable connection in the bifurcation diagram.  This phenomenon, known as ``mode-jumping,'' was studied by \cite{Hunt_Everall} and \cite{Everall_Hunt}.

Due to the high degree of symmetry present in perfect structures such as the one considered here, the use of standard ``\emph{imperfection methods}'' for bifurcation problems is insufficient to discover and organize its rich solution set.  Accordingly, we introduce a rigorous and systematic group-theoretic framework that provides analytical characterization of all bifurcations and bifurcating paths as well as a systematic classification of every equilibrium branch based on its geometric symmetry group.  This approach draws on ``\emph{equivariant bifurcation theory}'' \citep{GolStew88,chossat:lauterbach:2000,ikeda:murota:2010} and is integrated with efficient numerical branch-following algorithms \citep{Keller, allgower:georg:2003,gatermann:hohmann:1991:ICSE} to create a robust, consistent methodology for the theoretical and numerical study of highly-symmetric bifurcation problems. Through this work, we show that highly localized deformation stable equilibrium solutions exist as long wavelength secondary bifurcation paths.  These paths bifurcate in a cascading fashion from the short wavelength, wrinkled, primary bifurcation paths.  These primary paths in turn emerge from the system's symmetry-breaking bifurcations points on its flat configuration.  Later in this work, we demonstrate that a small imperfection in the linear foundation stiffness yields highly spatially localized stable solutions on the primary bifurcation path which is almost indistinguishable with secondary path for the perfect system.

We review the non-dimensional model and discuss its numerical implementation in Sections~\ref{sec:Section_2} and \ref{sec:Section_3}, respectively.  A parametric study is used to explore the effects of different types of foundations and the results are presented in Section~\ref{sec:Section_4}.  Bifurcation diagrams and stability results are presented and analyzed in each case.  In Section~\ref{sec:Section_4} we compare the imperfect structure with its perfect counterpart.  The conclusions of this work are presented in Section \ref{sec:Section_5}.  A brief presentation of the pertaining group theory and how it applies to the problem at hand is provided in Sec.~\ref{sec:symmetry} and Appendix~\ref{appendix-A}.

\section{Theory}
\label{sec:Section_2}

Section~\ref{sec:Section_2} presents the theoretical aspects of the beam model.  The perfect case and its imperfect counterpart are defined in Subsection~\ref{sec:perfect} and Subsection~\ref{sec:imperfect}, respectively.  The perfect model's symmetry group is presented in Subsection~\ref{sec:symmetry}, followed by Subsection~\ref{sec:stability} which discusses how to determine the stability of each orbit.
\subsection{Model---Perfect Case}
\label{sec:perfect}

\begin{figure*}
\begin{minipage}{0.5\textwidth}
\begin{center}
\includegraphics[width=\textwidth]{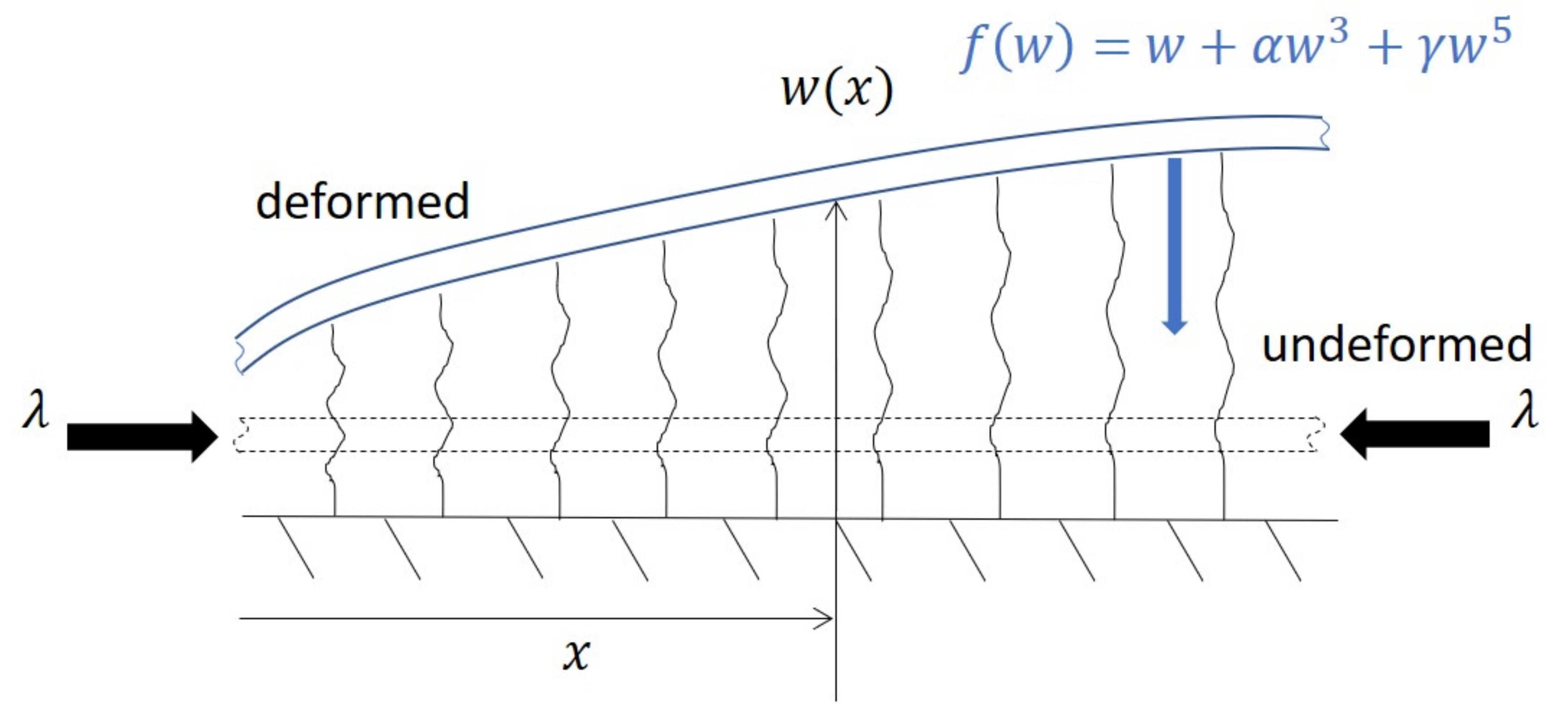}\\
(a)\\
\end{center}
\end{minipage}%
\begin{minipage}{0.55\textwidth}
\begin{center}
\includegraphics[width=\textwidth]{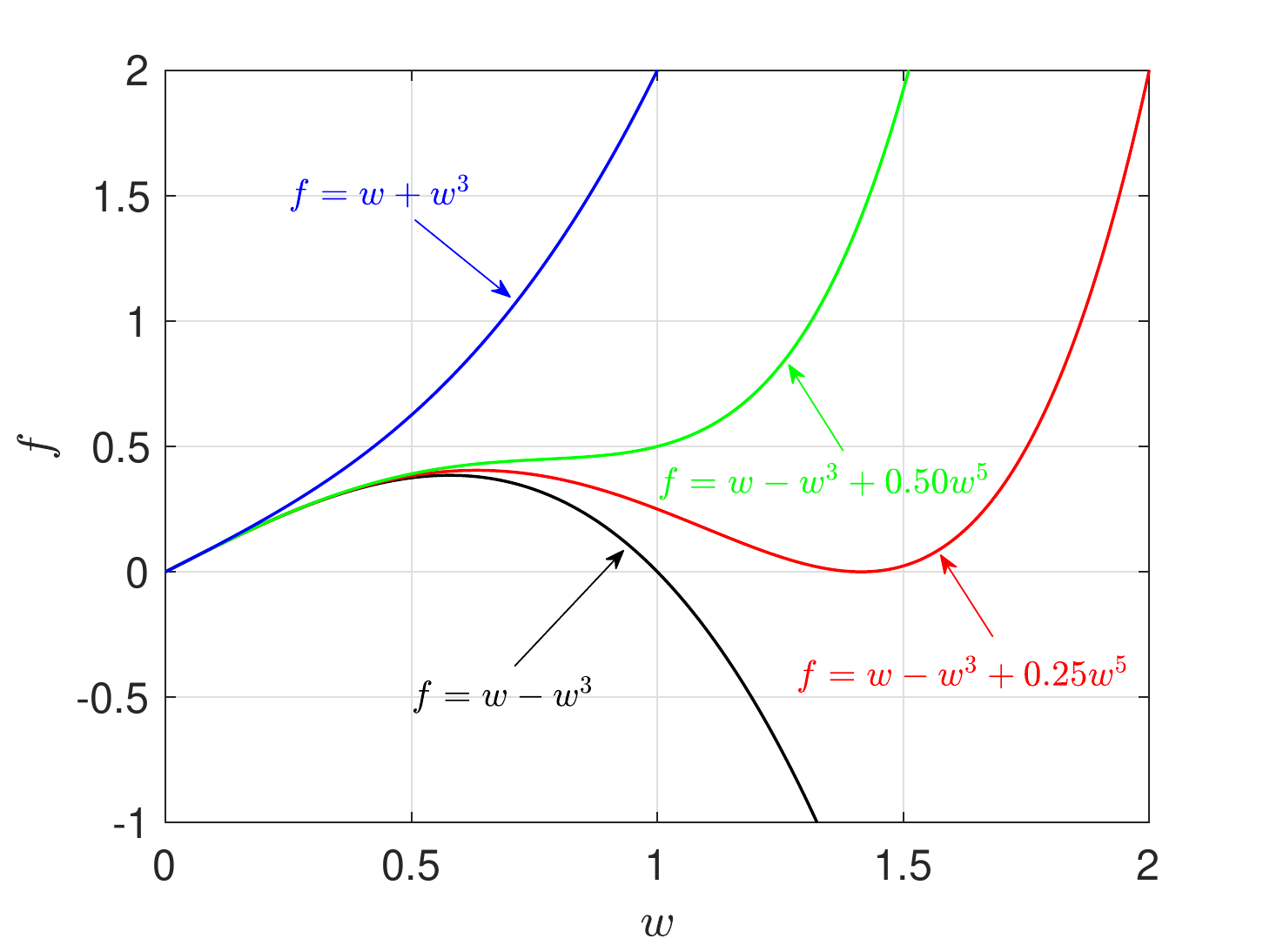}\\
(b)
\end{center}
\end{minipage}
\caption{(a)~Schematic of the undeformed and deformed configuration of the axially compressed beam on a nonlinear foundation.  (b)~Foundation force--displacement curves for the different nonlinear laws considered.}
\label{Fig:schematics}
\end{figure*}

The model adopted here is that of an inextensible, linear elastic beam resting on a nonlinear elastic foundation (with a cubic--quintic nonlinearity).  In order to make the problem manageable---and deal with compact symmetry groups---we will consider the $\bar{L}$-periodic solutions of an infinite beam and then choose $\bar{L} \gg 1$.  The beam has bending stiffness $EI$ and its undeformed centerline coincides with the $\bar{x}$-axis.  The beam is subjected to an axially compressive load $P$ and has the lateral deformation $\bar{w}(\bar{x})$.  The total potential energy density (per unit reference length) of the system is given by
\begin{equation}
\mathcal{\bar{E}}(\bar{w}(\bar{x});P) = \frac{1}{\bar{L}} \int_{-\bar{L}/2}^{\bar{L}/2}\left[\frac{1}{2}EI\left(\frac{d^2 \bar{w}}{d \bar{x}^2}\right)^2-\frac{1}{2}P\left(\frac{d\bar{w}}{d\bar{x}}\right)^2+\frac{1}{2}k_{2}\bar{w}^2+\frac{1}{4}k_{4}\bar{w}^4 +\frac{1}{6}k_{6}\bar{w}^6\right]d\bar{x} \; ;
\  \bar{w}(\bar{x}) \in \bar{H} \; ,
\label{eq:energy_dim}
\end{equation}
where the $\bar{L}$-periodic displacement functions $\bar{w}(\bar{x})$ belong to the space of admissible functions $\bar{H}$, here identified with the Hilbert space $\bar{H} := H_{\bar{L}}^2(\mathbb{R})$, consisting of all $\bar{L}$-periodic functions, say, $w(x)$, such that $w$ and its first two distributional derivatives are each square integrable functions on $(-\bar{L}/2,\bar{L}/2)$.  Identifying the characteristic length $\bar{L}_{c} :=(EI/k_{2})^{1/4}$, Eq.~\eqref{eq:energy_dim} can be non-dimensionalized by setting $x := \bar{x}/\bar{L}_{c}$, $w := \bar{w}/\bar{L}_{c}$ and $L := \bar{L}/\bar{L}_{c}$, leading to a potential energy density of the beam
\begin{equation}
\mathcal{E}(w(x);\lambda) =\frac{1}{L} \int_{-L/2}^{L/2}\left[\frac{1}{2}\left(\frac{d^2 w}{d x^2}\right)^2-\frac{1}{2}\lambda\left(\frac{dw}{dx}\right)^2+\frac{1}{2}w^2+\frac{\alpha}{4}w^4 +\frac{\gamma}{6}w^6\right]dx \; ;
\  w(x) \in H \; ,
\label{eq:energy}
\end{equation}
where $\mathcal{E} := \bar{\mathcal{E}}(\bar{L}_{c})^2/EI$, $\lambda := P (\bar{L}_{c})^2/EI = {P}/{\sqrt{k_{2}EI}}$, $\alpha := k_{4}(\bar{L}_{c})^6/EI = {k_{4} \sqrt{EI}}/({{k_{2}})^{3/2}}$, $\gamma := k_{6} (\bar{L}_{c})^8/EI = {k_{6}EI}/{ (k_{2})^2}$ and $H := H_{L}^2(\mathbb{R})$.  A schematic of the system is shown in Fig.~\ref{Fig:schematics}(a).

As inferred from Eq.~\ref{eq:energy} and depicted in Fig.~\ref{Fig:schematics}, the foundation force per unit length opposing a displacement $w$ is $f = w + \alpha w^3 + \gamma w^5$, with corresponding results labeled by $f_{(\alpha, \gamma)}$.  Two different nonlinear foundation types will be considered: a ``{\it softening foundation}'' with $\alpha = -1$ and a ``{\it hardening foundation}" with $\alpha = +1$.  From the first category ($\alpha = -1$) we distinguish three different variations: the ``{\it no re-hardening}'' foundation where $\gamma = 0$, the ``{\it mild re-hardening}'' foundation where $\gamma = 0.25$ and the ``{\it strong re-hardening}'' foundation where $\gamma = 0.50$, resulting in a monotonically increasing foundation force--displacement response.  The no re-hardening foundation, although interesting from a mathematical point of view, is unrealistic as it produces foundation forces in the same direction as $w$ above a certain displacement threshold.

The equilibrium solutions of the model extremize the energy density given in Eq.~\eqref{eq:energy} and are found by setting to zero the first variation $\mathcal{E}_{,w} \delta w$ of $\mathcal{E}$ with respect to $w$.  The corresponding fourth-order Euler--Lagrange ordinary differential equation (per period $L$) is
\begin{equation}
\mathcal{E}_{,w} \delta w = 0 \ \Longrightarrow \ \frac{d^4 w}{d x^4} + \lambda \frac{d^2 w}{d x^2}
+ w + \alpha w^3 +\gamma w^5 = 0\; ;
\quad x \in (-{L/ 2},\; {L/ 2}) \; ,
\label{eq:equilibrium}
\end{equation}
subjected to the four periodicity-induced boundary conditions\footnote{Assuming adequate continuity, the fourth-order Euler--Lagrange equation in $[-L/2, L/2]$ requires four boundary conditions to be satisfied.}
\begin{equation}
\frac{d^p w}{d x^p} (-{L/ 2}) = \frac{d^p w}{d x^p}({L/ 2})\; ; \quad p = 0, 1, 2, 3\; .
\label{eq:boundary}
\end{equation}
By embedding, any $w \in H_{L}^2(\mathbb{R})$ can be associated with a $C^1$ function.  We then see directly from Eq.~\eqref{eq:equilibrium} that all weak solutions in $H_{L}^2(\mathbb{R})$ are, in fact, classical.

Of interest here are the equilibrium solutions of this structure as a function of either the load $\lambda$ or its work conjugate quantity from Eq.~(\ref{eq:energy}), the (average) axial strain $\Delta$
\begin{equation}
\Delta := \displaystyle \frac{1}{2 L} \int_{-L/2}^{L/2}\left[\left(\frac{d w}{d x}\right)^2\right] dx\; .
\label{eq:displacement}
\end{equation}

The nonlinear boundary value problem defined by Eqs.~\eqref{eq:equilibrium},~\eqref{eq:boundary} admits a large symmetry group resulting in a complex structure with an infinite number of equilibrium paths.  Equilibrium paths related by symmetry form {\it orbits}, which is a more appropriate term to describe the equilibrium solutions of the problem at hand and thus the two terms are used here interchangeably\footnote{In fact, for systems with Lie symmetry groups (continuous infinite groups), such as the one considered here, the continuous orbits of solutions are known as \emph{relative equilibria}.  In general, relative equilibria of the system's equations of motion can correspond to an orbit of equilibria, traveling or rotating waves, dynamic trajectories that appear time-periodic in a suitable moving frame, or other more complicated motions.  For a discussion of the theory of relative equilibria and their stability see \cite{chossat:lauterbach:2000}.  In this work, we explore only orbits of equilibria, and so the distinction between a \emph{relative equilibrium} of a system with continuous symmetry and an \emph{orbit of equilibria} of a system with discrete symmetry is immaterial.}.  From the trivial, flat principal solution ($w = 0$) an infinity of primary equilibrium paths (more precisely continuous orbits) emerge. From each one of these, secondary orbits with different periods also emerge. Tertiary branches emerge from the secondary paths and so on, although the symmetry group of each orbit is reduced at each bifurcation, enough symmetries remain to admit further bifurcations.

The focus here is on following each one of these orbits, away from the bifurcation point of their origin and studying their stability. The symmetry group of this problem and its subgroups that explain the structure of the bifurcated equilibrium paths are well known \cite[e.g.][]{ikeda:murota:2010}.  However, for reasons of clarity and completeness, the interested reader can find a brief presentation of the pertaining theory and how it applies to the problem at hand in Sec.~\ref{sec:symmetry} and Appendix~\ref{appendix-A}.

The numerical solution of the boundary value problem given by Eqs~\eqref{eq:equilibrium},~\eqref{eq:boundary} is presented in Subsection~\ref{sec:PerfectModel}.

\subsection{Model---Imperfect Case}
\label{sec:imperfect}
It is also of interest to consider a more realistic, imperfect structure---slightly perturbed from the perfect one---and compare the corresponding equilibrium solutions. To this end we assume that the infinite beam of period $L$ has a foundation with a perturbed stiffness zone (softer when $\zeta < 0$, and stronger when $\zeta > 0$) of length $2x_0$ about the origin. The corresponding energy density and admissible displacement space are
\begin{equation}
\begin{gathered}
\mathcal{E}(w(x);\lambda) = \frac{1}{L} \int_{-L/2}^{L/2}\left[\frac{1}{2}\left(\frac{d^2 w}{d x^2}\right)^2-\frac{1}{2}\lambda\left(\frac{dw}{dx}\right)^2+\frac{1+z(x)}{2}w^2+\frac{\alpha}{4}w^4 +\frac{\gamma}{6}w^6\right]dx \; ;
\\ \\  z(x) = 0\ {\rm for}\ |x| > x_0\; ,\quad z(x) = \zeta \ {\rm for}\ |x| \le x_0 \; ;
\quad w(x) \in H \; ,
\label{eq:impenergy}
\end{gathered}
\end{equation}
where the admissible $L$-periodic displacement functions $w(x)$ belong to the same Sobolev space as for the perfect case in Eq.~\eqref{eq:energy}. The corresponding Euler--Lagrange equilibrium equations are
\begin{equation}
\mathcal{E}_{,w} \delta w = 0 \ \Longrightarrow \ \frac{d^4 w}{d x^4} + \lambda \frac{d^2 w}{d x^2} + [1+z(x)]w + \alpha w^3 +\gamma w^5 = 0\; ;
\quad x \in (-{L/ 2},\; {L/ 2}) \; ,
\label{eq:impequilibrium}
\end{equation}
subjected to the periodicity conditions given in Eq.~\eqref{eq:boundary}.

The numerical solution of the boundary value problem given by Eqs.~\eqref{eq:impequilibrium},~\eqref{eq:boundary} is presented in Subsection~\ref{sec:ImperfectModel}.


\subsection{Symmetries of the Perfect Model}
\label{sec:symmetry}

For the perfect, elastic beam--foundation model at hand, there exists a group $G$ of transformations that leave its energy $\mathcal{E}(w;\lambda)$---defined in Eq.~\eqref{eq:energy}---invariant under the action of all transformations $g \in G$.  For practical purposes, we want to deal with a compact symmetry group $G$. To this end a maximum fundamental period $L_d$ of all sought equilibrium solutions must be selected.  In principle, the choice of $L_d$ is limited only by the available computational resources.  It is also desirable to choose $L_d$ so that it is commensurate with (i.e.\ an integer multiple of) the fundamental period $L_c$ of the primary bifurcation orbit\footnote{In Sec.~\ref{sec:perfect:princ} it is shown that $L_c = 2\pi$ for the perfect beam--foundation system.}: $L_d = L_c q$ and $q \in \mathbb{N}$.  Thus, the appropriate Hilbert space of interest is $H = H^2_{L_d}(\mathbb{R})$.  Finally, it is important to point out that the selection of $q > 1$ facilitates the inclusion of ``period-extending'' (period-doubling, -tripling, etc.) solutions.\footnote{Indeed, with (as below) $q=20$, the primary bifurcation branch will have period $L_c$.  Then each secondary bifurcating branch may be associated with one of the periods $L_m = m L_c$ with $m \in \{1,2,\dots,q\}$.}

The symmetry group $G$ of the $L_d$-periodic, perfect beam--foundation system with energy density given by Eq.~\eqref{eq:energy}, is the infinite, compact group $G = D_{\infty h}$, a faithful representation on $H$ of which is generated by the following three linear operators:
\begin{itemize}
\item Phase-shift $c(\phi)$ by phase angle $\phi \in [-\pi,\pi)$:
\begin{align*}
T_{c(\phi)} :\; H &\rightarrow H,\quad
w(x) \mapsto T_{c(\phi)}[w(x)] := w(x + \mathfrak{c}(\phi)),
\end{align*}
where $\mathfrak{c}(\phi) := \phi  {L_d} / {2\pi} \in [-L_d/2,L_d/2)$;
\item Reflection $\sigma_v$ through $x=0$:
\begin{align*}
T_{\sigma_v} :\; H &\rightarrow H,\quad
w(x) \mapsto T_{\sigma_v}[w(x)] := w(-x);
\end{align*}
\item Reflection $\sigma_h$ through $y=0$:
\begin{align*}
T_{\sigma_h} :\; H &\rightarrow H,\quad
w(x) \mapsto T_{\sigma_h}[w(x)] := -w(x).
\end{align*}
\end{itemize}
Indeed, for every element $g \in D_{\infty h}$, the energy density Eq.~\eqref{eq:energy} is invariant
\begin{equation}
\mathcal{E}(T_{g}[w(x)];\lambda) = \mathcal{E}(w(x); \lambda) ; \quad \forall w(x) \in H, \; \forall \lambda \in \mathbb{R}.
\label{eq:symmetry}
\end{equation}

The {\it fixed-point space} $\mathcal{S}_{ D_{\infty h}} := \{w(x) \in H \;|\; T_{g}[w(x)] = w(x), \; \forall g \in D_{\infty h}\}$ of this compact group consists of all configurations $w(x)$ of the infinite, $L_d$-periodic beam that remain unaltered by the action of the group.  It is easy to see that $\mathcal{S}_{D_{\infty h}}$ contains only one element: the beam's reference (straight) configuration $w(x) = 0$.  Along with the equivariance property Eq.~\eqref{eq:equivariance} of the equilibrium Eqs.~\eqref{eq:equilibrium} (which is inherited from Eq.~\eqref{eq:symmetry}), this implies that $\overset{0}{w}(x; \lambda) = 0$ is the \emph{principal solution}---also termed \emph{fundamental solution}---for the beam--foundation system.

\begin{figure*}
\begin{center}
\begin{tabular}{ccm{0.75\textwidth}}
(a) & $D_{4d}$ & \includegraphics[width=0.75\textwidth]{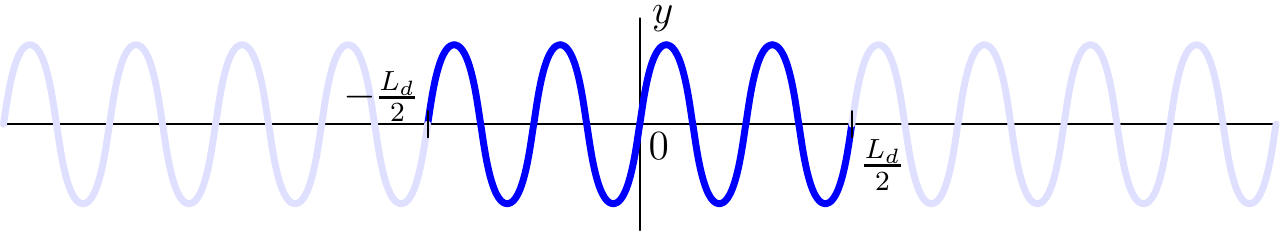}\\
(b) & $D_{4d}$ & \includegraphics[width=0.75\textwidth]{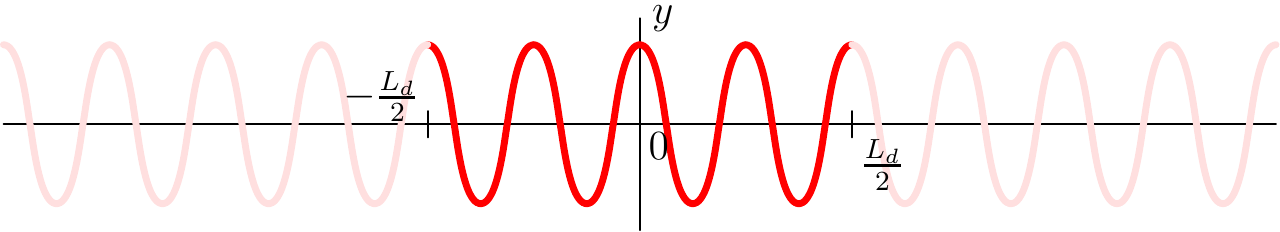}\\
(c) & $D_{4}$  & \includegraphics[width=0.75\textwidth]{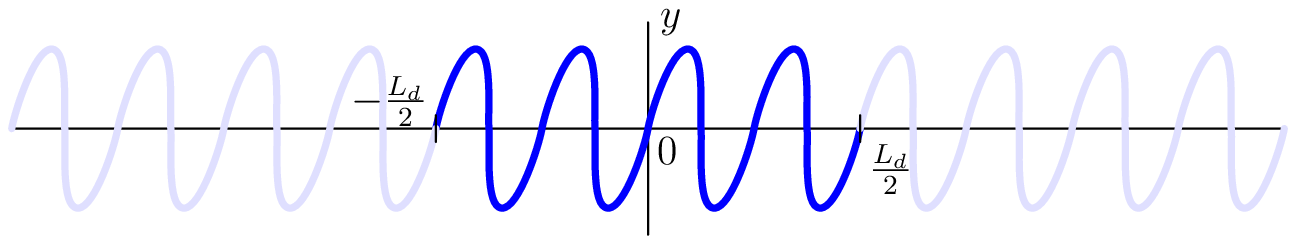}\\
(d) & $C_{4v}$ & \includegraphics[width=0.75\textwidth]{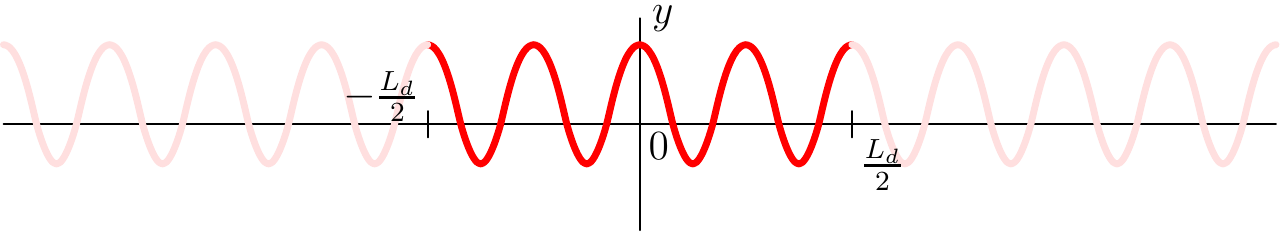}\\
(e) & $D_{2}$  & \includegraphics[width=0.75\textwidth]{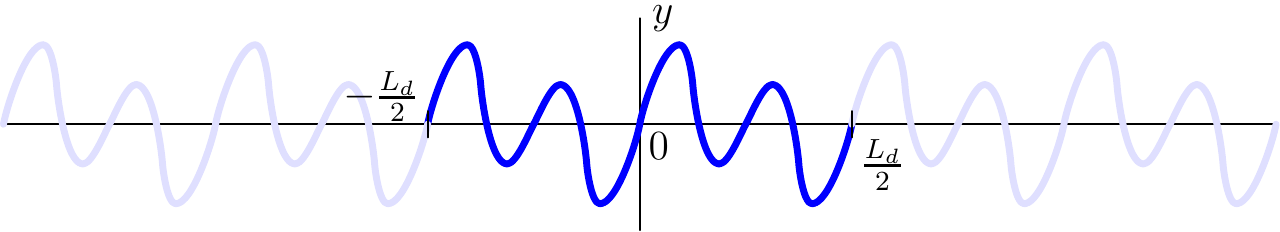}\\
(f) & $C_{2v}$ & \includegraphics[width=0.75\textwidth]{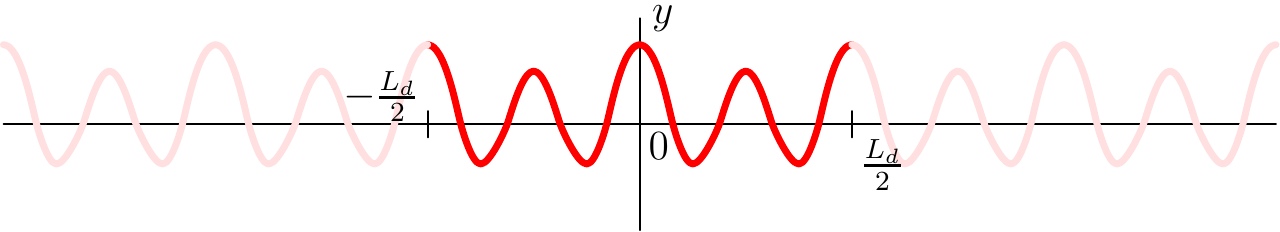}\\
(g) & $D_{1}$  & \includegraphics[width=0.75\textwidth]{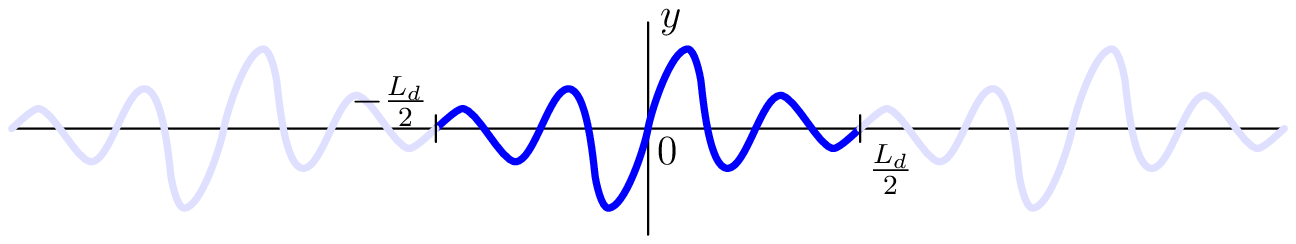}\\
(h) & $C_{1v}$ & \includegraphics[width=0.75\textwidth]{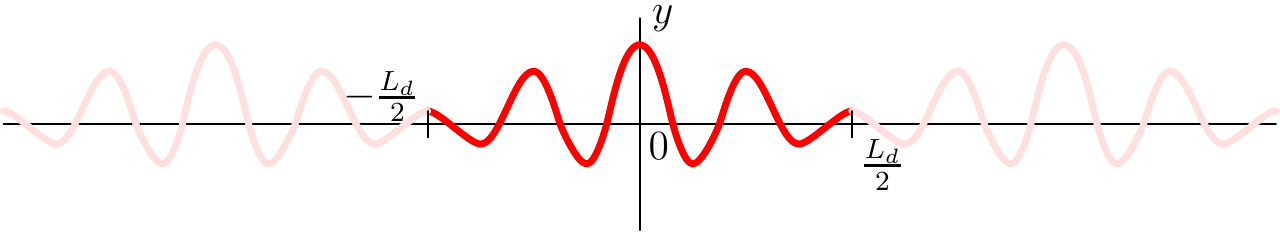}
\end{tabular}
\end{center}
\caption{$L_d$-periodic orbit configurations belonging to different symmetry groups.  Here $L_d = L_c q$, and $q=4$ is selected for illustrative purposes.  Subfigures~(a) and (b) present two members of a $D_{\infty h}$ orbit with $D_{4d}$ symmetry.  Subfigures~(c)--(h) present examples of configurations with symmetries that are subgroups of $D_{4d}$: (c) $D_4$, (d) $C_{4v}$, (e) $D_{2}$, (f) $C_{2v}$, (g) $D_1$, and (h) $C_{1v}$.}
\label{Fig:symmetric-configs}
\end{figure*}

In Sec.~\ref{sec:Section_4} and Appendix~\ref{appendix-A} we show that for this system the primary bifurcation orbits have symmetry group $D_{qd}$ (containing $4q$ elements) which is generated by the two symmetry elements $T_{\sigma_h c(\pi/q)}$ and $T_{\sigma_h \sigma_v}$ corresponding to \emph{phase-shift plus horizontal reflection}, and \emph{inversion through $x=0$}, respectively.  Note that $D_{qd}$ is a finite subgroup of $D_{\infty h}$.  A sample of $L_d$-periodic orbit configurations belonging to different symmetry groups is depicted in Fig.~\ref{Fig:symmetric-configs} for the special case $L_d = L_c q$, $q = 4$, with Fig.~\ref{Fig:symmetric-configs}(a) showing an element with $D_{4d}$ symmetry.
There are four inversions (through $x=0, L_d/8, L_d/4, 3L_d/8$; the inversions at $x=-L_d/2, -3L_d/8, -L_d/4$, respectively, are equivalent to the first four due to the $L_d$-periodicity of $w(x)$); four vertical mirrors (through $x = L_d/16, 3L_c/16, 5L_d/16, 7L_d/16$; the vertical mirrors at $x=-7L_d/16, -5L_d/16, -3L_d/16, -L_d/16$, respectively, are equivalent to the first four due to the $L_d$-periodicity of $w(x)$); and there are four phase-shifts ($\mathfrak{c} = -L_d/2, -L_d/4, 0, L_d/4$).  It is worth noting that the inversion through $x=0$ implies that $w(x) = -w(-x)$, i.e.\ $w(x)$ is an odd function.  Given a configuration $v(x) \in H$ there is an associated orbit of configurations $\{ w(x) \;|\; w(x) = T_g[v(x)] \text{ for some } g\in D_{\infty h}\}$.  Fig.~\ref{Fig:symmetric-configs}(b) shows an example of a member of the orbit of the configuration of Fig.~\ref{Fig:symmetric-configs}(a).  This configuration also has $D_{4d}$ symmetry\footnote{Actually, its symmetry is a subgroup of $D_{\infty h}$ that is \emph{conjugate} to $D_{4d}$.  Here, we will not be concerned with this distinction.}, and is obtained by a phase-shift of $\mathfrak{c} = L_d/16$.  Bifurcation from a primary orbit with $D_{qd}$ symmetry will break symmetries and result in bifurcating orbits with symmetry groups that are subgroups of $D_{qd}$.

In Sec.~\ref{sec:Section_4} and Appendix~\ref{appendix-A} we find secondary bifurcation orbits, bifurcating from the primary orbit with $D_{qd}$ symmetry, which have either $C_{rv}$ or $D_{r}$ symmetry (each containing $2r$ elements), where $r$ is a divisor of $q$.  Configurations with $C_{rv}$ symmetry have phase-shift and vertical mirror symmetries, whereas configurations with $D_r$ symmetry have phase-shift and inversion symmetries.  Figs.~\ref{Fig:symmetric-configs}(c)--(h) show examples of configurations with symmetry groups $D_4$, $C_{4v}$, $D_2$, $C_{2v}$, $D_1$, and $C_{1v}$, respectively.  Notice that, as presented here, configurations with $D_{r}$ symmetry are {\it odd}: $w(x) = -w(-x)$, and those with $C_{rv}$ symmetry are {\it even}: $w(x) = w(-x)$.  We refer to Appendix~\ref{appendix-A} for details.


\subsection{Stability}
\label{sec:stability}

To determine the stability of an equilibrium solution $(w(x);\lambda)$, one has to check the positive definiteness of the self-adjoint bilinear operator ${\mathcal E}_{,ww}$, evaluated at the investigated solution $(w(x);\lambda)$, by finding its eigenvalues $\beta$
\begin{equation}
({\mathcal E}_{,ww}\Delta w)\delta w = \beta <\Delta w, \delta w> \; ; \quad \forall \delta w \in H\;  ;
\ {\mathcal E}_{,ww} := {\mathcal E}_{,ww}(w(x);\lambda) \; ,
\label{eq:stab_oper}
\end{equation}
where $\Delta w$ is the corresponding eigenmode and $< \cdot \; ,\; \cdot >$ denotes an inner product in $H$.  A stable solution corresponds to a positive minimum eigenvalue\footnote{\label{foot:zero:eigval}As mentioned above, in the problem at hand, we find continuous orbits of equilibria, all having the same energy.  This implies the existence of a zero eigenvalue of the stability operator $\mathcal{E}_{,ww}$ in Eq.~\eqref{eq:stab_oper}.  Thus, all equilibria are, at best, neutrally stable.  Accordingly, in this work we ignore the zero eigenvalue associated with the solution orbit and require that \emph{all other} eigenvalues be positive for stability.} $\beta_{min} > 0$ (the number of eigenvalues depends on the dimension of $H$).  For the stability of periodic solutions (of period $L_d$) one can take advantage of the Bloch-wave representation theorem, according to which any eigenmode $\Delta w$ of the stability operator ${\mathcal E}_{,ww}$ in \eqref{eq:stab_oper} admits the following representation
\begin{equation}
\Delta w (x) = \exp (i 2 \pi k x/L_d)  \; p(x) \; ,
\label{eq:Bloch_mode}
\end{equation}
where $i=\sqrt{-1}$ is the imaginary unit, $p(x)$ is $L_d$-periodic, and $k \in [0, 1)$ is the wavenumber.  Thus, the Bloch-wave representation reduces the eigenvalue problem Eq.~\eqref{eq:stab_oper} to a set of smaller dimensional ones (one such problem for each value of the wavenumber). By scanning all admissible values of $k$, one can find $\beta_{min}(k)$ for each value of $k$ and ultimately obtain
$\beta_{min} := {\displaystyle \min_{k}}(\beta_{min}(k))$.

For a well-posed problem, its stress-free (unloaded) configuration at $\lambda=0$ is stable; as the load increases stability will be lost at the first bifurcation point encountered along the loading path at some $\lambda_b$.


\section{Numerical Method}
\label{sec:Section_3}

The Finite Element Method (FEM) is used for the numerical solution of the boundary value problem defined by Eqs.~\eqref{eq:equilibrium},~\eqref{eq:boundary}. The calculation domain $[-L_d /2,\ L_d /2]$ is divided in $N_d$ equal length elements.  A cubic Hermite interpolation is used to approximate the displacement $w(x)$ within each two-node element.  The unknown degrees of freedom ${\bf q}$ are the values of the displacement and its derivative, i.e.\ ${\bf q}(x)=(w(x),dw/dx(x))$.  The discretization procedure results in a $2(N_d + 1)$ nonlinear system of algebraic equations for the same number of unknowns.  A three point Gauss integration scheme\footnote{This is adequate for correct integration of the higher-order gradient term in the stiffness matrix.} was found satisfactory for the numerical calculations, in view of the fine meshes required for the highly localized deformation solutions.  Element sizes of length $L_d/400$ are used in the reported calculations; smaller elements were also tested but made no difference for the required accuracy of the results.

Given that we are seeking solutions in $H_{L_d}^2 (\mathbb{R})$ approximated by Hermite cubics, we cannot apply Eq.~\eqref{eq:boundary} for lack of adequate continuity.  From group theoretic considerations (see Section~\ref{sec:symmetry} and Appendix~\ref{appendix-A}), we observe that the deformations on the primary bifurcation orbit should satisfy\footnote{As described in Sect.~\ref{sec:symmetry}, the primary bifurcation orbit consists of an infinite set of configurations generated by the symmetries of $D_{\infty h}$; here we select one specific representative of the orbit.} $w(x)=-w(-x)$ corresponding to the $D_{qd}$ symmetry group, whereas on the secondary bifurcation orbits they should satisfy $w(x)=w(-x)$ and $w(x)=-w(-x)$ corresponding to $C_{rv}$ and $D_{r}$ symmetry groups, respectively.  These conditions enable the application of the following essential boundary conditions for the finite element model, while the work conjugate natural boundary conditions remain unconstrained\footnote{The higher-order derivatives that must vanish by symmetry are thus only approximately satisfied.}:
\\

\indent {\it Primary bifurcated equilibrium orbits:}
\begin{equation}
\begin{aligned}
w(-{L_d/2}) &= w({L_d/2})=0,\\
\frac{d w}{d x}(-L_d/2) &= \frac{d w}{d x}(L_d/2),
\end{aligned}
\qquad
(D_{qd}\text{ symmetry});
\label{eq:primary_BC}
\end{equation}
\indent {\it Secondary bifurcated equilibrium orbits:}
\begin{equation}
\begin{gathered}
\begin{aligned}
w(-{L_d/2}) &= w({L_d/2}),\\
\frac{d w}{d x} (-{L_d/2}) &= \frac{d w}{d x}({L_d/2}) = 0,
\end{aligned}
\qquad
(C_{rv}\text{ symmetry});\\
\\
\begin{aligned}
w(-{L_d/2}) &= w({L_d/2})=0,\\
\frac{d w}{d x}(-L_d/2) &= \frac{d w}{d x}(L_d/2),
\end{aligned}
\qquad
(D_{r}\text{ symmetry}).
\end{gathered}
\label{eq:secondary_BC}
\end{equation}
A standard Newton--Raphson method is employed to solve the nonlinear system with an arc-length continuation method \citep[see][]{Keller} to handle local extrema in the dimensionless axial load $\lambda$ or its work-conjugate (average) strain $\Delta$.

Dispersion relations for an equilibrium path of period $L_d$ are based on examination of the eigenvalues of the stiffness matrix (discretized second derivative of the energy density $\mathcal{E}_{,ww}$) appropriately modified by using the Bloch-wave theorem, which couples the degrees of freedom $\bf q$ at the two ends of the domain with the help of the dimensionless wavenumber $k$
\begin{equation}
{\bf q}(L_d/2) = \exp(i 2\pi k)\; {\bf q}(-L_d/2) \;  ; \quad k \in [0, 1).
\label{eq:blochboundary}
\end{equation}

The resulting complex, Hermitian stiffness matrix has real eigenvalues. As discussed in Section~\ref{sec:stability}, stability of the equilibrium path at hand is guaranteed only if the minimum eigenvalue of this stiffness matrix is positive for all values of the wavenumber $k$, i.e.\ $\beta_{min}(k) \ > 0, \ k \in [0,1)$.

Additional information is available from the stiffness matrix of an equilibrium path: bifurcation points appearing on that path are obtained by examining the vanishing of the real eigenvalues of the path's stiffness matrix, while the corresponding eigenmodes are used to start the emerging higher-order bifurcated equilibrium paths.


\section{Results}
\label{sec:Section_4}

In this section we present the principal solution, the primary and higher-order bifurcated equilibrium orbits for the perfect structure, the equilibrium orbits for the imperfect structure, as well as their stability for the nonlinear beam--foundation model described in Section~\ref{sec:Section_2}. Solid and dashed lines in figures correspond, respectively, to stable and unstable solutions, based on Bloch-wave analysis of the largest supercell, here taken to be $L_d = L_c q = 40\pi$.


\subsection{Perfect Model}
\label{sec:PerfectModel}

Here we analyze the equilibrium solutions of the perfect model described in Subsection~\ref{sec:perfect}, starting with the principal solution for the initial (and largest) symmetry group $D_{\infty h}$.  We study its stability and find the continuous orbit of the primary bifurcation paths, emerging from the lowest critical load, which has a $D_{qd}$ symmetry group. We subsequently follow the emerging secondary bifurcation branches of even lower symmetry groups away from the primary branch. It should be noted at this point that due to symmetry, several equilibrium paths emerge from each bifurcation point, usually belonging to different orbits.  For the graphical projections employed here, only different orbits result in different graphs. Moreover, by abuse of language we use the word ``{\it equilibrium path}'' for any path belonging to the same orbit.


\subsubsection{Principal Solution and its Stability}
\label{sec:perfect:princ}

As discussed in Subsection~\ref{sec:symmetry}, the unique element of the model's fixed-point space $\mathcal{S}_{ D_{\infty h}}$ is ${\stackrel{0}{w}}(x; \lambda) = 0$, the trivial principal solution of the system given by Eqs.~\eqref{eq:equilibrium},~\eqref{eq:boundary}.  To investigate its stability, i.e.\ whether it corresponds to a local energy minimizer, we must find the minimum eigenvalue $\beta_{min}(\lambda)$ of the corresponding stability operator $\mathcal{E}^0_{,ww} := \mathcal{E}_{,ww}({\stackrel{0}{w}}(x;\lambda);\lambda)$ evaluated on the principal solution, as indicated in Eq.~\eqref{eq:stab_oper}.  The minimum eigenvalue $\beta_{min}(\lambda)$ can be found by minimizing the quadratic form $(\mathcal{E}^0_{,ww} \delta w)\delta w$ over all admissible $\delta w$ with a unit norm
\begin{equation}
\beta_{min}(\lambda) = \min_{\delta w} \left\{(\mathcal{E}^0_{,ww}\delta w)\delta w \right\},
\quad  <\delta w, \delta w>  = \parallel \delta w \parallel ^2 = 1 \; .
\label{eq:raleighquotient}
\end{equation}
Using the Fourier series representation\footnote{Note, as written it is necessary to take $\delta w_{s0} := 0$.} of $\delta w(x)$
\begin{equation}
\delta w = \sum_{n=0}^\infty \left[ \delta w_{cn} \cos \Big(\frac{2n \pi}{L_d} x \Big) +
\delta w_{sn} \sin \Big(\frac{2n \pi}{L_d} x \Big) \right] \; ,
\label{eq:deltawfourier}
\end{equation}
we obtain by substituting Eq.~\eqref{eq:deltawfourier} into Eq.~\eqref{eq:raleighquotient}
\begin{equation}
\beta_{min}(\lambda) = \min_{\parallel \delta w \parallel =1} \left\{ (\delta w_{c0})^2 + \sum_{n=1}^\infty {\frac{1}{2}}\Big[ \Big(\Big(\frac{2n \pi}{L_d}\Big)^2-1\Big)^2+ (2-\lambda) \Big(\frac{2n \pi}{L_d}\Big)^2\Big] \Big[(\delta w_{cn})^2+ (\delta w_{sn})^2\Big] \right\} \; .
\label{eq:primary-bmin}
\end{equation}

A simple inspection of Eq.~\eqref{eq:primary-bmin} reveals that $\beta_{min}(\lambda) > 0$ for $0 \leq \lambda < 2$ and hence the principal solution $\stackrel{0}{w}$ is always stable for $0 \leq \lambda < 2$. The critical bifurcation, i.e.\ the one corresponding to the lowest possible root of $\beta_{min}(\lambda)=0$, occurs at $\lambda = \lambda_c = 2$. Moreover, taking $L_c = 2\pi$ and $L_d = L_c q,\; q \in \mathbb N$ where $q \gg 1$, one can see that this bifurcation corresponds to a double eigenvalue with eigenmodes
\begin{equation}
\lambda_c = 2\; ,\ L_c = 2\pi\; ,\ n_c = q \; ; \quad {\stackrel{c1}{w}}(x) = \sin(x)\; ,\ {\stackrel{c2}{w}}(x) = \cos(x) \; .
\label{eq:critical}
\end{equation}

Additional, double bifurcations can be found from Eq.~\eqref{eq:stab_oper} corresponding to the next higher roots of $\beta(\lambda_n) = 0$ with $\lambda_n > \lambda_c = 2$, $L_n = (q/n)L_c$ and eigenmodes given by
\begin{equation}
\lambda_{n} =  \left[ \left(\frac{n}{q}\right)^2 + \left(\frac{q}{n}\right)^2\right] \; ,\ q \neq n \in \mathbb N\; ; \quad
{\stackrel{n1}{w}}(x) = \sin\left(\frac{n}{q} x \right)\; ,\ {\stackrel{n2}{w}}(x) = \cos\left(\frac{n}{q} x \right) \; .
\label{eq:highercritical}
\end{equation}
Each $\lambda_n$ corresponds to a bifurcation point, as guaranteed by the transversality conditions \citep[see][]{Iooss_Joseph}
\begin{equation}
\mathcal{E}^n_{,w \lambda}{\stackrel{ni}{w}}=0, \quad  \; i = 1,2 ; \quad\text{and}\quad
\det\left[\left( \left[ {\frac{d\mathcal{E}^0_{,ww}}{d\lambda}} \right]_{\lambda=\lambda_n}  {\stackrel{ni}{w}} \right) {\stackrel{nj}{w}} \right] \neq 0 \; ,
\label{eq:transversality}
\end{equation}
where the superscript $n$ denotes evaluation of the quantity at hand at a point $ ({\stackrel{0}{w}},\; \lambda_n)$.  These results are consistent with group theory, as discussed in Appendix~\ref{appendix-A}.

\subsubsection{Primary Bifurcation Orbit and its Stability}
\label{sec:Stability}

\begin{figure*}
\centering
\includegraphics[width=0.5\textwidth]{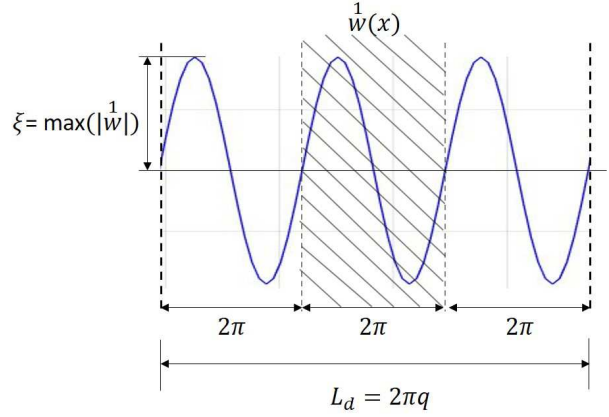}
\caption{A primary bifurcated equilibrium path ${\stackrel{1}{w}}(x;\xi)$ is a $2\pi$-periodic function with extremum values $\pm \xi$, where $\xi := \max_{-L_d/2 \le x \le L_d/2}{| {\stackrel{1}{w}}(x) |}$ is the bifurcation amplitude parameter. It is a representative of the continuous orbit of the $D_{qd}$-symmetric bifurcated paths emerging at $\lambda_c=2$.}
\label{Fig:primary-w}
\end{figure*}

We focus next on the bifurcated equilibrium paths which are the {\it uniformly wrinkled} periodic configurations of the beam--foundation system.  These paths emerge from the lowest critical load $\lambda_c$, where---according to Eqs.~\eqref{eq:critical} and \eqref{eq:transversality}---a double, pitchfork bifurcation occurs (see Appendix~\ref{appendix-A}).  At this point a continuous orbit of bifurcated equilibrium paths can be constructed starting from any linear combination of the eigenmodes: $a \cos(x) + b \sin(x)$.  A representative element of this orbit that also belongs to the fixed-point space ${\mathcal{S}}_{D_{qd}}$ (see Appendix~\ref{appendix-A}), is the $2\pi$-periodic solution ${\stackrel{1}{w}}(x;\xi)$, plotted in Fig.~\ref{Fig:primary-w}, and parameterized using the bifurcation amplitude $\xi := \max_{-L_d/2 \le x \le L_d/2}{| {\stackrel{1}{w}}(x) |}$.
It is obtained by solving numerically the Euler--Lagrange Eq.~\eqref{eq:equilibrium} subjected to the boundary conditions Eq.~\eqref{eq:primary_BC} for the full nonlinear problem in the interval $(-L_d/2, L_d/2)$, with $L_d = 2\pi$ in this case.  The reason for the $\xi$ parameterization\footnote{In fact, even the employed parameterization is problematic, since it is restricted to $\xi \geq 0$, and is therefore unable to distinguish between the two ``halves'' of the bifurcated path.} is that the applied axial load $\lambda(\xi)$ is not a monotonic function of the bifurcation amplitude $\xi$.

\begin{figure*}
\begin{minipage}{0.53\textwidth}
\begin{center}
\includegraphics[width=\textwidth]{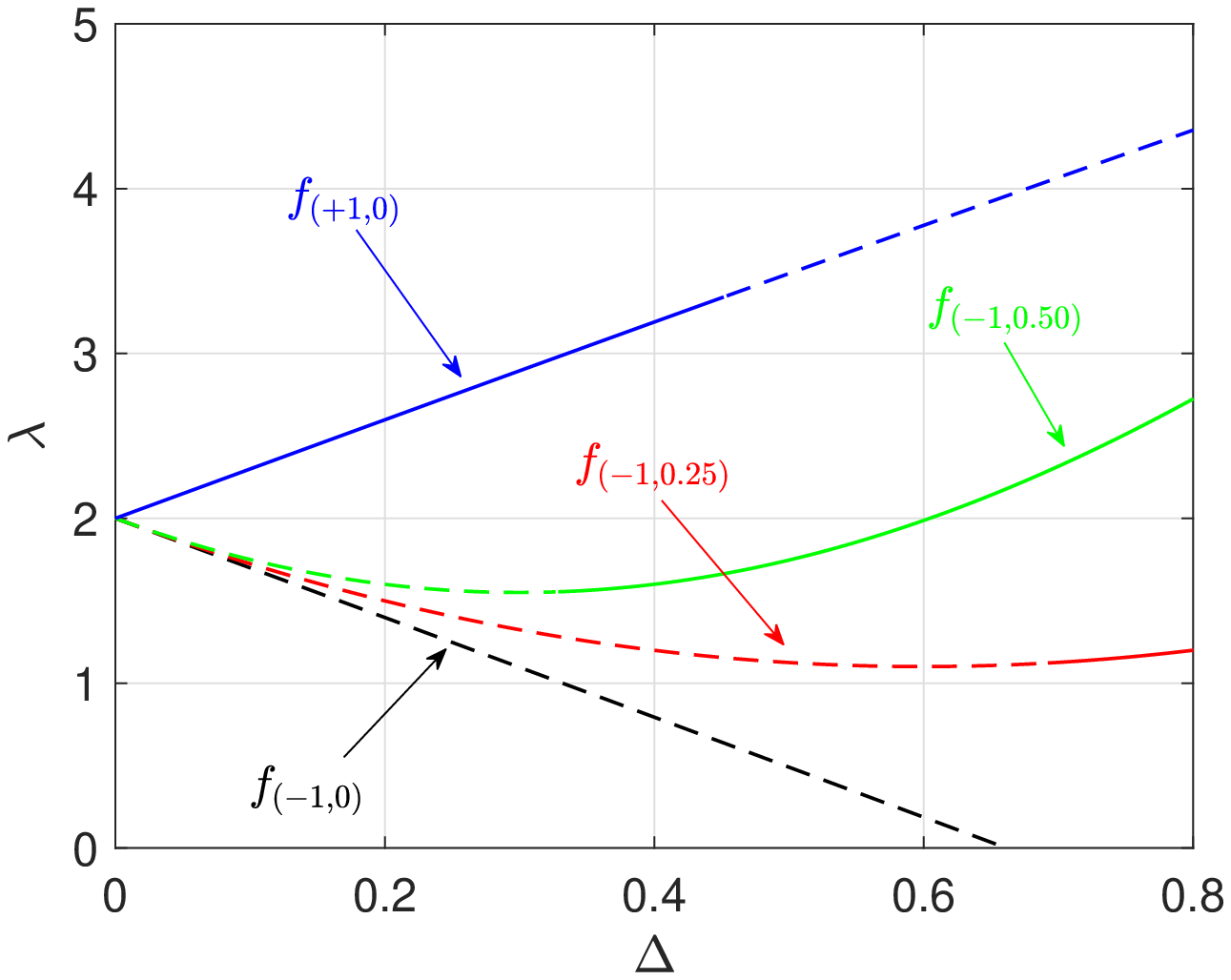}\\%
(a)\\%
\includegraphics[width=\textwidth]{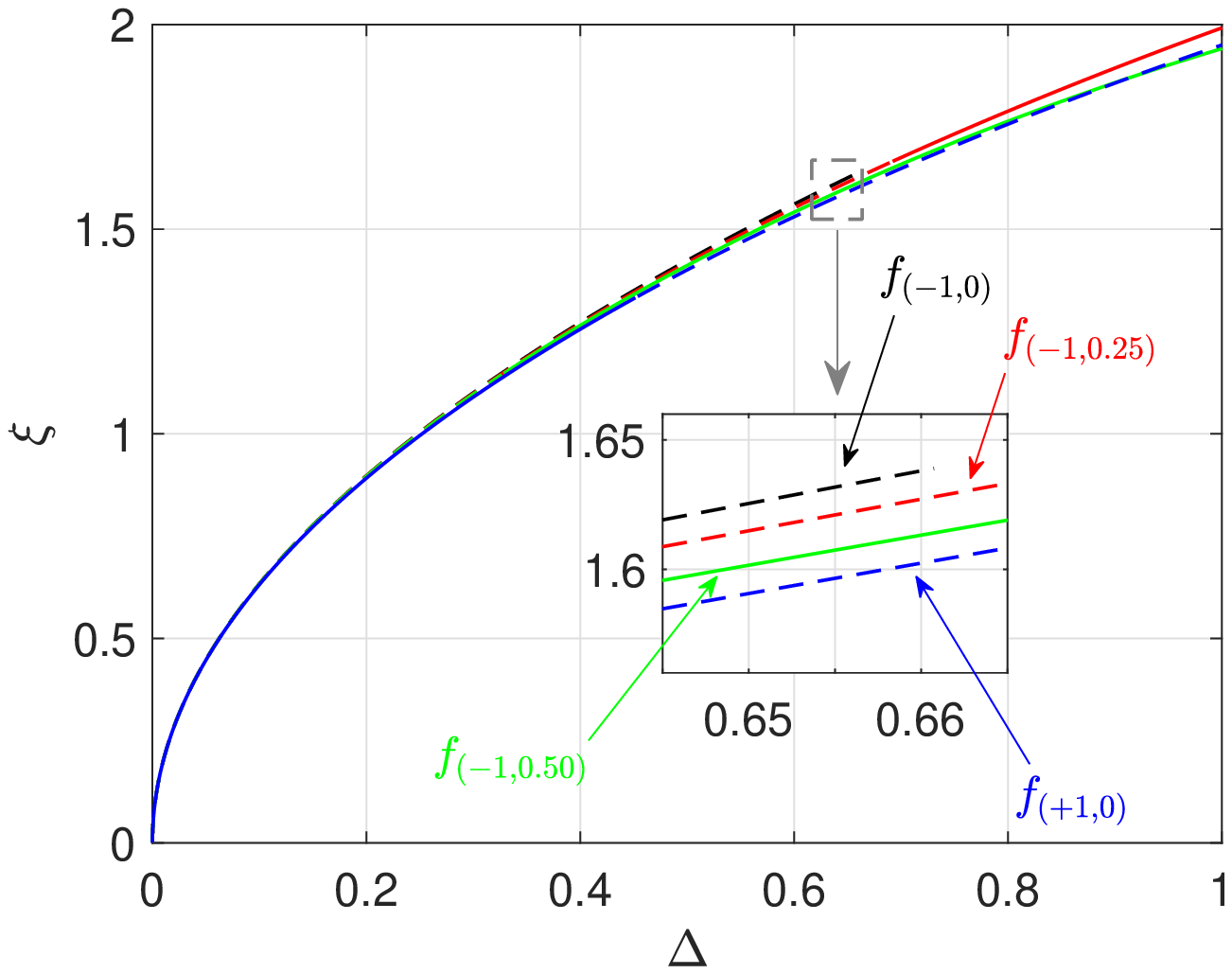}\\
(c)%
\end{center}
\end{minipage}%
\hspace*{-0.5cm}%
\begin{minipage}{0.53\textwidth}
\begin{center}
\includegraphics[width=\textwidth]{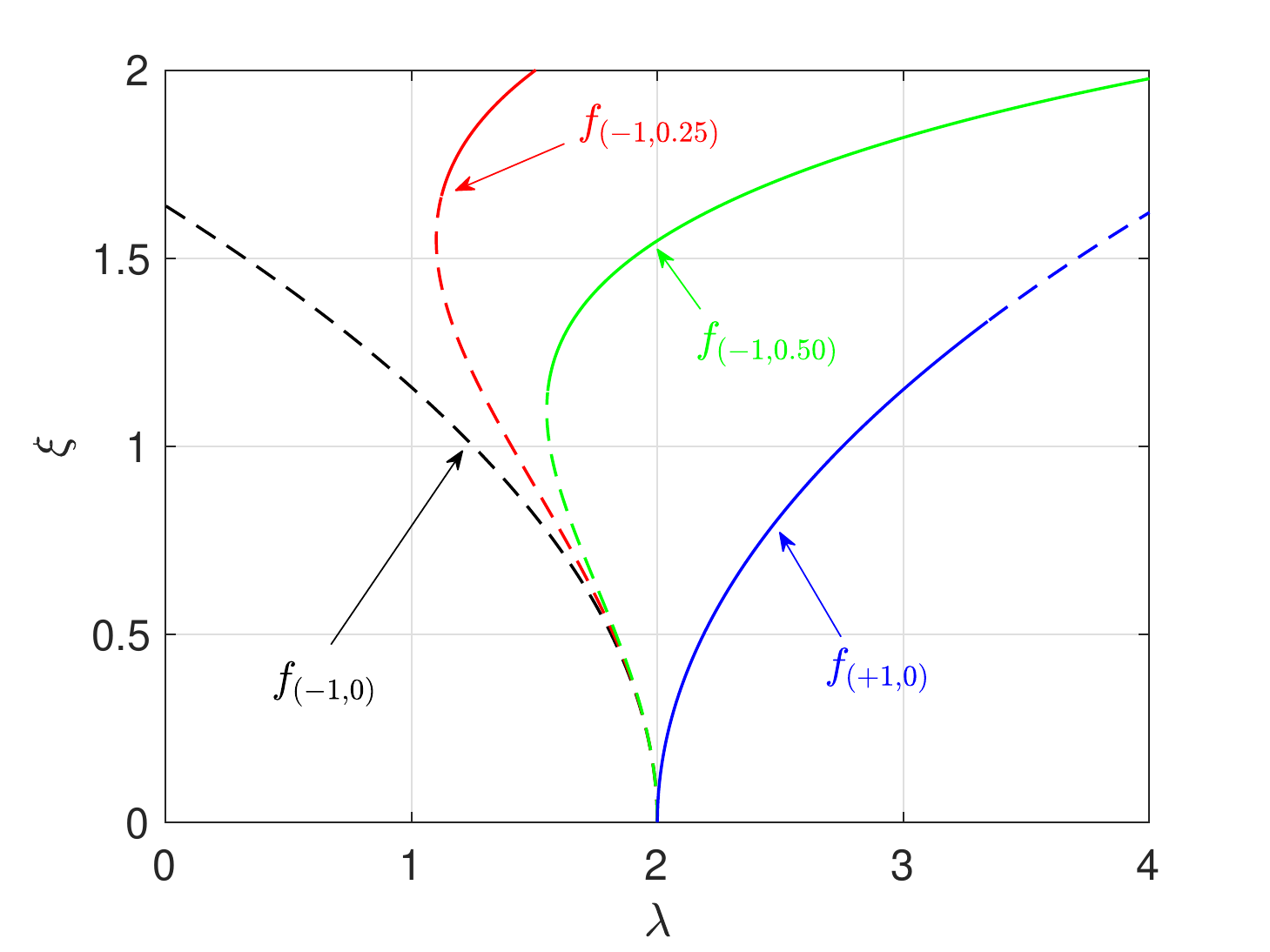}\\%
(b)\\%
\includegraphics[width=\textwidth]{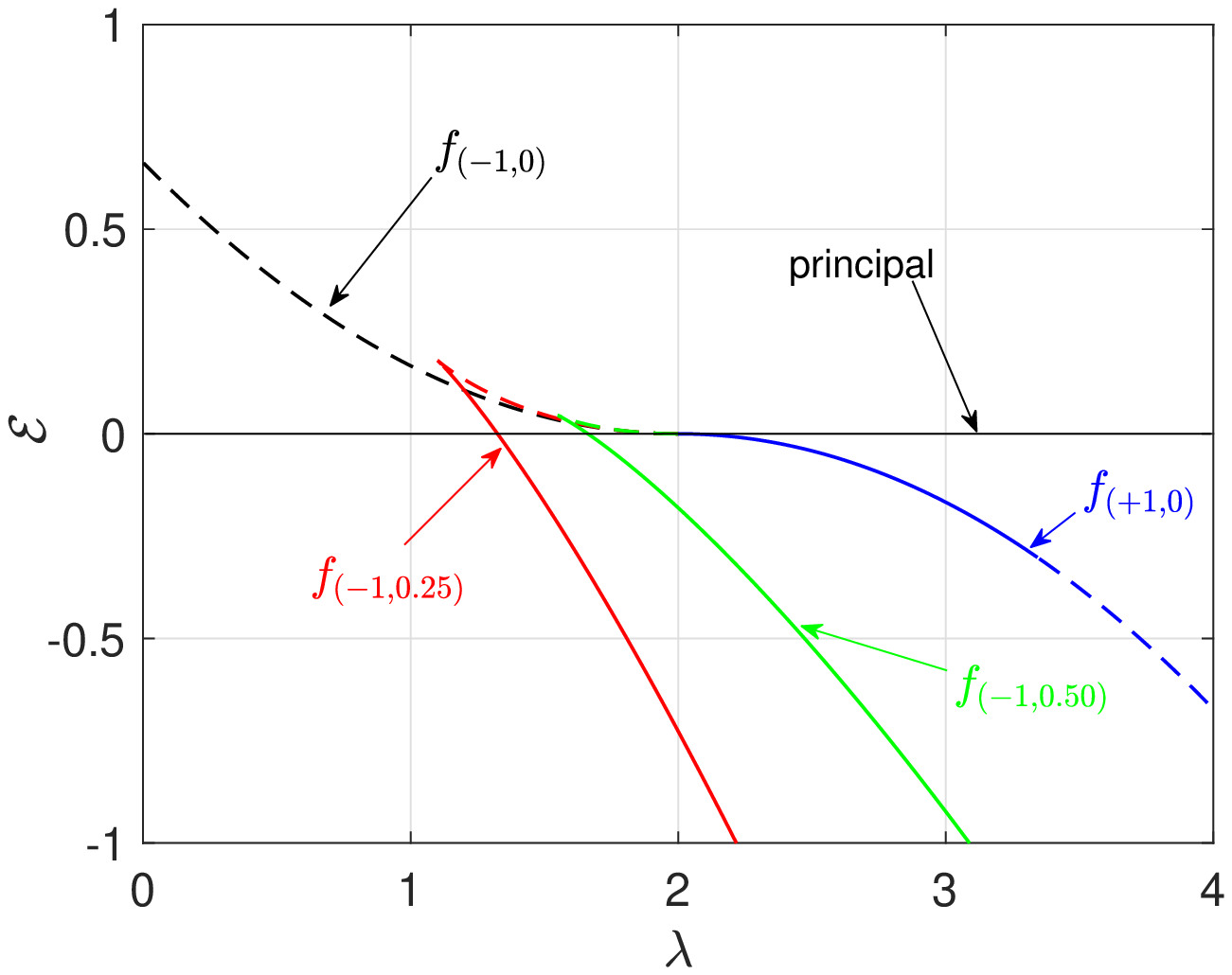}\\%
(d)%
\end{center}
\end{minipage}
\caption{Plots showing the emergence of the primary bifurcation orbit ${\stackrel{1}{w}}(x;\xi)$ from the principal solution (${\stackrel{0}{w}}(x; \lambda)=0$) of a perfect structure for different types of foundations. Bifurcation diagrams showing (a)~load $\lambda$ vs strain $\Delta$, (b)~amplitude $\xi$ vs load $\lambda$, (c)~amplitude $\xi$ vs strain $\Delta$, and (d)~energy density $\mathcal{E}$ vs load $\lambda$.  Solid and dashed lines correspond, respectively, to the stable and unstable parts of the primary bifurcation orbit, based on Bloch-wave analysis of a $L_d = 2\pi$ unit cell.}
\label{Fig:primary-diagrams}
\end{figure*}

The nonlinear foundation plays an important role on the primary bifurcated equilibrium orbit.  This is shown in Fig.~\ref{Fig:primary-diagrams}, which depicts the load $\lambda$ vs strain $\Delta$, bifurcation amplitude $\xi$ vs load $\lambda$, bifurcation amplitude $\xi$ vs strain $\Delta$, and energy density $\mathcal{E}$ vs load $\lambda$ for the continuous orbit of the primary bifurcation path ${\stackrel{1}{w}}(x;\xi)$.
Note that for the softening foundations $\alpha = -1$ the load $\lambda$ is initially a decreasing function of the strain $\Delta$ and the bifurcation amplitude $\xi$. This decrease is monotonic for the no re-hardening foundation $\gamma = 0$, but reverses itself for the mild ($\gamma = 0.25$) and strong ($\gamma = 0.5$) re-hardening foundations.  In contrast, for the hardening foundation $\alpha = + 1$, the load $\lambda$ is a monotonically increasing function of the strain $\Delta$ and of the bifurcation amplitude $\xi$.  These trends are shown in Fig.~\ref{Fig:primary-diagrams}(a) and Fig.~\ref{Fig:primary-diagrams}(b).  In Fig.~\ref{Fig:primary-diagrams}(c) one can also observe that the amplitude $\xi$ vs strain $\Delta$ relation is practically independent of the foundation, with only small differences emerging for large values of the strain.  The energy increases as we move along the primary bifurcation orbit for the softening foundations $\alpha = -1$, monotonically for $\gamma = 0$, but eventually reversing the trend for $\gamma = 0.25$ and $\gamma = 0.5$.  The energy decreases monotonically as we move along the primary bifurcation orbit for the hardening foundation $\alpha = +1$.  The energy trends can be observed in Fig.~\ref{Fig:primary-diagrams}(d).

The stability of the primary bifurcation orbit ${\stackrel{1}{w}}(x;\xi)$ is also recorded in Fig.~\ref{Fig:primary-diagrams}; solid and dashed lines correspond, respectively, to stable and unstable parts.  Notice that for the softening with no re-hardening foundation ($\alpha = -1, \gamma = 0$) the primary bifurcation orbit is always unstable.  For the re-hardening foundations ($\alpha = -1, \gamma \neq 0$), in the neighborhood of $\xi=0$, the primary bifurcation paths are unstable.  Stability is regained a little past the point where the load $\lambda$ reaches its minimum as a function of the strain $\Delta$ (the limit point).  Also at the limit point, the energy reaches its maximum value and starts decreasing as the bifurcation amplitude increases.  The energy trends are recorded in Fig.~\ref{Fig:primary-diagrams}(d).  The stability results for the hardening foundation ($\alpha = +1$) run opposite to the softening case: the primary bifurcated equilibrium orbit is stable in the neighborhood of $\xi=0$ but loses stability at about $\xi \approx 1.3$.

\begin{figure*}
\begin{center}
\includegraphics[width=0.55\textwidth]{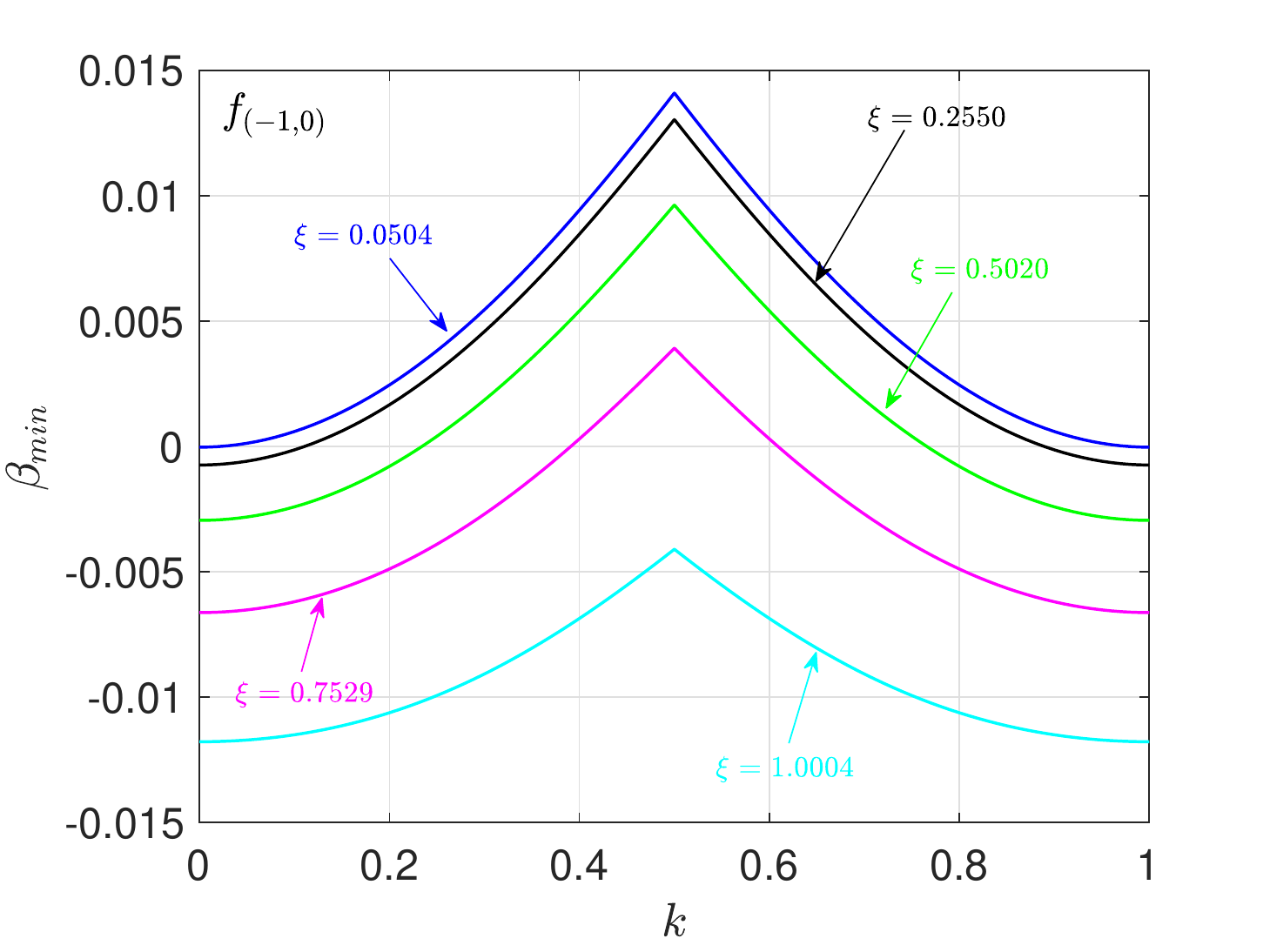}\\
(a)\\
\includegraphics[width=0.55\textwidth]{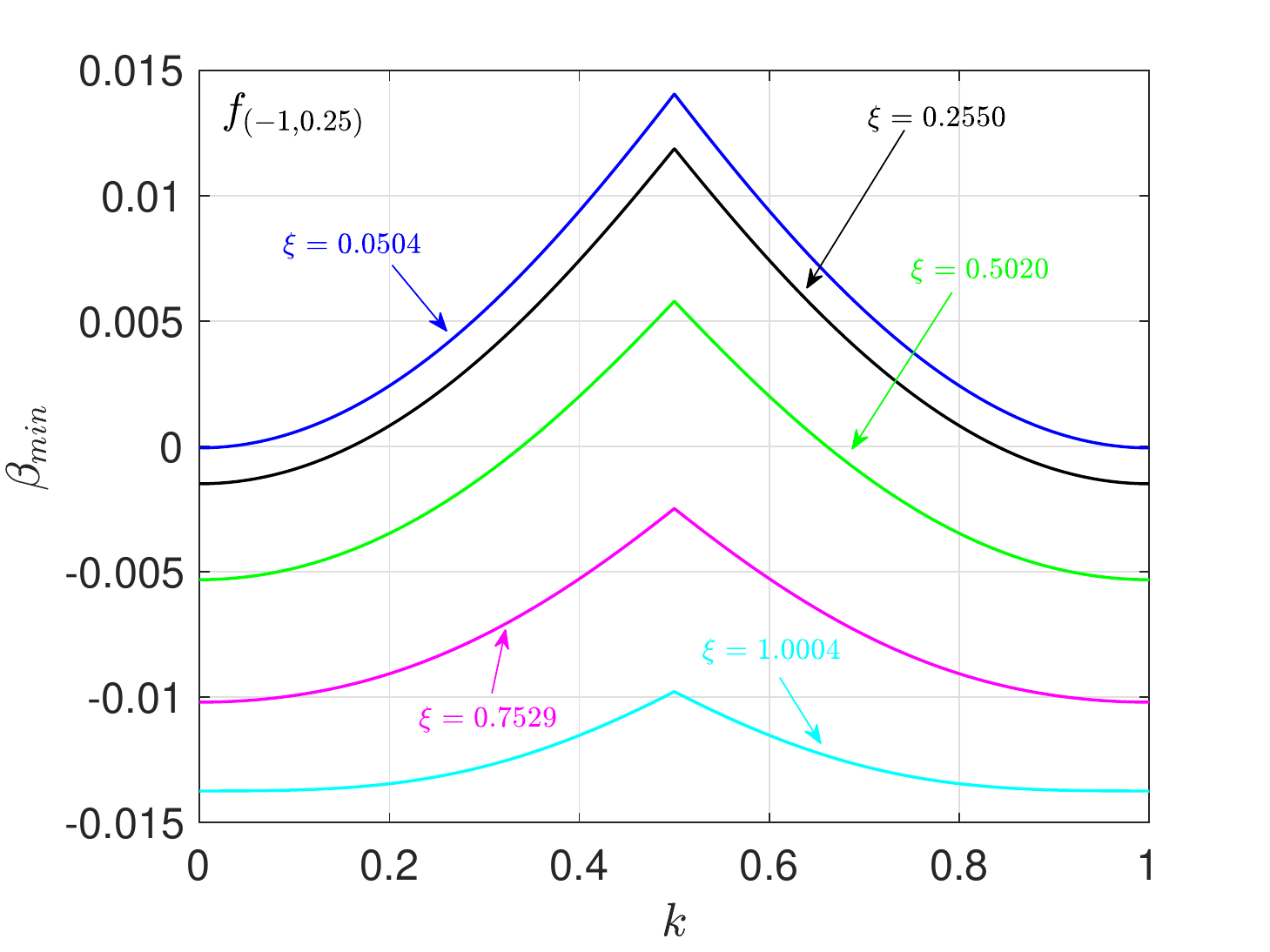}\\
(b)\\
\includegraphics[width=0.55\textwidth]{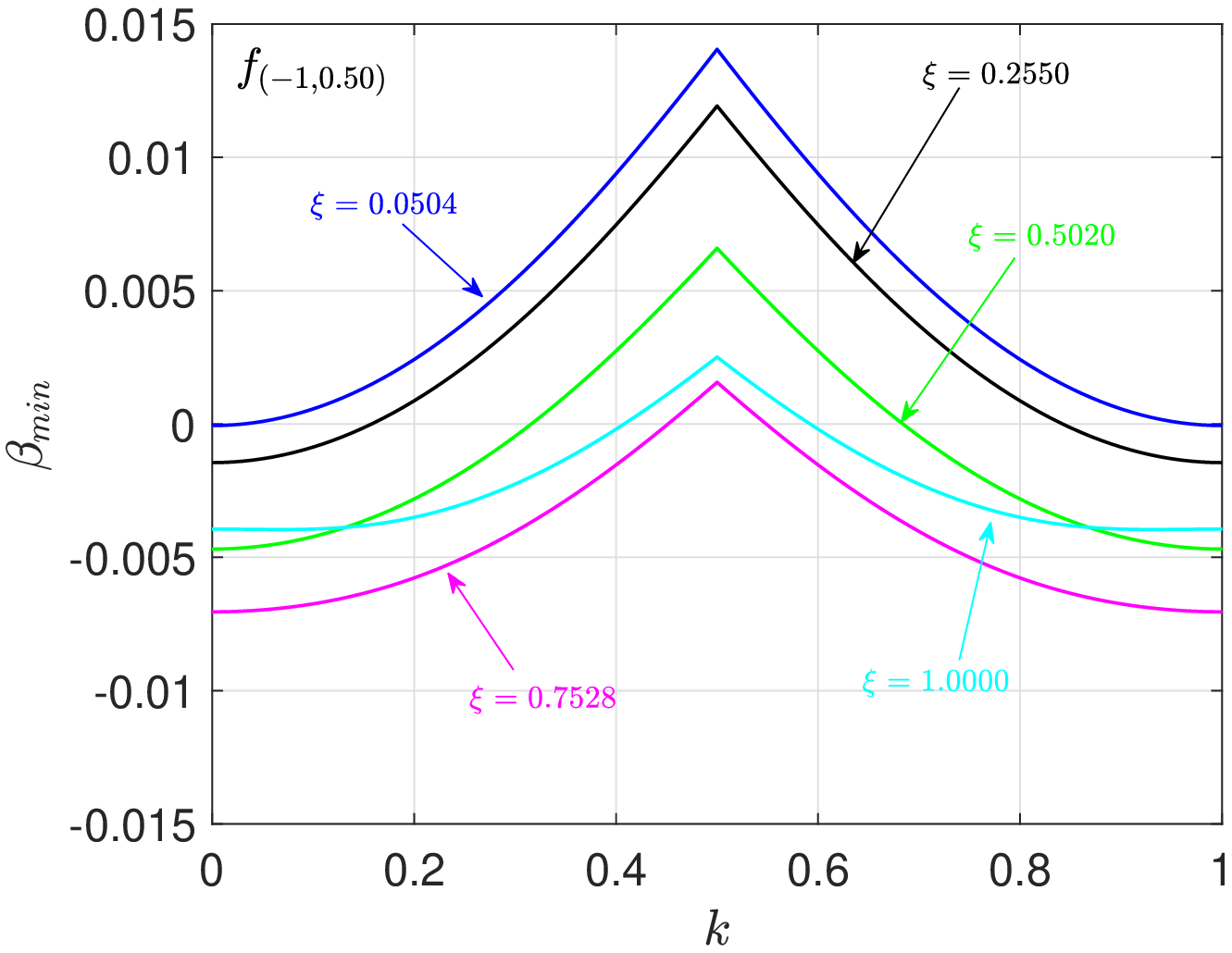}\\
(c)
\end{center}
\caption{Dispersion relations (minimum eigenvalue of the stability operator $\beta_{min}(k)$ vs wavenumber $k$) for the primary bifurcation paths of the three different soft foundation models: (a)~no re-hardening $(\alpha = -1,\ \gamma = 0)$, (b)~mild re-hardening $(\alpha = -1,\ \gamma = 0.25)$, and (c)~strong re-hardening $(\alpha = -1,\ \gamma = 0.50)$.}
\label{Fig:dispersion-soft}
\end{figure*}

\begin{figure*}
\centering
\includegraphics[width=0.55\textwidth]{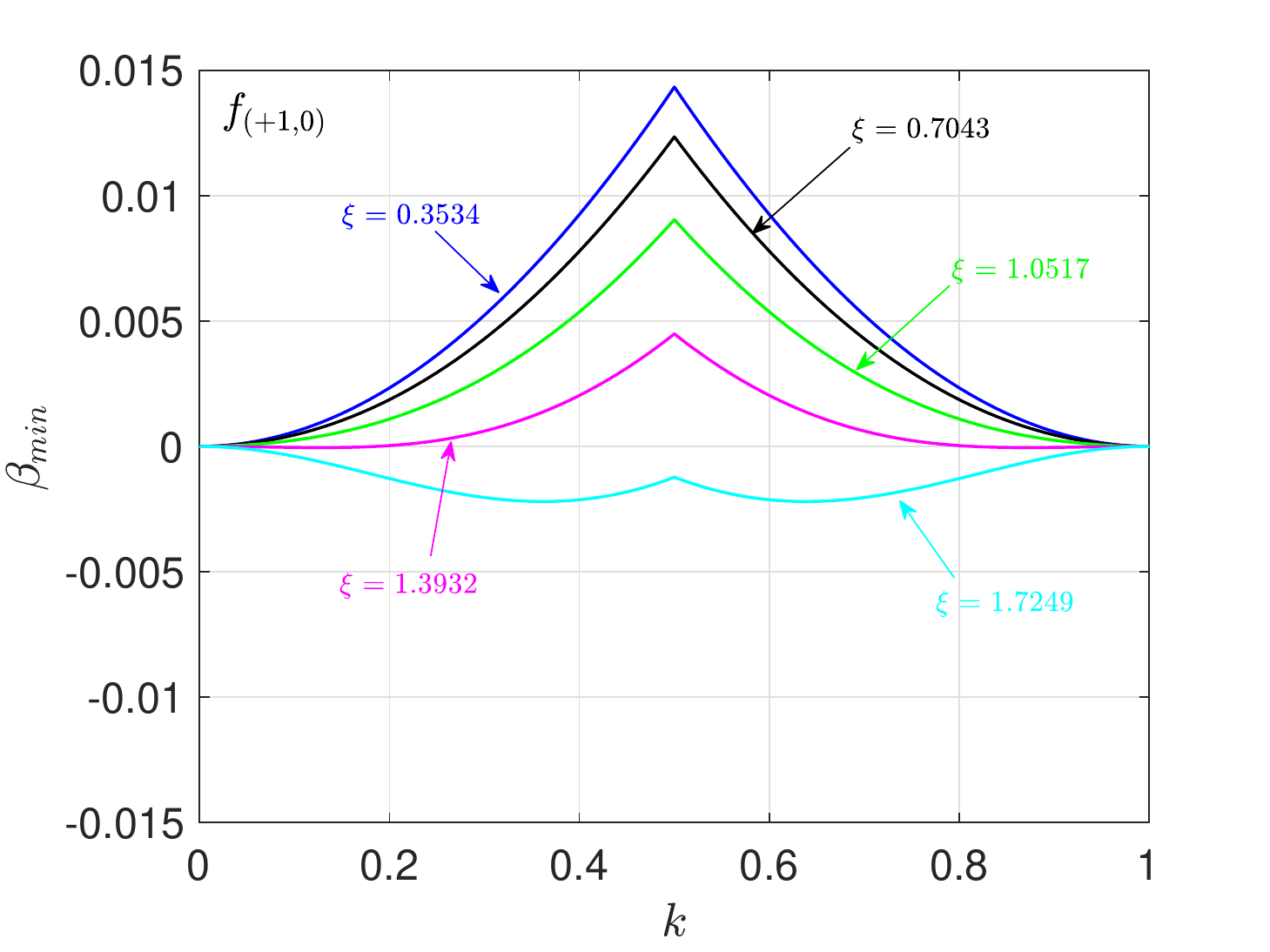}
\caption{Dispersion relations (minimum eigenvalue of the stability operator $\beta_{min}(k)$ vs wavenumber $k$) for the primary bifurcation path in the case of a hardening foundation ($\alpha = +1, \gamma = 0$).}
\label{Fig:dispersion-hard}
\end{figure*}

A more detailed stability study of the primary bifurcation orbit ${\stackrel{1}{w}}(x;\xi)$ is provided by the dispersion relation, giving its minimum eigenvalue $\beta_{min}(k)$ as a function of the wavenumber $k$ at different values of the bifurcation amplitude $\xi$.  Results for the softening foundations ($\alpha = -1$) are recorded in Fig.~\ref{Fig:dispersion-soft} and for the hardening foundation ($\alpha = +1$) in Fig.~\ref{Fig:dispersion-hard}.
Since the stability operator ${\mathcal E}_{,ww}$ is self-adjoint, the eigenvalues of the corresponding discretized Hermitian stiffness matrix are symmetric with respect to $k=0$, i.e.\ $\beta(k) = \beta(-k)$.  This property, combined with the periodicity of the eigenvalue, i.e.\  $\beta(k+1) = \beta(k)$, following from Eq.~\eqref{eq:blochboundary}, results in the mirror symmetry of the graphs in Fig.~\ref{Fig:dispersion-soft} and Fig.~\ref{Fig:dispersion-hard} with respect to $k=1/2$. Their intersection with the $\beta_{min}=0$ line indicates bifurcation points.

For the softening foundations, as the bifurcation amplitude $\xi$ increases the graphs in Fig.~\ref{Fig:dispersion-soft} the minimum eigenvalue curve $\beta_{min}(k)$ shifts downward and the curve flattens.  For the re-hardening case, as the primary bifurcation orbit ${\stackrel{1}{w}}(x;\xi)$ passes the limit point, the graph starts moving upward, as seen in Fig.~\ref{Fig:dispersion-soft}(c) corresponding to the strong re-hardening foundation ($\alpha = -1, \gamma = 0.5$).

The dispersion curves for the hardening foundation are plotted in Fig.~\ref{Fig:dispersion-hard}.
Notice that for bifurcation amplitudes up to $\xi \approx 1.3$, $\beta_{min}(k) > 0$ as expected from the stability results in Fig.~\ref{Fig:primary-diagrams}. For a fixed wavenumber $k$, the minimum eigenvalue $\beta_{min}(k)$ is a decreasing function of the bifurcation amplitude $\xi$.  Moreover, and in contrast to the curves for the softening foundation case in Fig.~\ref{Fig:dispersion-soft} which shift in the negative $y$ direction as the bifurcation amplitude increases, all the dispersion curves are pinned at the ends $\beta_{min}(0) = \beta_{min}(1) = 0$.  As the bifurcation amplitude increases further ($\xi \approx 1.725$), the primary bifurcation path is unstable over the entire range of wavenumbers, $\beta_{min}(k) < 0$.

Finally a remark that applies to the existence of $\beta_{min}(0)=0$ in the dispersion curves is in order.  At $k=0$, as seen in the Fig.~\ref{Fig:dispersion-soft} graphs $\beta_{min}=0$ for the principal solution ($\xi=0$), while $\beta=0$ is an eigenvalue for the primary bifurcation orbit ($\xi>0$), but not the minimum one.  In the Fig.~\ref{Fig:dispersion-hard} graphs $\beta_{min}=0$ for the principal solution and the primary bifurcation orbit ($\forall \xi \geq 0$).  The reason stems from the fact that any solution---principal, primary, etc.---of the equilibrium equations which is $2\pi$-periodic, and hence corresponds to $k=0$ according to Eq.~\eqref{eq:blochboundary}, has a zero eigenvalue. Indeed differentiating the equilibrium equations for an equilibrium solution $\mathcal{E}_{,w}(w(x+\mathfrak{c});\lambda) \delta w = 0$ with respect to an arbitrary phase-shift of $\mathfrak{c}$ we obtain
\begin{equation}
{\frac{d}{d\mathfrak{c}}} \left[ {\cal E}_{,w}(w(x+\mathfrak{c});\lambda)\delta w \right] =0 \quad \Longrightarrow \quad
\left({\cal E}_{,ww} \left[{\frac{dw}{dx}}(x;\lambda)\right] \right)\delta w = 0 \; ,
\label{eq:zeroofbeta}
\end{equation}
indicating that zero is always an eigenvalue of the stability operator ${\cal E}_{,ww}$ with corresponding eigenmode $dw/dx$.

\subsubsection{Secondary Bifurcation Orbits for Softening Foundations ($\alpha = -1$)}

We are interested next in the secondary bifurcated equilibrium orbits emerging from the primary one, i.e.\ solutions of the Euler--Lagrange Eq.~\eqref{eq:equilibrium} subjected to the boundary conditions Eq.~\eqref{eq:secondary_BC} for a beam of length $L_d = L_c q$ resting on a softening foundation ($\alpha = -1$).  According to the dispersion results in Fig.~\ref{Fig:dispersion-soft}, the primary orbit is always unstable for long wavelengths, i.e.\ in the neighborhood of $k=0$; as the bifurcation amplitude $\xi$ increases, it becomes unstable for shorter wavelengths.  To capture the longest possible wavelength mode, a $q \gg 1$ is needed and hence our numerical calculations use $q =20$ along with boundary conditions Eq.~\eqref{eq:secondary_BC} for the $C_{rv}$ or $D_r$ secondary bifurcating paths.

\begin{figure*}
\begin{minipage}{0.53\textwidth}
\begin{center}
\includegraphics[width=\textwidth]{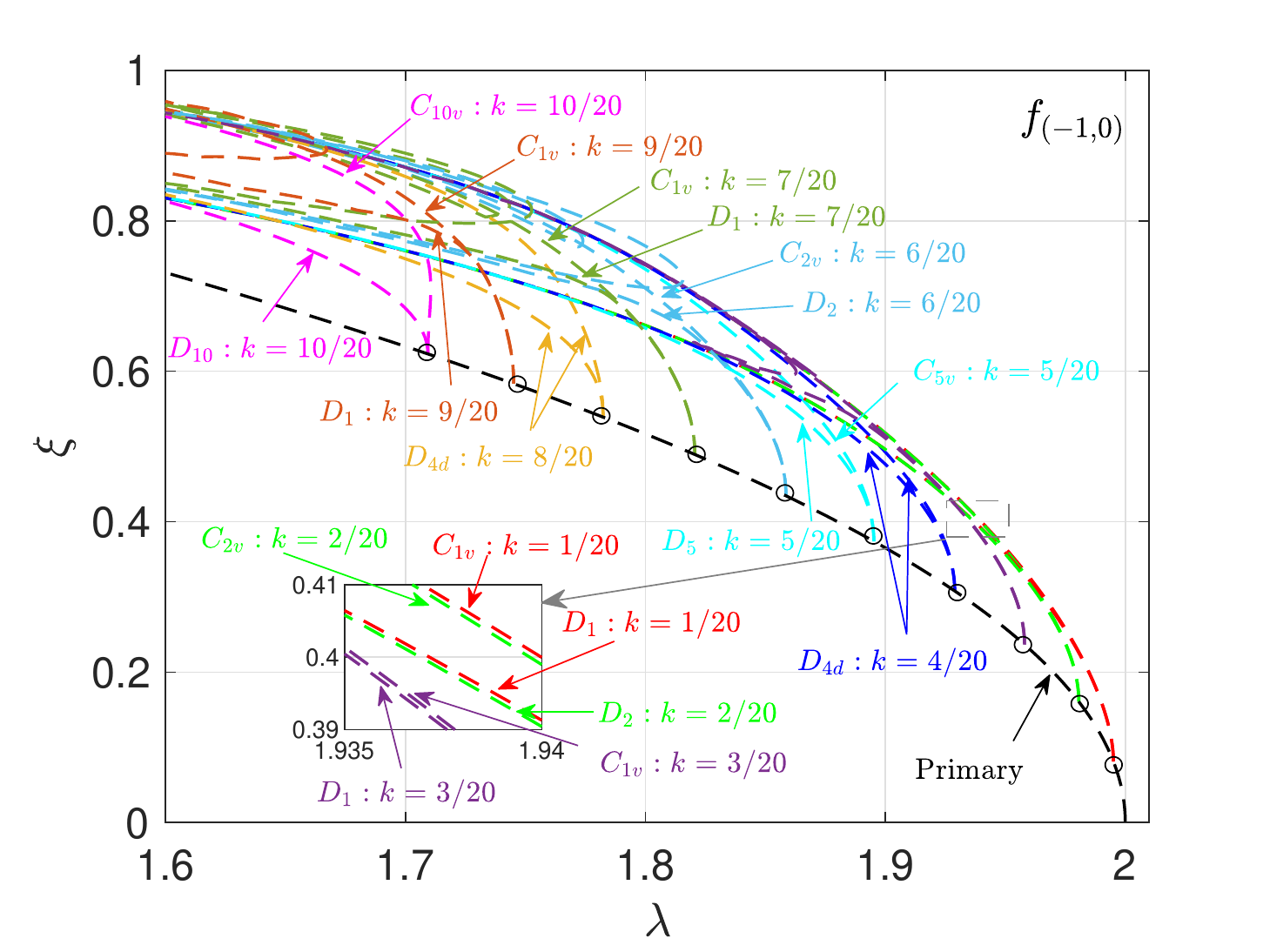}\\%
(a)\\%
\includegraphics[width=\textwidth]{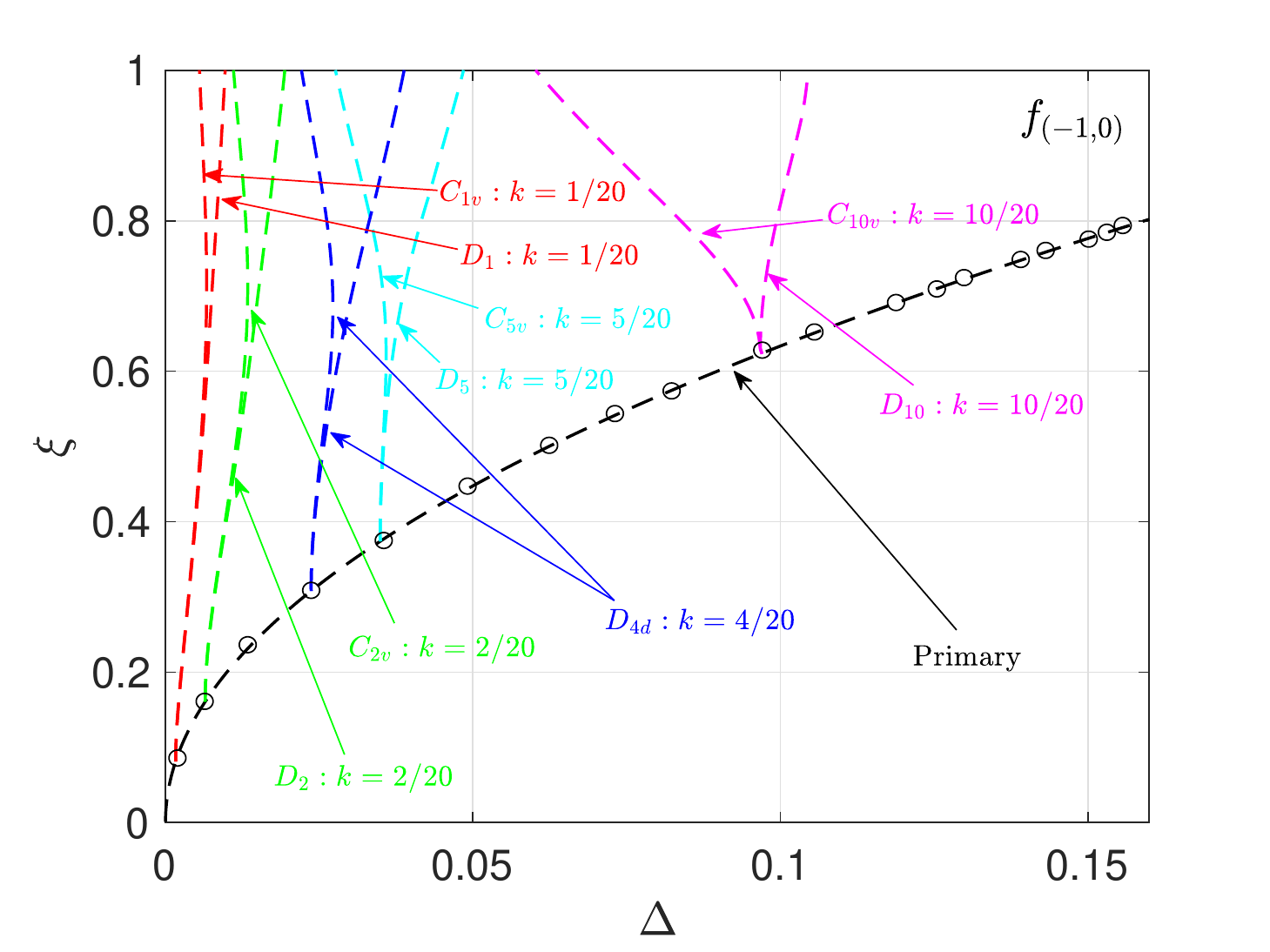}\\
(c)%
\end{center}
\end{minipage}%
\hspace*{-0.5cm}%
\begin{minipage}{0.53\textwidth}
\begin{center}
\includegraphics[width=\textwidth]{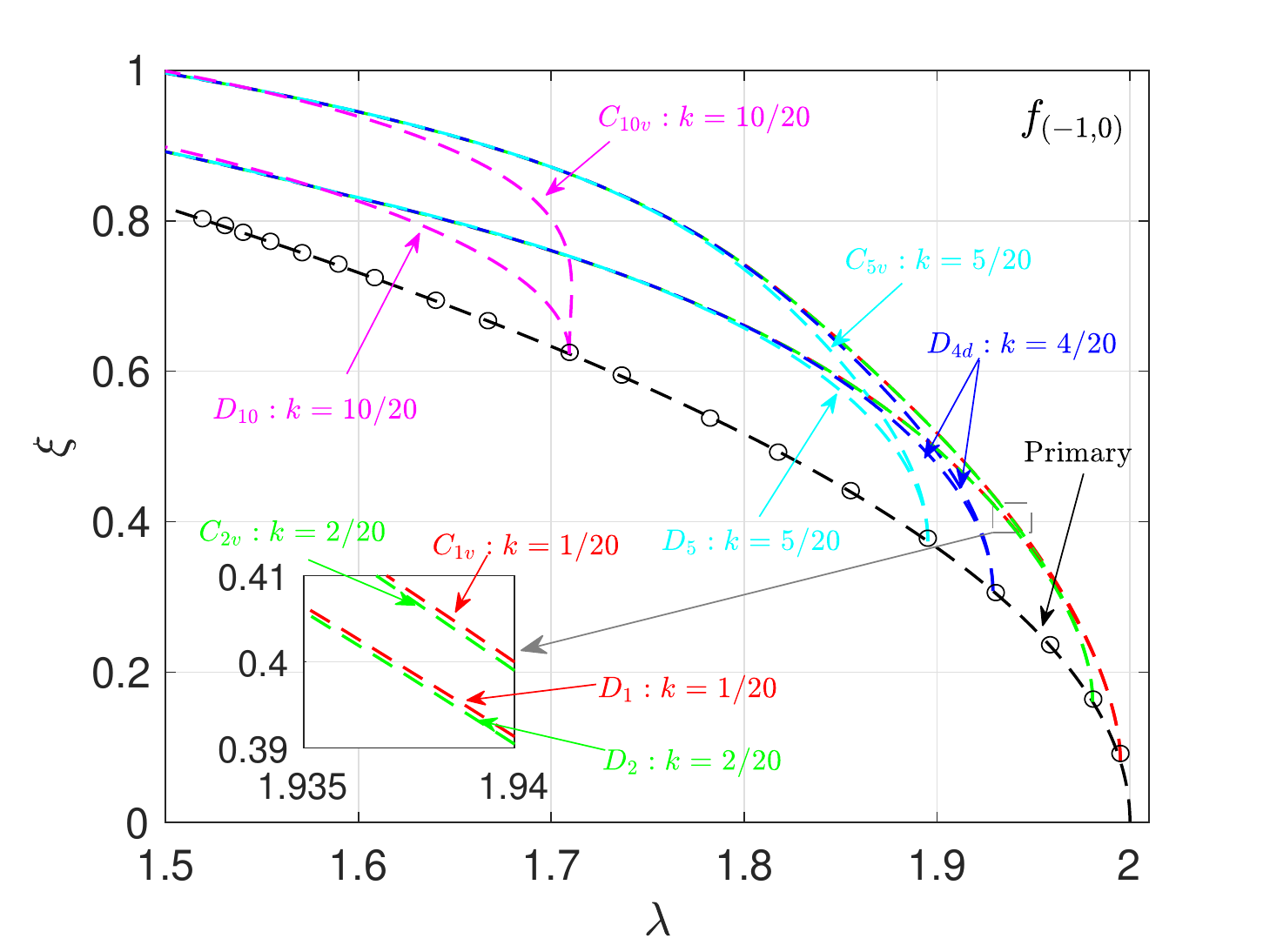}\\%
(b)\\%
\includegraphics[width=\textwidth]{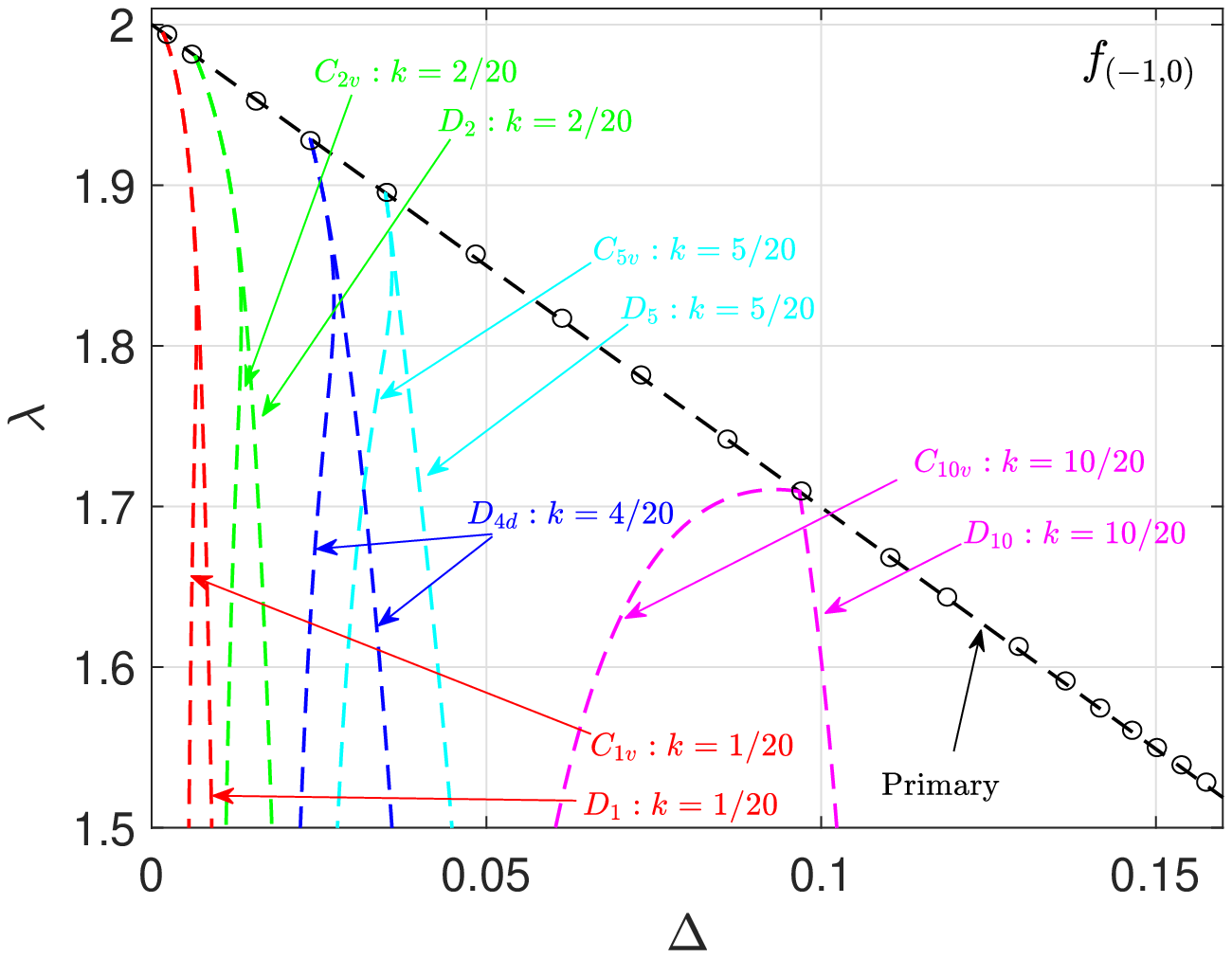}\\%
(d)%
\end{center}
\end{minipage}
\caption{Emergence of secondary bifurcations from the primary orbit in the case of the softening foundation model with no re-hardening ($\alpha = -1,\ \gamma = 0$).  All bifurcation points are indicated by a small circle.  Notice that all paths are unstable (plotted in dashed lines), based on Bloch-wave analysis of a $q=20$ ($L_d=40 \pi$) supercell.  (a)~Amplitude $\xi$ vs load $\lambda$ bifurcation diagram for \emph{all} secondary orbits emerging from the primary orbit.  (b)~Amplitude $\xi$ vs load $\lambda$, (c)~amplitude $\xi$ vs strain $\Delta$, and (d)~load $\lambda$ vs strain $\Delta$ bifurcation diagrams for the \emph{simple harmonic} secondary bifurcated orbits $k=n/q, n=1,2,4,5,10$.}
\label{Fig:soft-none-dia}
\end{figure*}

The secondary bifurcation orbits for the case of the softening foundation with no re-hardening ($\alpha = -1,\ \gamma = 0$) are recorded in Fig.~\ref{Fig:soft-none-dia}.
On the primary bifurcation orbit, we mark all the (double) bifurcation points found for a $L_d=40\pi$ supercell.  The secondary bifurcation orbits plotted are labeled two ways: the first label indicates the symmetry group of the orbit.  The second label pertains to the corresponding eigenmodes: from the symmetry of the dispersion curve with respect to $k=1/2$ follows that each bifurcation point is double, since it corresponds to two different modes; for example $k=7/20$ and $k=1-7/20=13/20$.  The value of each wavenumber is directly related to the number of periods in the supercell; for example $k=7/20$ means a solution with 7 periods for the 20 unit supercell. For simplicity, only the lower of the two wavenumbers corresponding to each bifurcation point is recorded. A group-theoretic explanation of the results is given in Appendix~\ref{appendix-A}.  The top two graphs in Fig.~\ref{Fig:soft-none-dia} show the amplitude $\xi$ vs load $\lambda$ in the interval $1.6 \le \lambda \le 2$; more specifically, all the secondary orbits emerging from the primary one are plotted in Fig.~\ref{Fig:soft-none-dia}(a) while---to avoid clutter---only the orbits corresponding to the simple harmonic modes $k=n/q, \ n = 1, 2, 4, 5,10$ are plotted in Fig.~\ref{Fig:soft-none-dia}(b).  Results for these simple harmonic secondary bifurcated orbits are presented as bifurcation amplitude $\xi$ vs strain $\Delta$ in Fig.~\ref{Fig:soft-none-dia}(c) and as load $\lambda$ vs strain $\Delta$ in Fig.~\ref{Fig:soft-none-dia}(d).  Observe that the secondary bifurcated orbit that emerges closest to the critical load is the one with longest wavelength ($k=1/20$), while orbits with shorter wavelength modes emerge at lower loads.  Notice that all secondary bifurcation orbits are, like the primary one, unstable.

Although all bifurcation points are double, the number of emerging orbits can be either one, for a globally transverse bifurcation, or two, for the case of a globally pitchfork bifurcation.  As seen from Fig.~\ref{Fig:soft-none-dia}, the only two single-orbit transverse bifurcations are found at $k=4/20$ and $k=8/20$; from all other bifurcation points there emerge two different globally pitchfork bifurcation orbits.  Another interesting feature of this problem is that all bifurcations---even the globally transverse ones---are locally pitchfork since symmetry dictates that the coefficients of the bifurcation equations are all zero: ${\mathcal E}_{ijk}=0$.  A group theory justification of this, due to symmetry, as well as an explanation for the transverse/pitchfork orbits for the different wavenumber $k$ is given in Appendix~\ref{appendix-A}.  The two secondary bifurcation branches emerging from each bifurcation point (corresponding either to different orbits in a pitchfork case or the two halves of the transverse orbit) are nearly coincident initially, but subsequently grow apart; the further the bifurcation point is from the origin of the principal branch, the sooner the secondary branches separate from each other.  This is shown in Fig.~\ref{Fig:soft-none-dia}.

\begin{figure*}
\begin{center}
\includegraphics[width=0.55\textwidth]{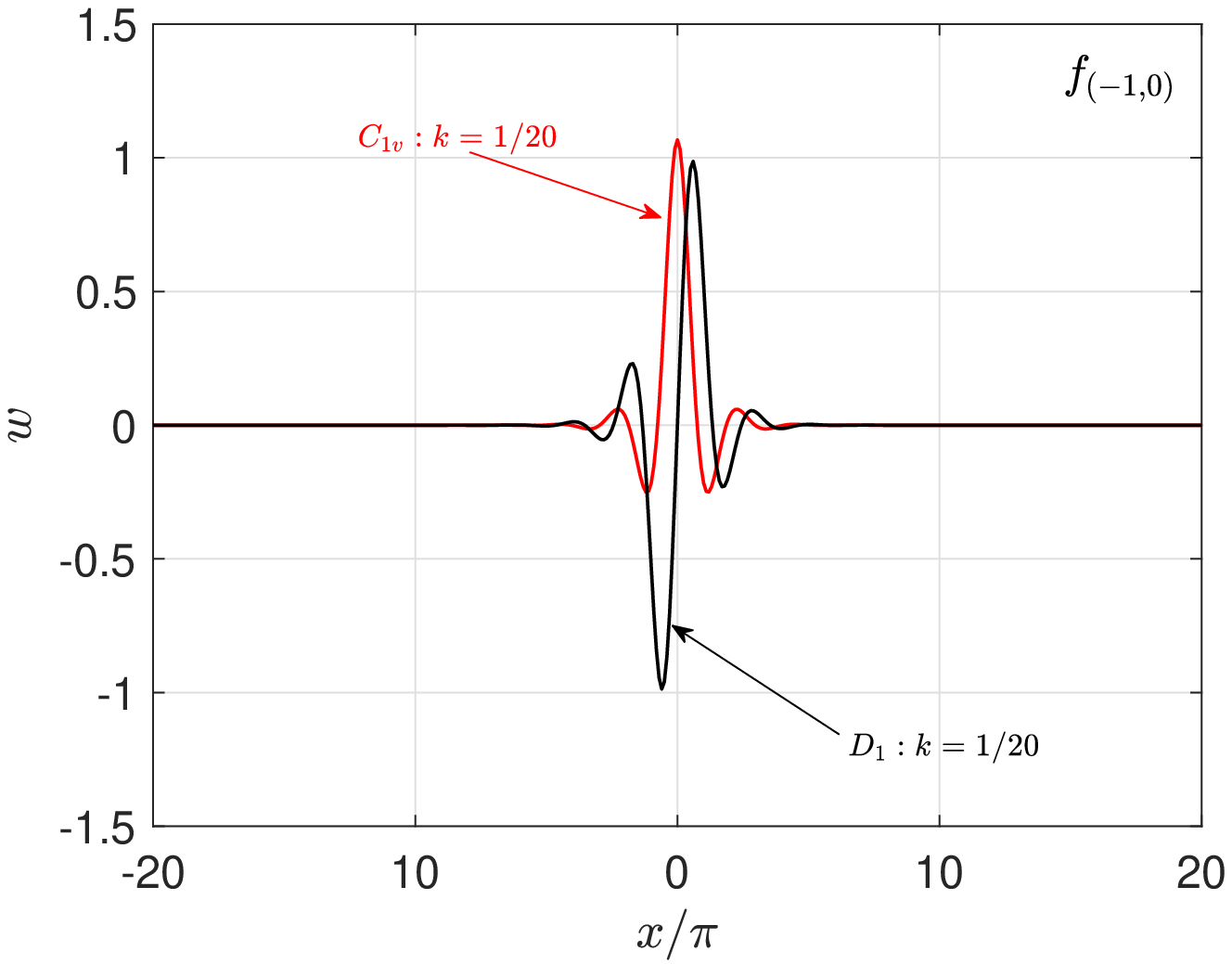}\\
(a)\\
\includegraphics[width=0.55\textwidth]{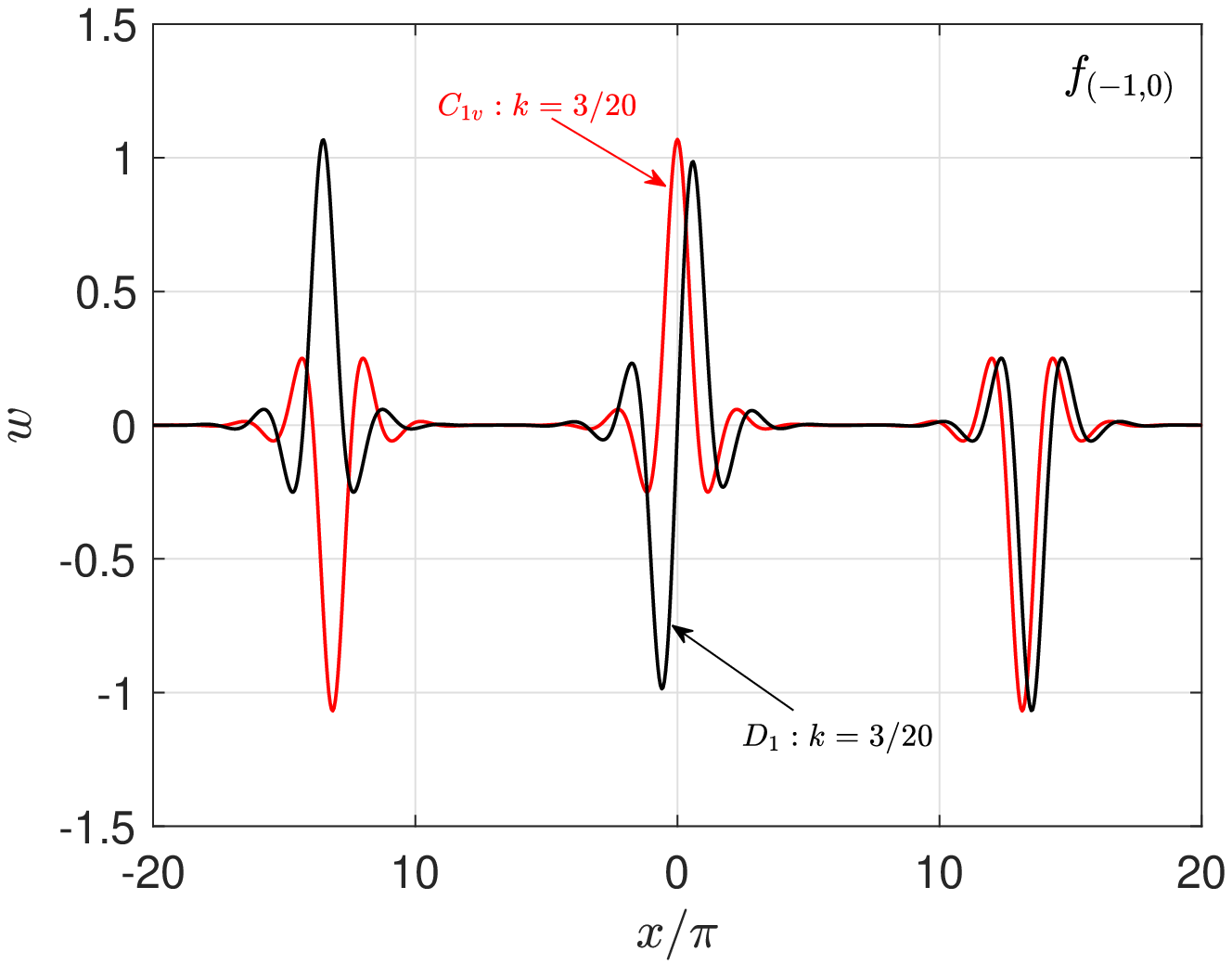}\\
(b)\\
\includegraphics[width=0.55\textwidth]{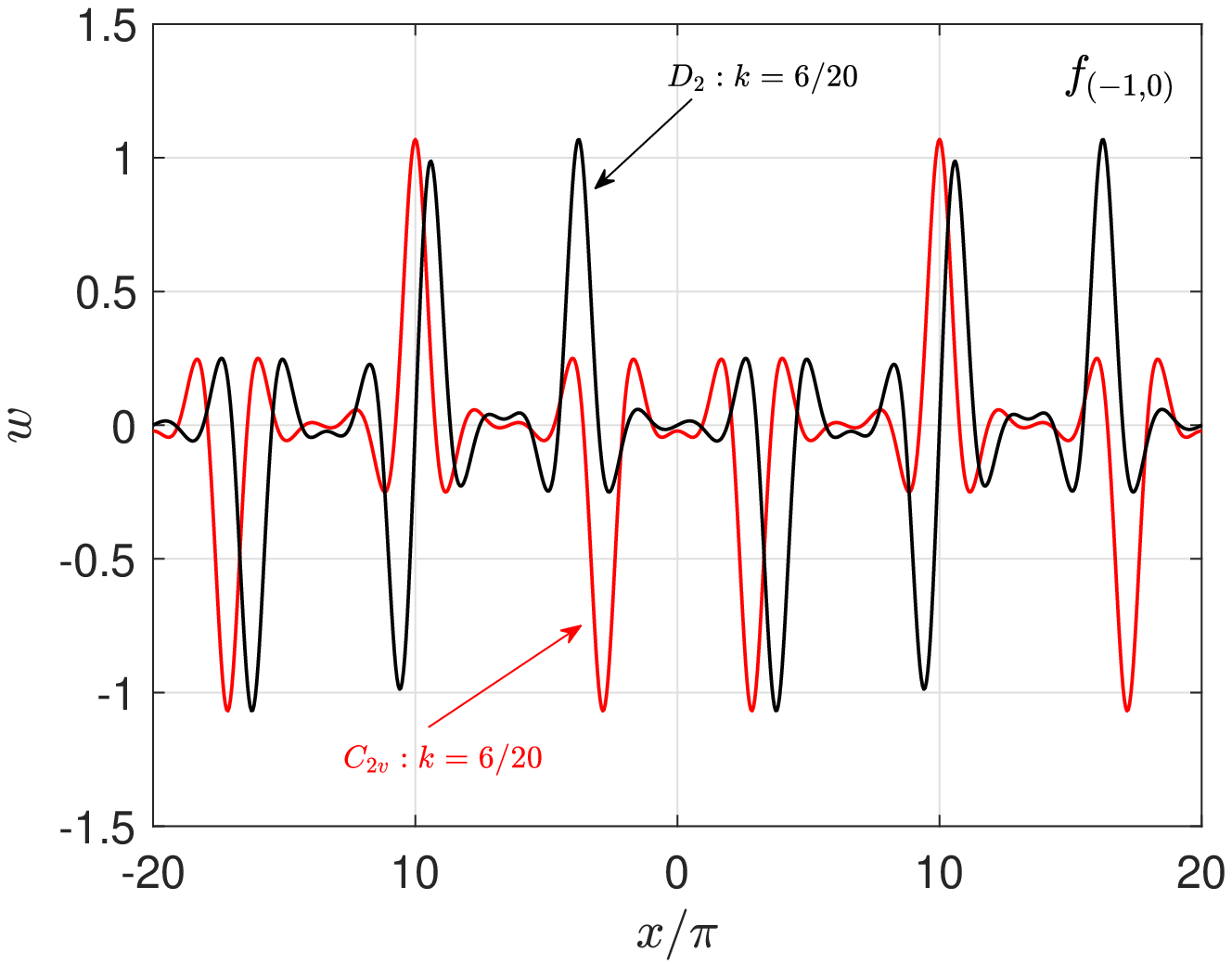}\\
(c)
\end{center}
\caption{Symmetric and antisymmetric solutions $w(x)$ for the softening foundation model with no re-hardening ($\alpha = -1,\ \gamma = 0$), calculated for a $q=20\ (L_d=40 \pi)$ supercell at a load $\lambda=1.3$ for secondary bifurcation orbits corresponding to different values of the wavenumber (a)~$k=1/20$, (b)~$k=3/20$, and (c)~$k=6/20$.  The bifurcated orbits depicted here are all unstable.}
\label{Fig:soft-none-mode}
\end{figure*}

The next question to be addressed pertains to the shape, i.e.\ spatial dependence of these secondary bifurcated orbits.  Symmetric (even with respect to $x=0$) and antisymmetric (odd with respect to $x=0$) equilibrium solutions for the beam on a softening foundation with no re-hardening ($\alpha = -1,\ \gamma = 0$), calculated for a $q=20\ (L_d=40\pi)$ supercell at load $\lambda=1.3$ for different orbits, are plotted in Fig.~\ref{Fig:soft-none-mode}.
These orbits emerge from pitchfork bifurcation points with wavenumbers: $k=1/20$ and $3/20$ (corresponding to $C_{1v}$ orbits for the symmetric case and $D_1$ orbits for the antisymmetric case) and $k=6/20$ (corresponding to a $C_{2v}$ orbit for the symmetric case and a $D_2$ orbit for the antisymmetric case).  These secondary bifurcation paths are plotted for a load $\lambda=1.3$, well away from their origin point on the primary bifurcation orbit.  In each case the solution is localized about one (for $k=1/20$) or more (for $k=3/20$ and $6/20$) locations, distributed uniformly along the supercell domain, as expected from the wavenumber of the corresponding eigenmode.  Recall, from the results in Fig.~\ref{Fig:soft-none-dia}, that all solutions depicted in Fig.~\ref{Fig:soft-none-mode} are unstable.  It is noteworthy that the equilibrium paths for $k= 3/20$ and $6/20$ consist of equally spaced repetitions of the symmetric and antisymmetric $k= 1/20$ shapes as their fundamental building blocks.  We conclude that the equilibrium solutions that are most localized are secondary bifurcated paths emerging near the critical point and correspond to the eigenmode with the smallest wavenumber; however they are always unstable due to the monotonically softening foundation.

\begin{figure*}
\begin{center}
\includegraphics[width=0.55\textwidth]{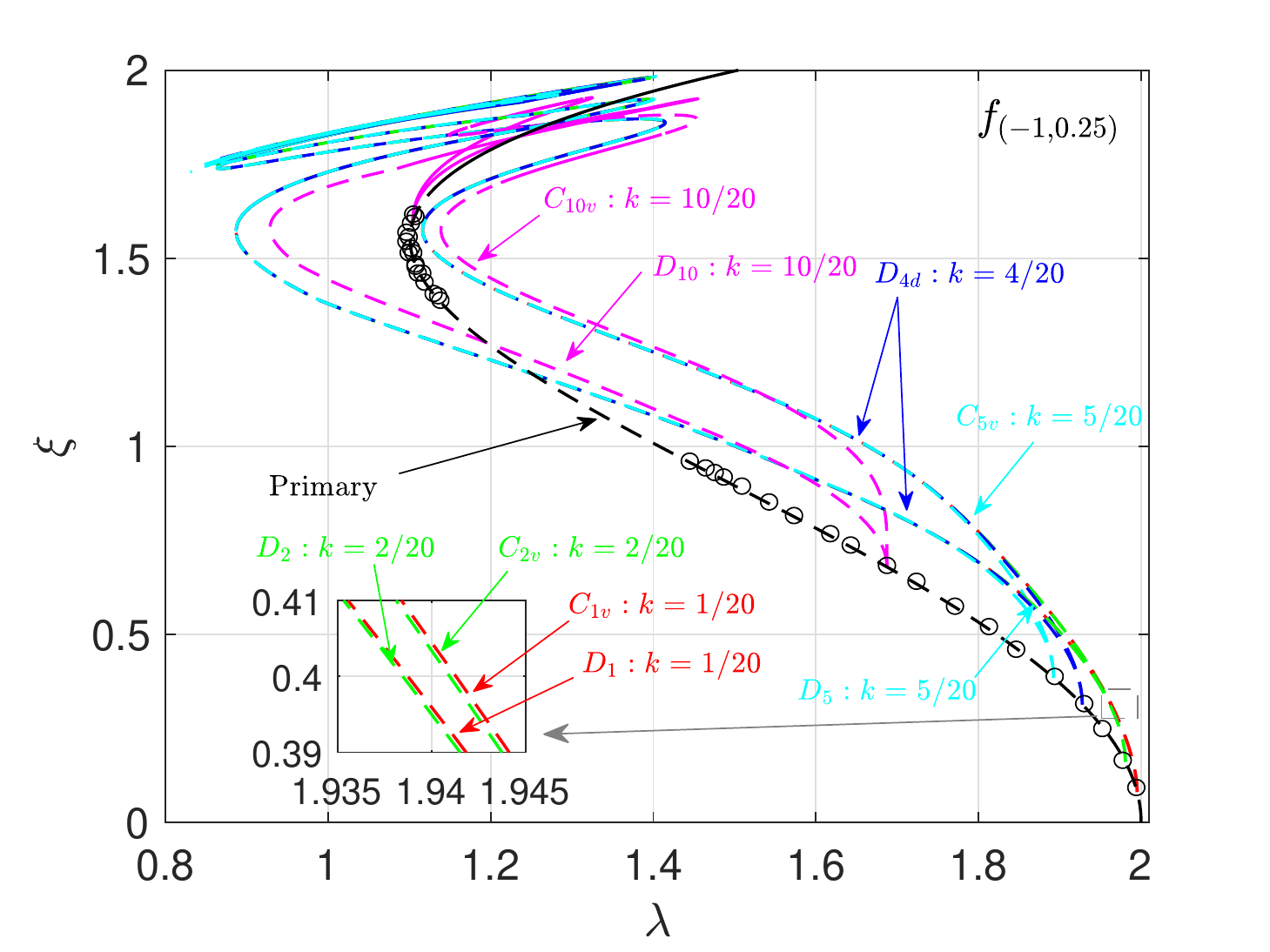}\\
(a)\\
\includegraphics[width=0.55\textwidth]{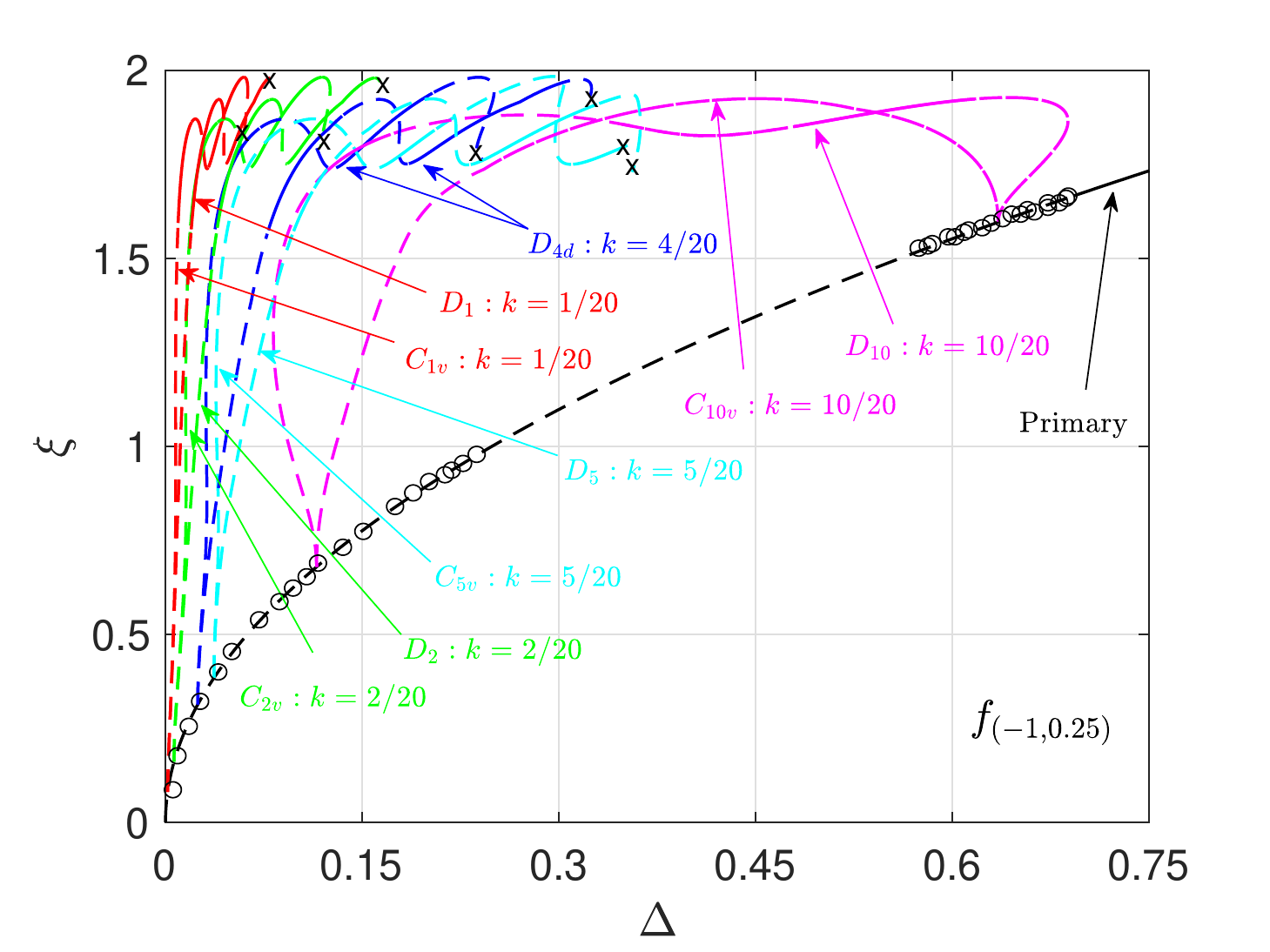}\\
(b)\\
\includegraphics[width=0.55\textwidth]{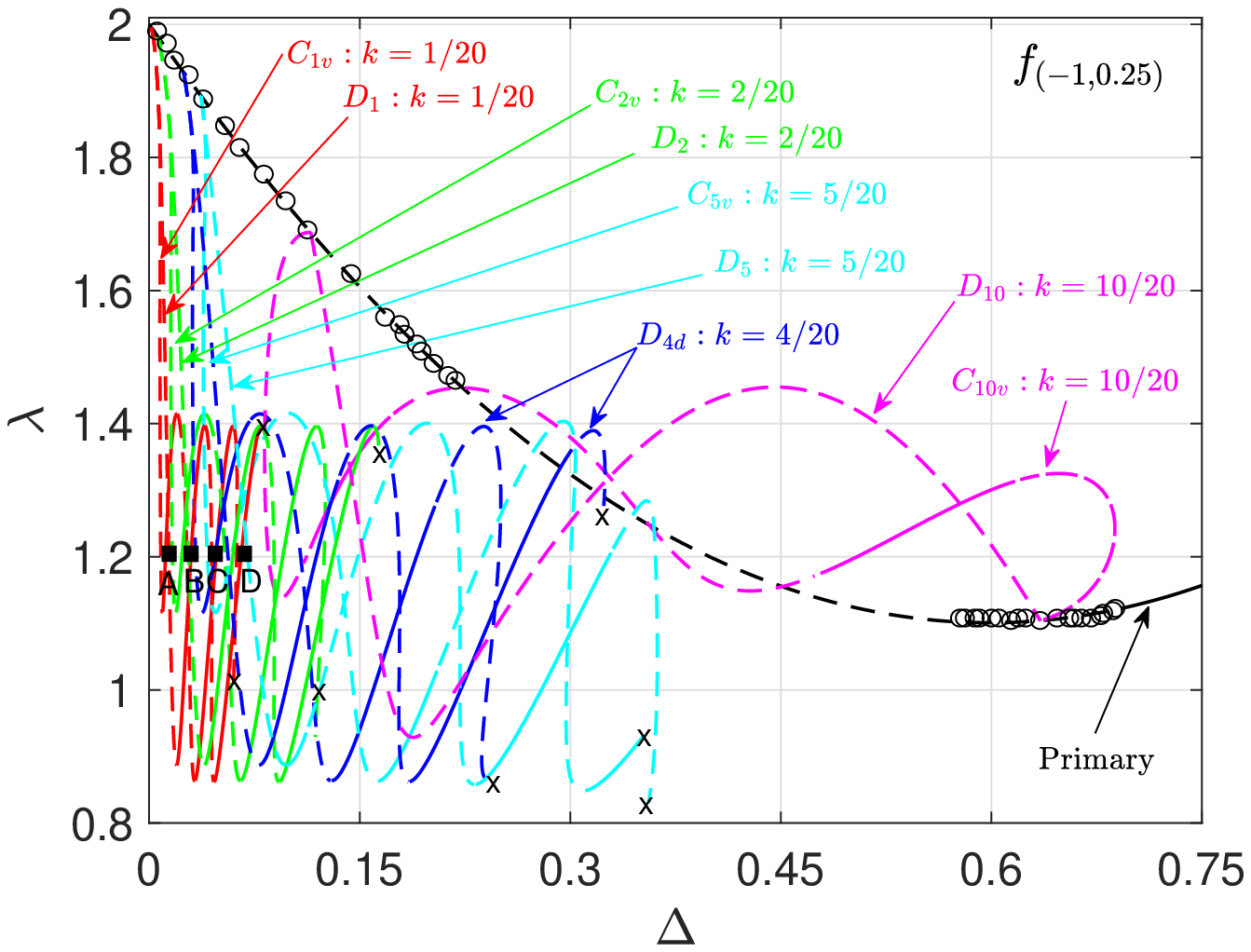}\\
(c)
\end{center}
\caption{Emergence of secondary bifurcations from the primary bifurcation orbit in the case of the softening foundation with mild re-hardening ($\alpha = -1,\ \gamma = 0.25$).  All bifurcation points are indicated by a small circle.  Solid and dashed lines correspond, respectively, to the stable and unstable parts of these equilibrium paths, based on Bloch-wave analysis of a $q=20\ (L_d=40 \pi)$ supercell.  Notice that all bifurcated orbits have stable portions, due to the re-hardening foundation.  Moreover these secondary orbits eventually join the primary branch.  Results are presented only for orbits corresponding to the \emph{simple harmonic} modes $k =n/q,\ n = 1, 2, 4, 5, 10$ and show the (a)~amplitude $\xi$ vs load $\lambda$, (b)~amplitude $\xi$ vs strain $\Delta$, and (c)~load $\lambda$ vs strain $\Delta$ bifurcation diagrams.}
\label{Fig:soft-mild-dia}
\end{figure*}

Calculations for the secondary bifurcation orbits for the softening foundation with mild re-hardening ($\alpha = -1,\ \gamma = 0.25$) are presented next in Fig.~\ref{Fig:soft-mild-dia}.
From the results in Fig.~\ref{Fig:primary-diagrams}, the primary bifurcation orbit is initially unstable, but stabilizes upon reaching a limit load for a large value of the bifurcation amplitude parameter ($\xi >1.5$).  Moreover, from the corresponding dispersion results in Fig.~\ref{Fig:dispersion-soft} and similarly to the no re-hardening case, the primary orbit is always unstable for long wavelengths, i.e.\ in the neighborhood of $k=0$; as the bifurcation amplitude $\xi$ increases, it becomes unstable for shorter wavelengths as well. However, and in contrast to the no re-hardening foundation in Fig.~\ref{Fig:soft-none-dia}, the bifurcation points appear concentrated in two clusters, one near $\lambda = 2$ where the secondary bifurcation paths emerge and the other near the limit load where the secondary paths reconnect to the primary path.  To avoid clutter, only orbits corresponding to the simple harmonic modes $k =n/q,\ n = 1, 2, 4, 5, 10$ are plotted; for the same reason orbits emerging closer to the critical load are stopped before reconnecting to the primary branch (ending point marked by a $\times$). For the purpose of demonstration, only the secondary bifurcation orbits with the $k=10/20$ eigenmode are continued all the way until they reconnect to the primary one.

The main difference from the results in Fig.~\ref{Fig:soft-none-dia}, for the no re-hardening foundation, is the considerably more complicated structure of the secondary bifurcation orbits.  This is seen by a comparison to the mild re-hardening case in Fig.~\ref{Fig:soft-mild-dia}.  The corresponding secondary bifurcated orbits, initially similar to their counterparts in the no re-hardening case, reach a state where they oscillate between two bounds of the load, in Fig.~\ref{Fig:soft-mild-dia}(a) and in Fig.~\ref{Fig:soft-mild-dia}(c), or of the amplitude, in Fig.~\ref{Fig:soft-mild-dia}(b) \citep[e.g.][]{Hunt_et_al}.  Even more interesting is the fact that these oscillating secondary bifurcated paths are initially unstable, but stabilize and destabilize repeatedly upon reaching either a limit load or another bifurcation point.

\begin{figure*}
\hspace*{-0.5cm}%
\begin{minipage}{0.55\textwidth}
\begin{center}
\includegraphics[width=\textwidth]{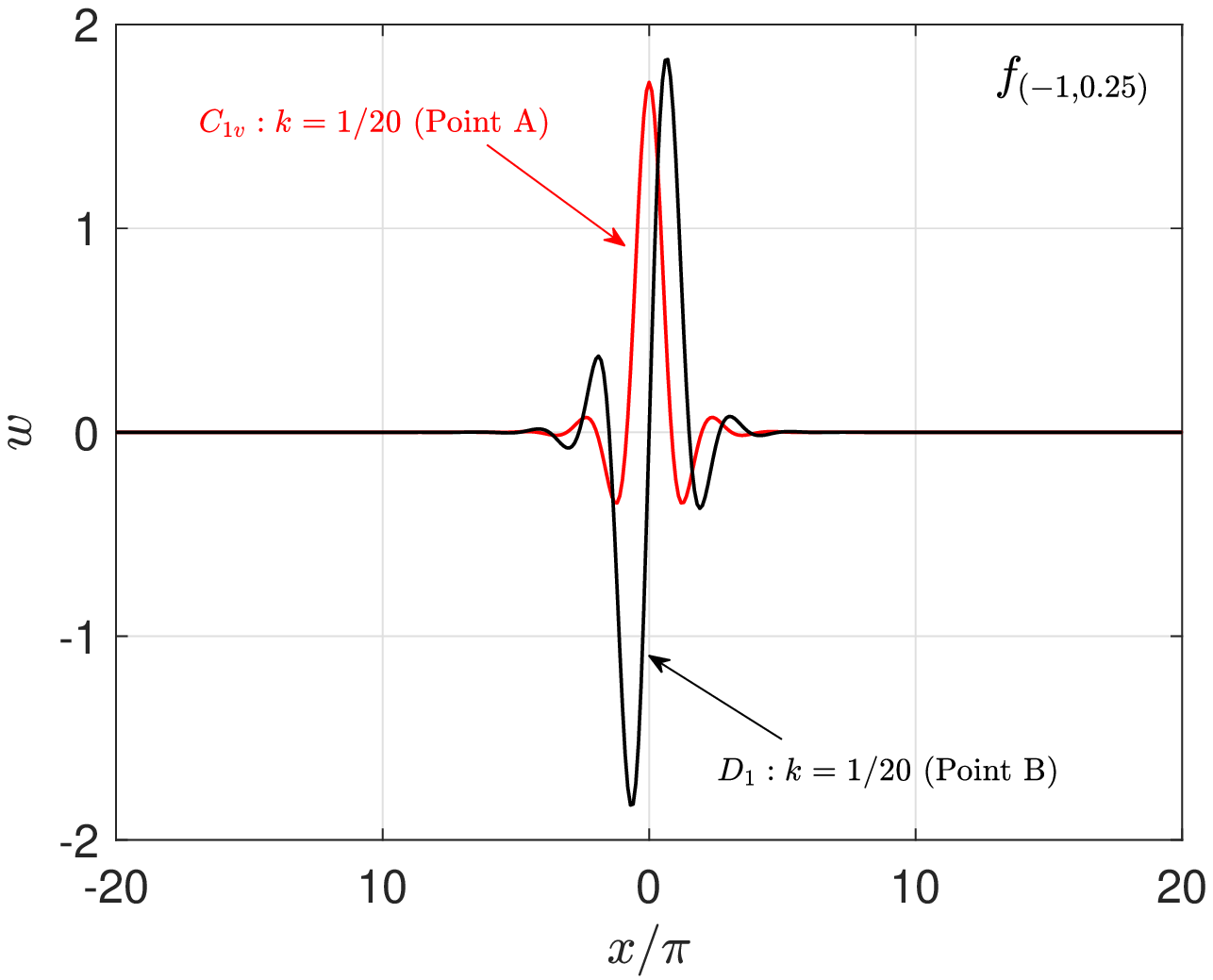}\\%
(a)%
\end{center}
\end{minipage}%
\hspace*{-0.5cm}%
\begin{minipage}{0.55\textwidth}
\begin{center}
\includegraphics[width=\textwidth]{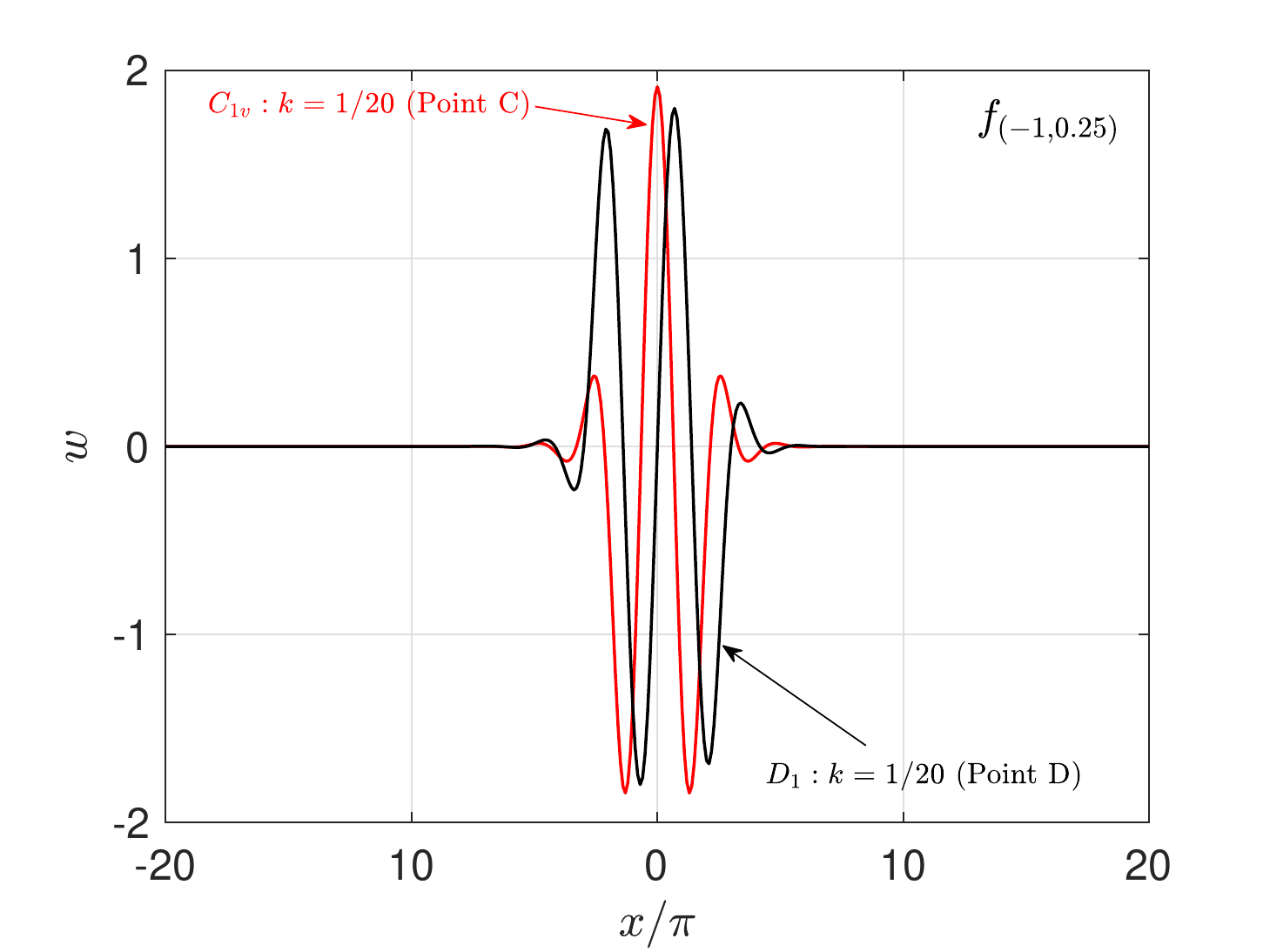}\\%
(b)%
\end{center}
\end{minipage}
\caption{Symmetric and antisymmetric stable solutions $w(x)$ for the softening foundation model with mild re-hardening ($\alpha = -1,\ \gamma = 0.25$), calculated for a $q=20\ (L_d=40\pi)$ supercell at a load $\lambda = 1.2$ but at different strains $\Delta$ for the secondary bifurcation orbits corresponding to the lowest value of the wavenumber $k=1/20$.  (a)~Points A and B, and (b)~Points C and D (see Fig.~\ref{Fig:soft-mild-dia}(c)).}
\label{Fig:soft-mild-mode}
\end{figure*}

The shape of the symmetric (even with respect to $x=0$) and antisymmetric (odd with respect to $x=0$) secondary bifurcated orbits calculated for a load $\lambda = 1.2$ and emerging from the bifurcation with the lowest wavenumber $k=1/20$, are plotted in Fig.~\ref{Fig:soft-mild-mode}.
These are all stable, with the solutions at points A and C corresponding to symmetric ($C_{1v}$) orbits and the ones at B and D to antisymmetric ($D_1$) ones.  Observe that the deformation is spatially localized at the middle of the supercell calculation domain; as the strain increases notice a corresponding increase in the number of peaks by comparing the symmetric solutions at points A and C or the antisymmetric ones at B and D.  In comparing the secondary bifurcation orbits corresponding to $k=1/20$ for the two different foundations in Fig.~\ref{Fig:soft-none-mode} and Fig.~\ref{Fig:soft-mild-mode} (at points A and B), it is noteworthy that the shape of the localized deformation equilibrium paths are very similar (as the corresponding loads are close: $\lambda = 1.3$ in the first and $\lambda = 1.2$ in the second) although the corresponding solutions are unstable and stable, respectively.  We have thus observed that a stable equilibrium path with a sharply localized deformation pattern in only one location occurs only when the foundation re-hardens.  An uncluttered bifurcation diagram showing only the primary and localized $C_{1v} : k=1/20$ secondary bifurcated equilibrium paths is shown later in Fig.~\ref{Fig:single-localized-dia} (where we also demonstrate that these stable solutions are insensitive to imperfections).

\begin{figure*}
\hspace{-0.5cm}
\begin{center}
\includegraphics[width=0.55\textwidth]{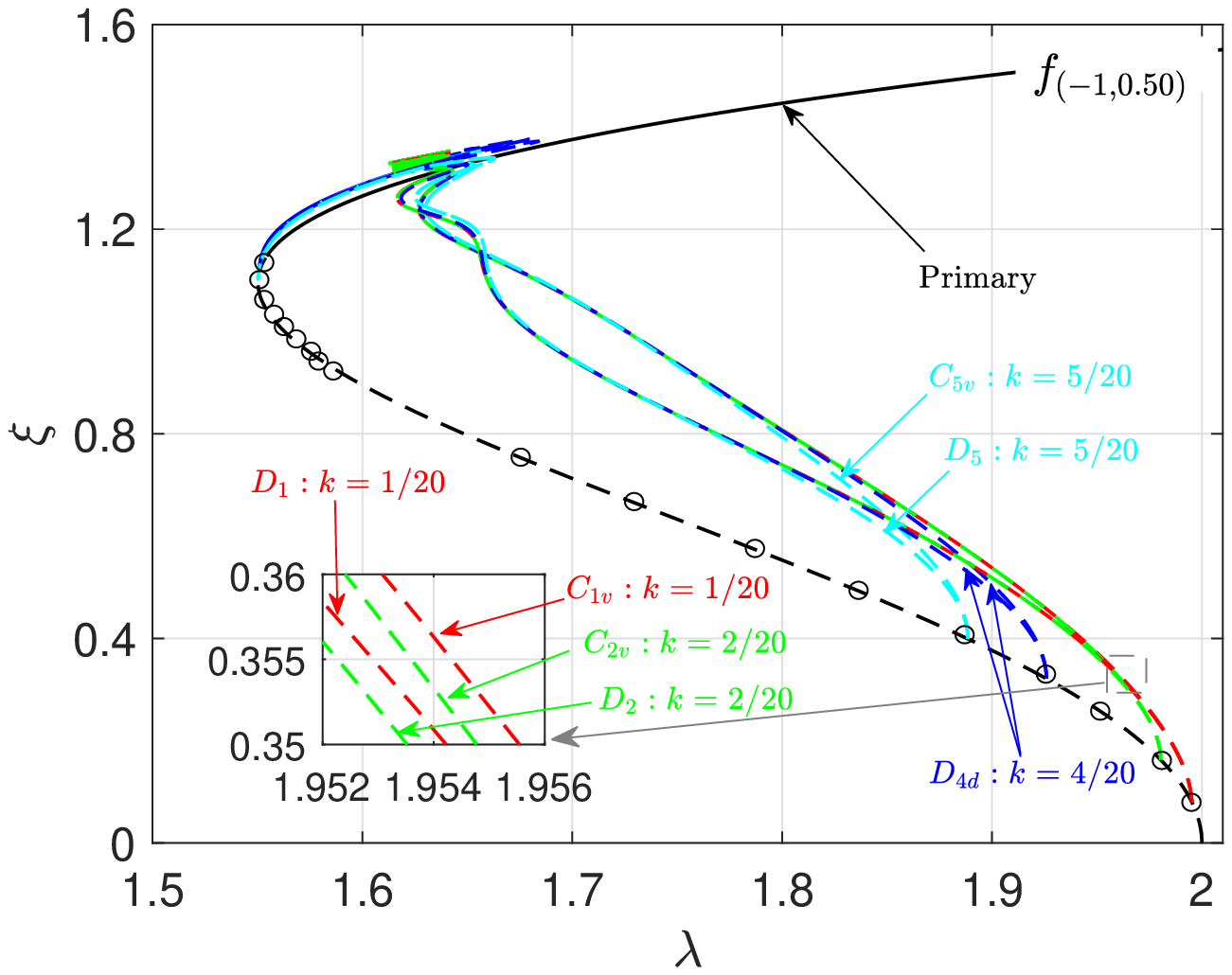}\\
(a)\\
\includegraphics[width=0.55\textwidth]{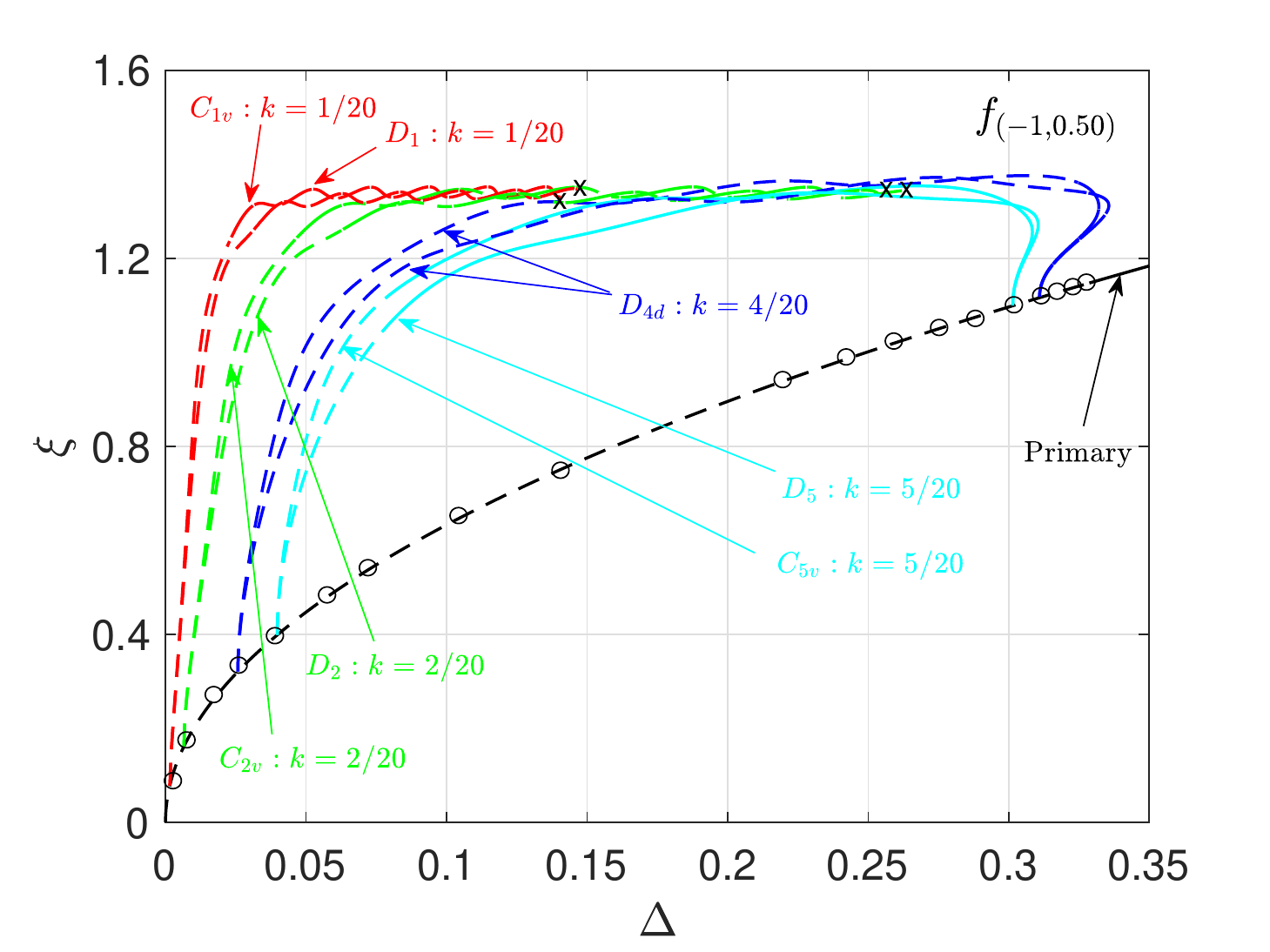}\\
(b)\\
\includegraphics[width=0.55\textwidth]{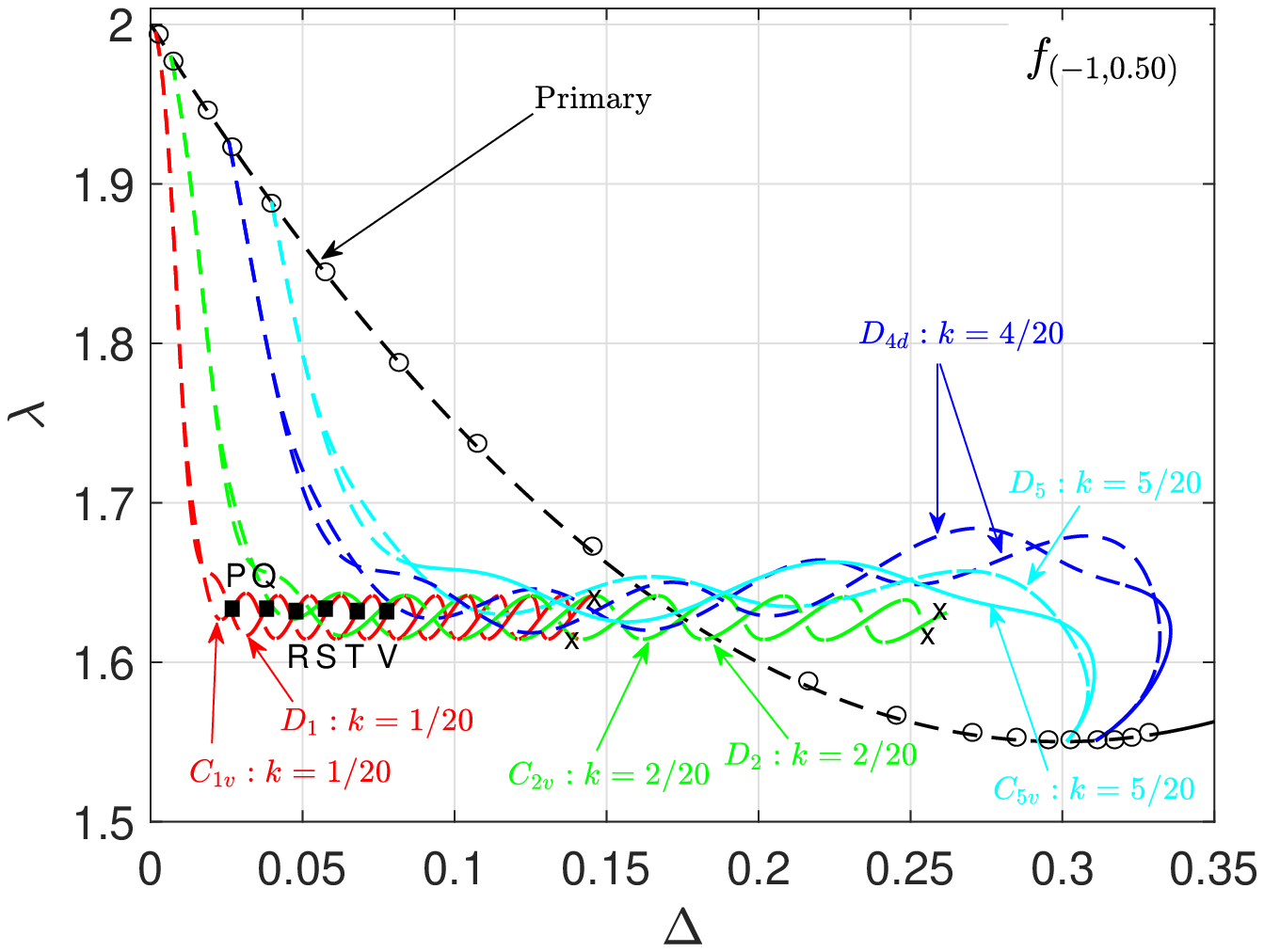}\\
(c)
\end{center}
\caption{Emergence of secondary bifurcations from the primary bifurcation orbit in the case of the softening foundation with strong re-hardening ($\alpha = -1,\ \gamma = 0.50$).  All bifurcation points are indicated by a small circle. Solid and dashed lines correspond, respectively, to the stable and unstable parts of these equilibrium paths, based on Bloch-wave analysis of a $q=20\ (L_d = 40\pi)$ supercell.  Notice that all bifurcated orbits have stable portions, due to the re-hardening foundation.  Moreover these secondary orbits eventually join the primary branch.  Results are presented only for orbits corresponding to the \emph{simple harmonic} modes $k =n/q,\ n = 1, 2, 4, 5, 10$ and show the (a)~amplitude $\xi$ vs load $\lambda$, (b)~amplitude $\xi$ vs strain $\Delta$, and (c)~load $\lambda$ vs strain $\Delta$ bifurcation diagrams.}
\label{Fig:soft-strong-dia}
\end{figure*}

Results for secondary bifurcation orbits of the beam with the softening foundation and strong re-hardening ($\alpha = -1,\ \gamma = 0.50$), calculated using a $q=20\ (L=40\pi)$ supercell, are plotted in Fig.~\ref{Fig:soft-strong-dia}.
Similarly to the mild re-hardening foundation results in Fig.~\ref{Fig:soft-mild-dia}, the secondary orbits for the strongly re-hardening foundation form loops that emerge and reconnect to the primary orbit.  Once again, to avoid clutter, only secondary orbits corresponding to the simple harmonic modes $k =n/q,\ n = 1, 2, 4, 5, 10$ are plotted with orbits emerging closer to the critical load stopped before reconnecting to the primary one (ending point marked by a $\times$).  The same features observed in Fig.~\ref{Fig:soft-mild-dia} are also present here, namely the upper and lower bounds for the bifurcation amplitude and the load as functions of the strain, as well as the alternating stable/unstable portions of the corresponding orbits.

\begin{figure*}
\hspace{-0.5cm}
\begin{center}
\includegraphics[width=0.55\textwidth]{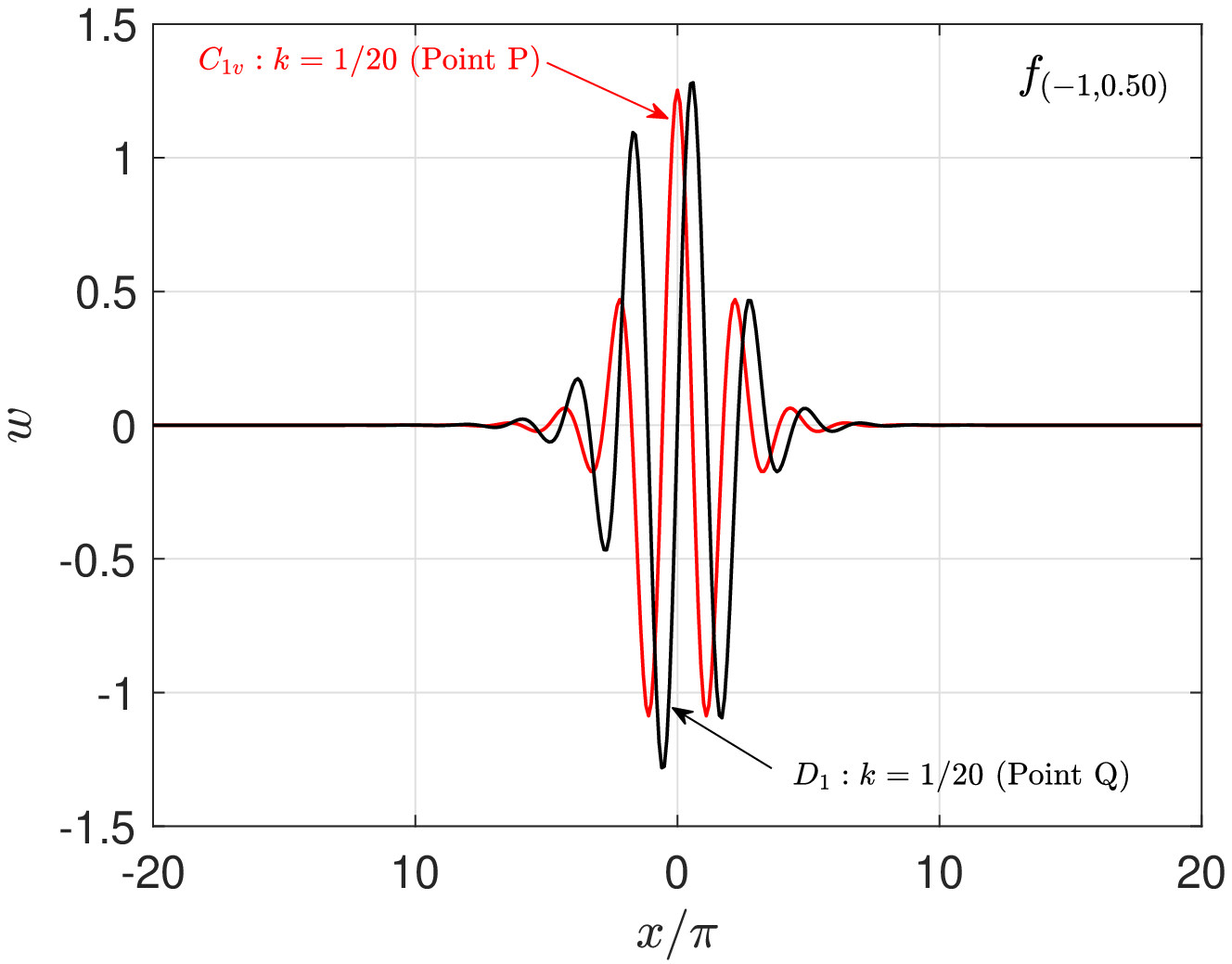}\\
(a)\\
\includegraphics[width=0.55\textwidth]{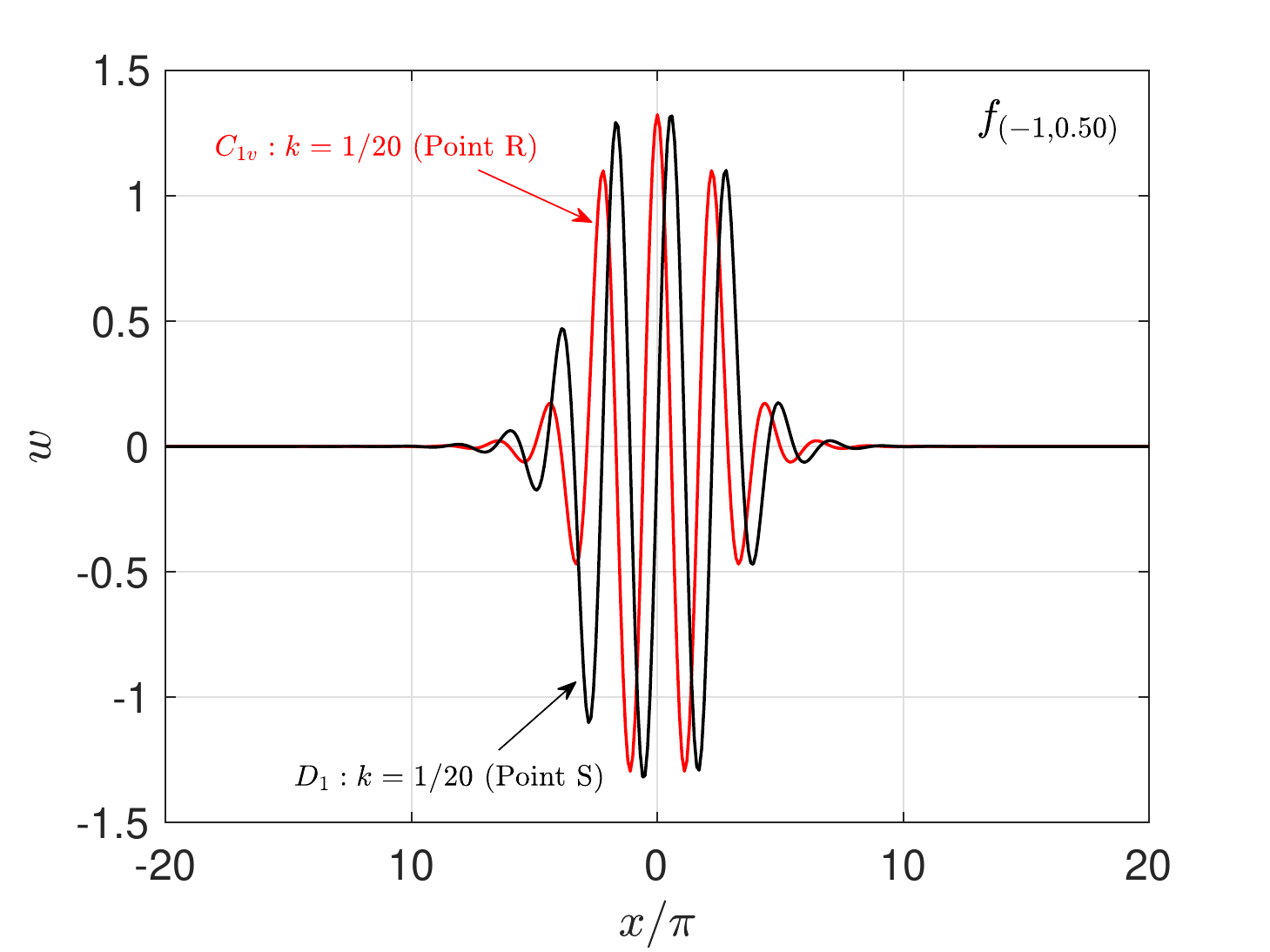}\\
(b)\\
\includegraphics[width=0.55\textwidth]{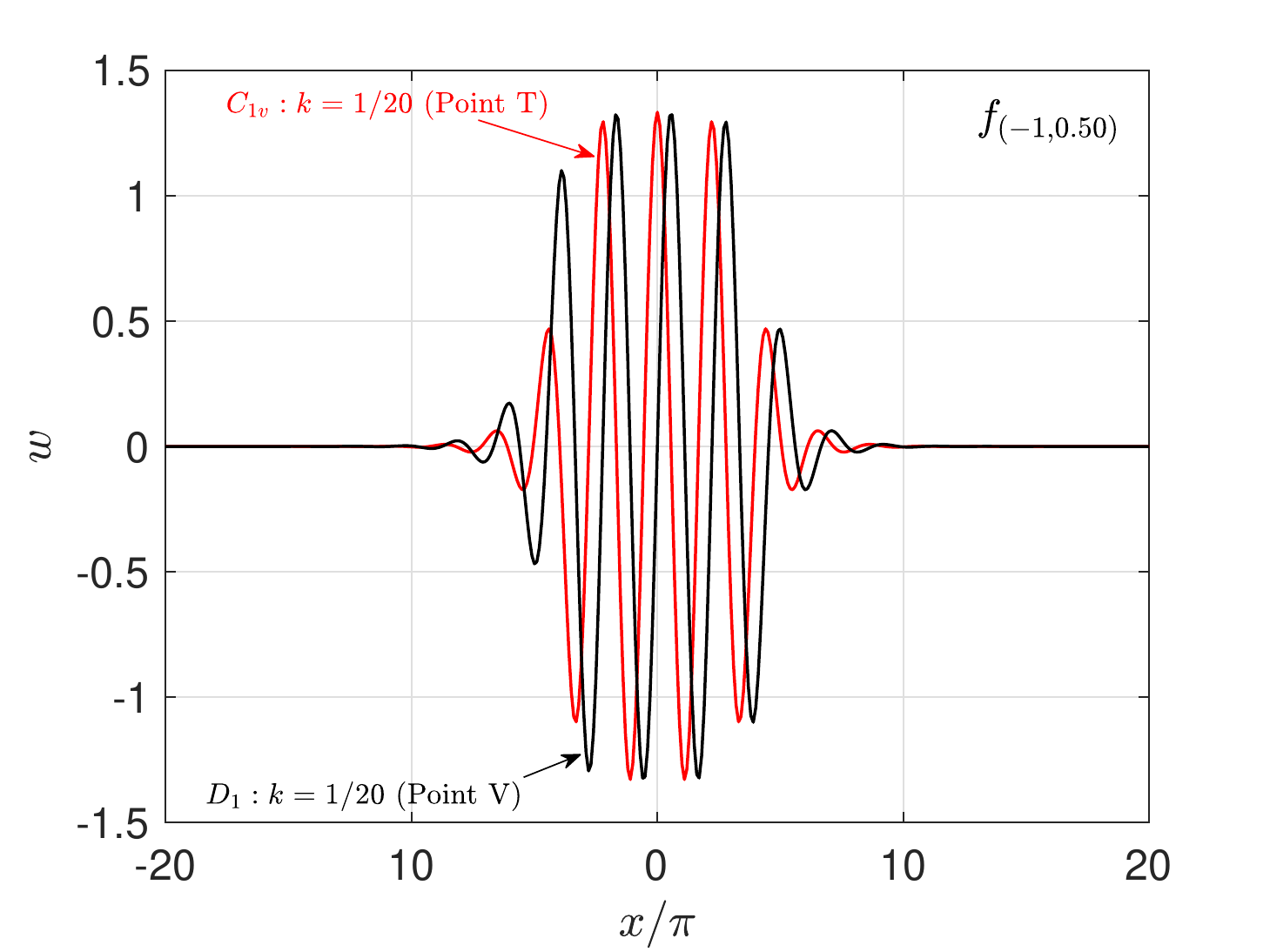}\\
(c)
\end{center}

\caption{Symmetric and antisymmetric stable solutions $w(x)$ for the softening foundation with strong re-hardening ($\alpha = -1,\ \gamma = 0.50$), calculated for a $q=20\ (L_d= 40 \pi)$ supercell at load $\lambda = 1.62$ but different strains $\Delta$ for the secondary bifurcation orbits corresponding to the lowest value of the wavenumber $k=1/20$.  (a)~Points P and Q, (b)~Points R and S, and (c)~Points T and V (see Fig.~\ref{Fig:soft-strong-dia}(c)).}
\label{Fig:soft-strong-mode}
\end{figure*}

The shape of the equilibrium paths for the symmetric and antisymmetric (with respect to $x=0$) secondary bifurcation orbits calculated for a load $\lambda = 1.62$ and corresponding to the bifurcation with the lowest wavenumber ($k=1/20$) are plotted in Fig.~\ref{Fig:soft-strong-mode}.
These are all stable, with the solutions at points P, R, and T corresponding to symmetric ($C_{1v}$) orbits and the ones at Q, S, and V corresponding to antisymmetric ($D_{1}$) ones.  Once again the deformation is localized at the middle of the supercell calculation domain.  However, the localized region for the strong re-hardening foundation consists of a concentrated packet of high amplitude oscillations as seen in Fig.~\ref{Fig:soft-strong-mode}(a) in contrast to the highly localized deformation observed in Fig.~\ref{Fig:soft-mild-mode}(a); although, due to the re-hardening behavior, the evolution of deformation (see Fig.~\ref{Fig:soft-mild-mode}(a)--(b) and Fig.~\ref{Fig:soft-strong-mode}(a)--(c)) in both the models show an increase in the number of oscillations with the applied strain.

Reviewing the results for the softening foundation models ($\alpha=-1$), we observe that the primary bifurcation orbit corresponding to the lowest (i.e.\ critical) load $\lambda_c = 2$ is pitchfork and subcritical---with the respect to the load---and unstable at least for small amplitudes $\xi$.  Moreover, with increasing values of $\xi$ on the primary path, the mode corresponding to the lowest wavenumber (i.e.\ $k=1/q$ for the supercell $L_d = 2\pi q$) becomes unstable first followed by higher wavenumber modes.  The critical load is an accumulation point for the cascading secondary bifurcation points.  From these bifurcation points we find the deformation with only one localization zone is obtained by following the secondary orbit with the lowest wavenumber $k=1/q$. Although this orbit is always initially unstable, we find that for the case of the re-hardening foundations it stabilizes, thus explaining the mechanism of generating stable---and hence observable---highly localized deformation solutions.  It should also be repeated at this point that the secondary bifurcation orbits have their own symmetry groups (albeit smaller than the group $D_{qd}$ of the primary orbit) and hence symmetry-breaking tertiary bifurcations are possible.  However, since the mechanism for the development of stable, highly localized deformation solutions has been established with the help of secondary bifurcation orbits, these higher-order bifurcation paths will not be explored.

\subsubsection{Secondary Bifurcation Orbits for the Stiffening Foundation ($\alpha=+1$)}

\begin{figure*}
\hspace{-0.5cm}
\begin{center}
\includegraphics[width=0.55\textwidth]{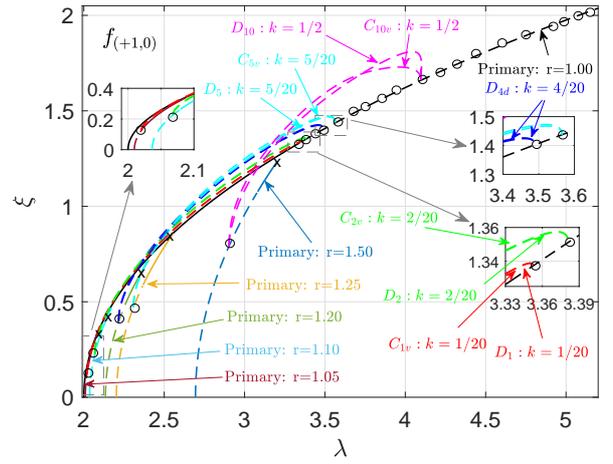}\\
(a)\\
\includegraphics[width=0.55\textwidth]{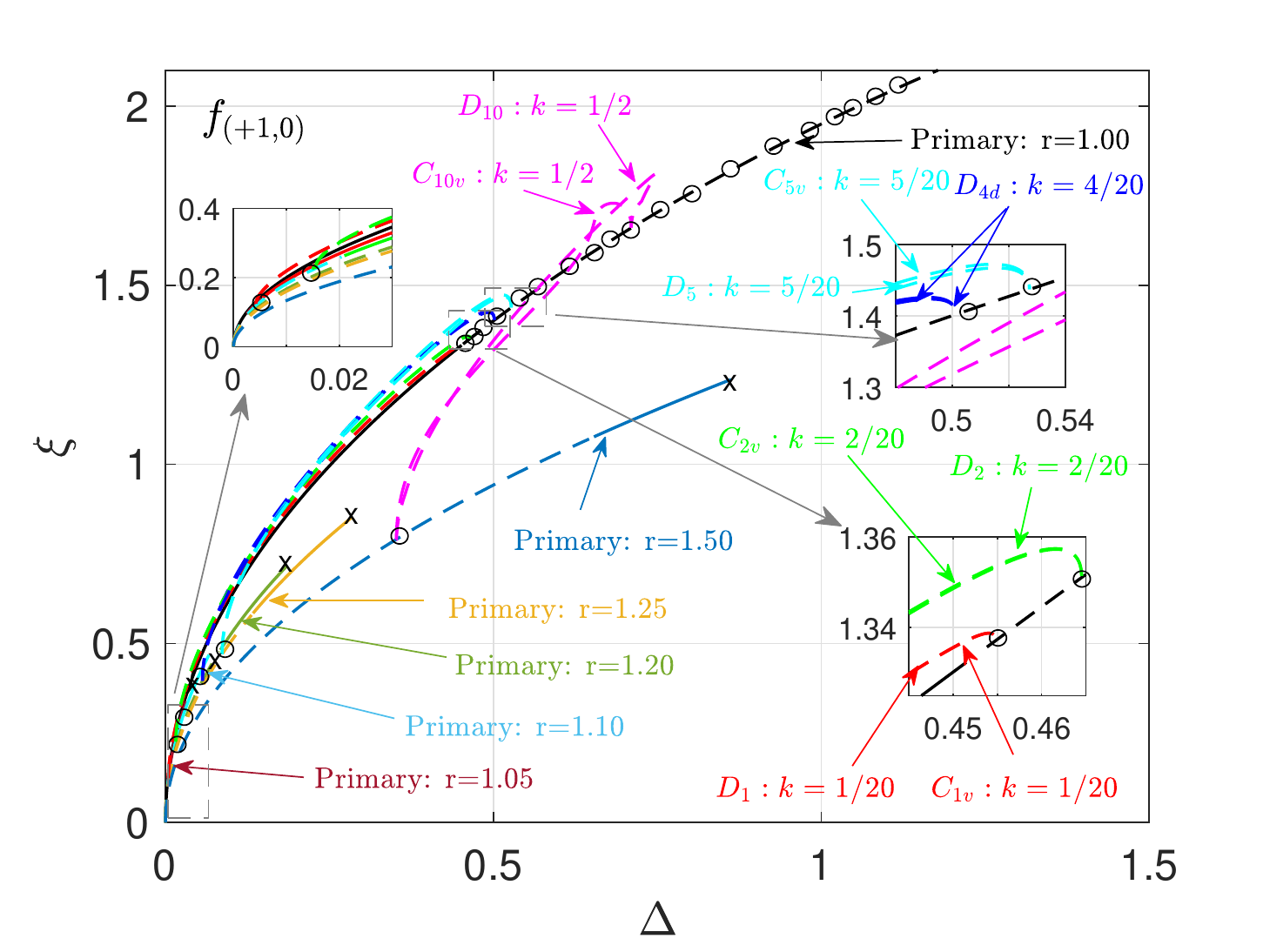}\\
(b)\\
\includegraphics[width=0.55\textwidth]{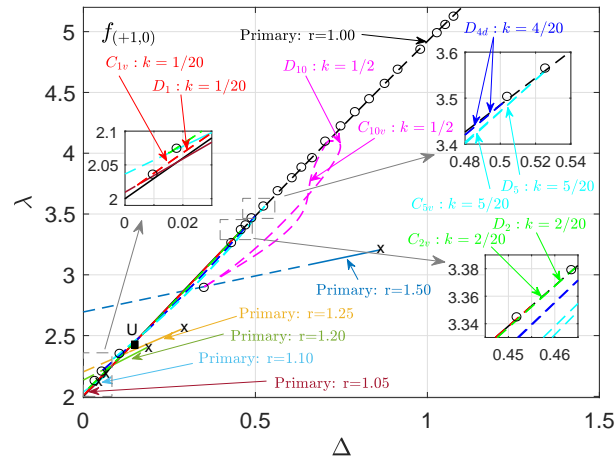}\\
(c)
\end{center}
\caption{Emergence of secondary bifurcations from the primary bifurcation orbit in the case of the monotonically hardening foundation ($\alpha = +1,\ \gamma = 0$).  All bifurcation points are indicated by a small circle.  Solid and dashed lines correspond, respectively, to the stable and unstable parts of these equilibrium paths, based on Bloch-wave analysis of a $q=20\ (L_d = 40\pi)$ supercell.  Notice that the primary orbit is stable from its origin at $\lambda_c = 2$ up to point $(\lambda, \Delta) \approx (3.4, 0.5)$, beyond which secondary bifurcation orbits emerge.  These eventually join primary orbits emerging from the principal solution at $\lambda > 2$ (orbits with unit cell periods $L_d = 2 \pi/r, r > 1$, see also Eq.~\eqref{eq:highercritical}).  Results are presented only for orbits corresponding to the \emph{simple harmonic} modes $k =n/q$, $n = 1, 2, 4, 5, 10$ and show the (a)~amplitude $\xi$ vs load $\lambda$, (b)~amplitude $\xi$ vs strain $\Delta$, and (c)~load $\lambda$ vs strain $\Delta$ bifurcation diagrams.}
\label{Fig:hard-dia}
\end{figure*}

Attention is now turned to the secondary bifurcated equilibrium orbits emerging from the primary one, i.e.\ solutions to Eqs.~\eqref{eq:equilibrium},~\eqref{eq:secondary_BC} for a beam of length $L_d = 2\pi q$ resting on a monotonically stiffening foundation ($\alpha = +1$).  As seen in Fig.~\ref{Fig:primary-diagrams}, the corresponding primary bifurcation orbit is supercritical, i.e.\ increasing with respect to both load $\lambda$ and strain $\Delta$ and---in contrast to the softening foundation case---stable in the neighborhood of the critical load $\lambda_c = 2$ (see also the dispersion results in Fig.~\ref{Fig:dispersion-hard}).  The secondary bifurcation orbits corresponding to the simple harmonic modes $k =n/q$, $n = 1, 2, 4, 5, 10$ are recorded in Fig.~\ref{Fig:hard-dia}.
They correspond to pitchfork bifurcations with two orbits emerging from each bifurcation point, except for the one with $k=1/4$ which is globally transverse with only one orbit emerging at the bifurcation.  These unstable secondary orbits eventually join the supercritical, globally pitchfork primary orbits emerging at loads $\lambda > 2$ (and hence with unit cell lengths $L_d = 2 \pi / r, r > 1$, see also Eq.~\eqref{eq:highercritical} and e.g., \cite{Hunt_et_al}).  It is also worth mentioning that the primary bifurcations emerging for $\lambda > 2$ are initially unstable for small values of the amplitude $\xi$, but at larger amplitudes they stabilize.  This can be as seen in Fig.~\ref{Fig:hard-dia} for the primary bifurcation paths with $r=1.5$, $1.25$, and $1.20$ (periods $L_d= 2\pi/ r = 4 \pi /3$, $8 \pi/5$, and $5 \pi/3$).

\begin{figure*}
\centering
\includegraphics[width=0.55\textwidth]{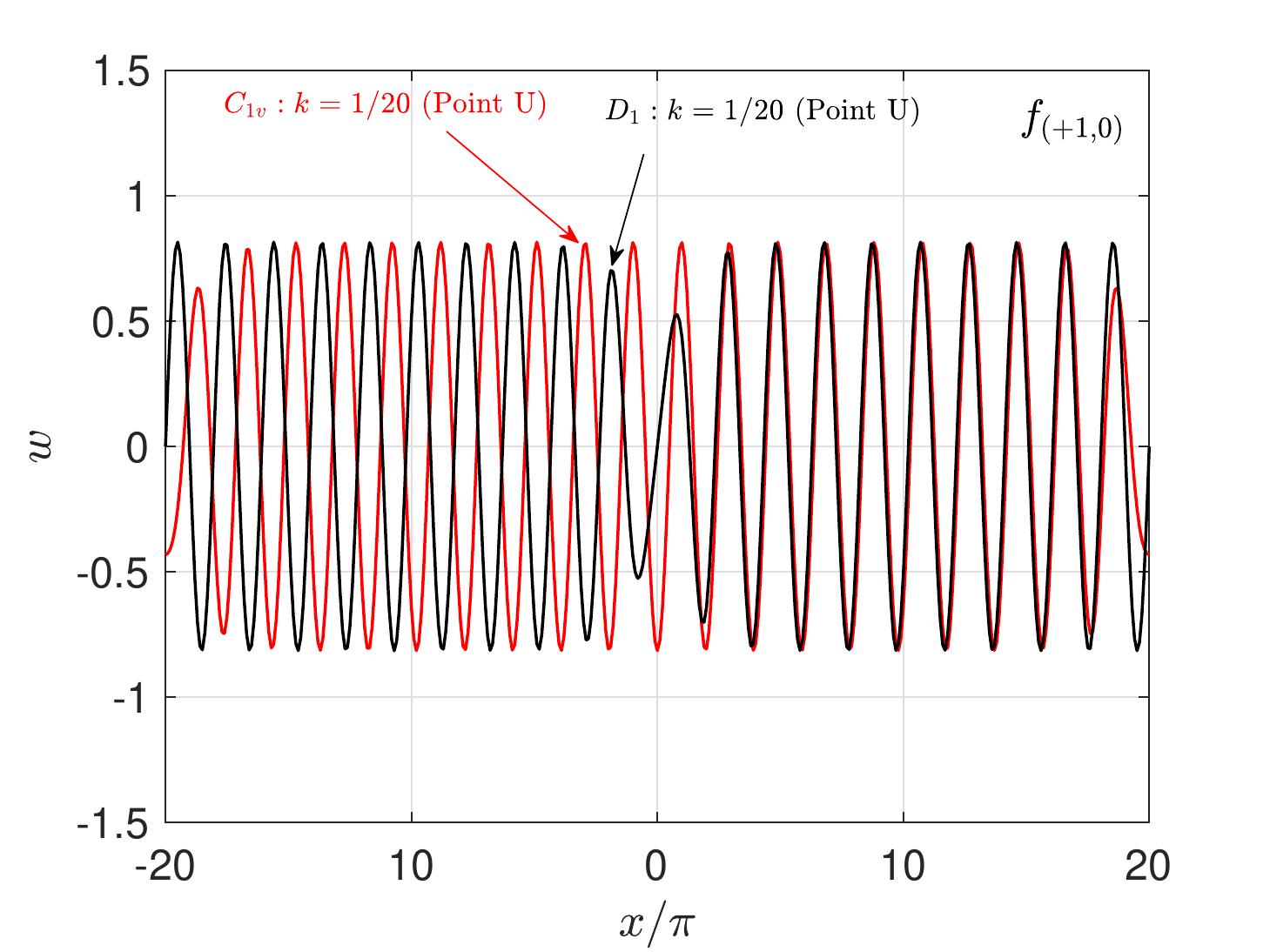}
\caption{Symmetric and antisymmetric unstable solutions $w(x)$ for the monotonically hardening foundation ($\alpha = +1,\ \gamma = 0$), calculated for a $q=20\ (L= 40 \pi)$ supercell at load $\lambda = 2.48$ for the secondary bifurcation orbits corresponding to the lowest value of the wavenumber $k=1/20$ (see Point~U in Fig.~\ref{Fig:hard-dia}(c)).}
\label{Fig:hard-mode}
\end{figure*}

The hardening foundation model presents an important difference with the softening foundation ones: the absence of equilibrium solutions with a single, highly localized deformation zone.  Typical results are shown in Fig.~\ref{Fig:hard-mode} depicting symmetric (even with respect to $x=0$) and antisymmetric (odd with respect to $x=0$) unstable solutions.
These configurations correspond to the monotonically hardening foundation ($\alpha = +1,\ \gamma = 0$), calculated for a $q=20 \ (L_d=40 \pi)$ supercell at a load $\lambda=2.48$ and are associated with the secondary orbit emerging from the bifurcation point with the lowest wavenumber ($k=1/20$, see point U in Fig.~\ref{Fig:hard-dia}(c)).  These unstable secondary bifurcated orbits eventually connect to an, also unstable, primary bifurcation orbit emerging from the principal solution at $\lambda > 2$.  This helps explain the shapes observed in Fig.~\ref{Fig:hard-mode}.  Only the critical primary orbit is stable near $\lambda_c = 2$.  However, once the load $\lambda$ becomes somewhat larger, $\lambda > \lambda_c$, there exist stable primary orbits.

\subsection{Imperfect Model}
\label{sec:ImperfectModel}

The imperfect model for $L_d = L_c q$ with $q =20$ has the same trivial solution ($w(x; \lambda)=0$) as its perfect counterpart.  Its considerably lower symmetry group ($D_{1h}$) results in simple bifurcations (of symmetry\footnote{There are also simple bifurcations of symmetry $D_1$ which occur at loads $\lambda_{cia}$, but $\lambda_{cia} > \lambda_{cis}$ (see Appendix~\ref{A-imp-mode}).  Here we are interested in the first (i.e.\ lowest load) bifurcation point, so we do not present the imperfect bifurcations with $D_1$ symmetry.} $C_{1v}$) at loads $\lambda_{cis}$ that depend on the imperfection amplitude $\zeta$ and size $2 x_0$.  This critical load and corresponding eigenmode can be found analytically and the corresponding calculations are given in Appendix~\ref{A-imp-mode}.  Starting at this critical load and using the corresponding eigenmode as an initial guess, we construct the equilibrium solutions for an imperfect beam and compare them to the corresponding solutions of its perfect counterpart of the same length.

\begin{figure*}
\begin{minipage}{0.50\textwidth}
\begin{center}
\includegraphics[width=\textwidth]{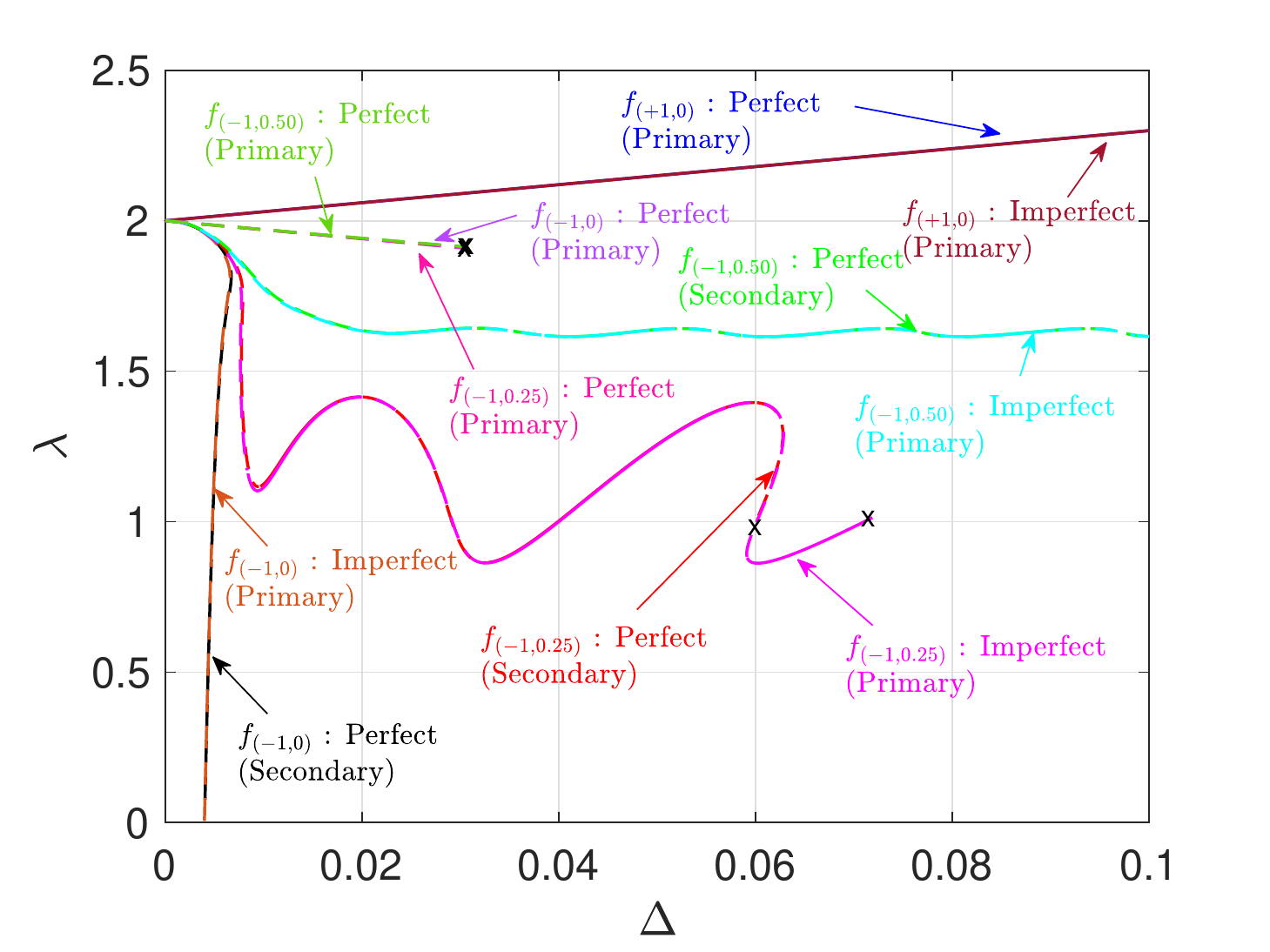}\\%
(a)\\%
\includegraphics[width=\textwidth]{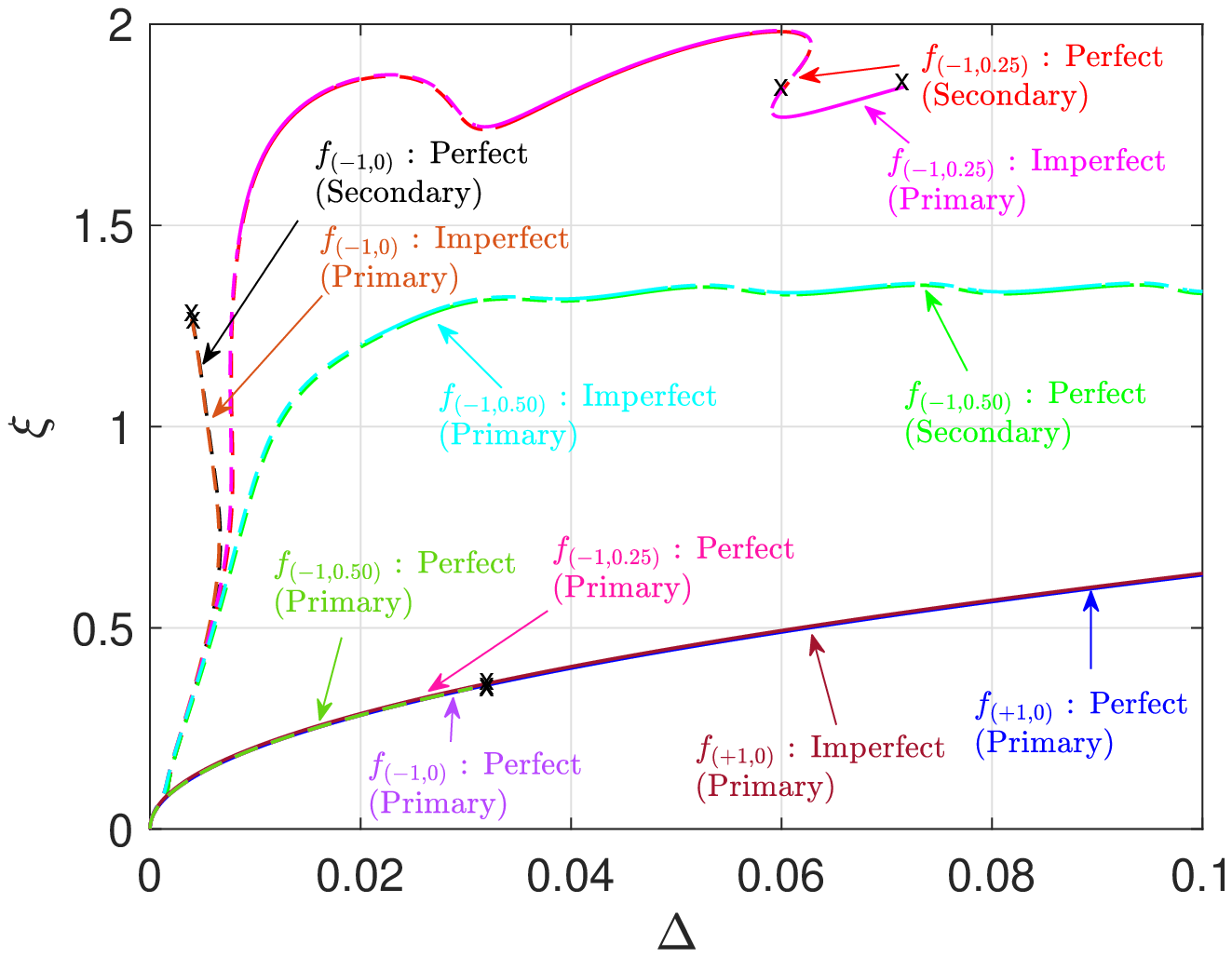}\\
(c)%
\end{center}
\end{minipage}%
\hspace*{-0.5cm}%
\begin{minipage}{0.50\textwidth}
\begin{center}
\includegraphics[width=\textwidth]{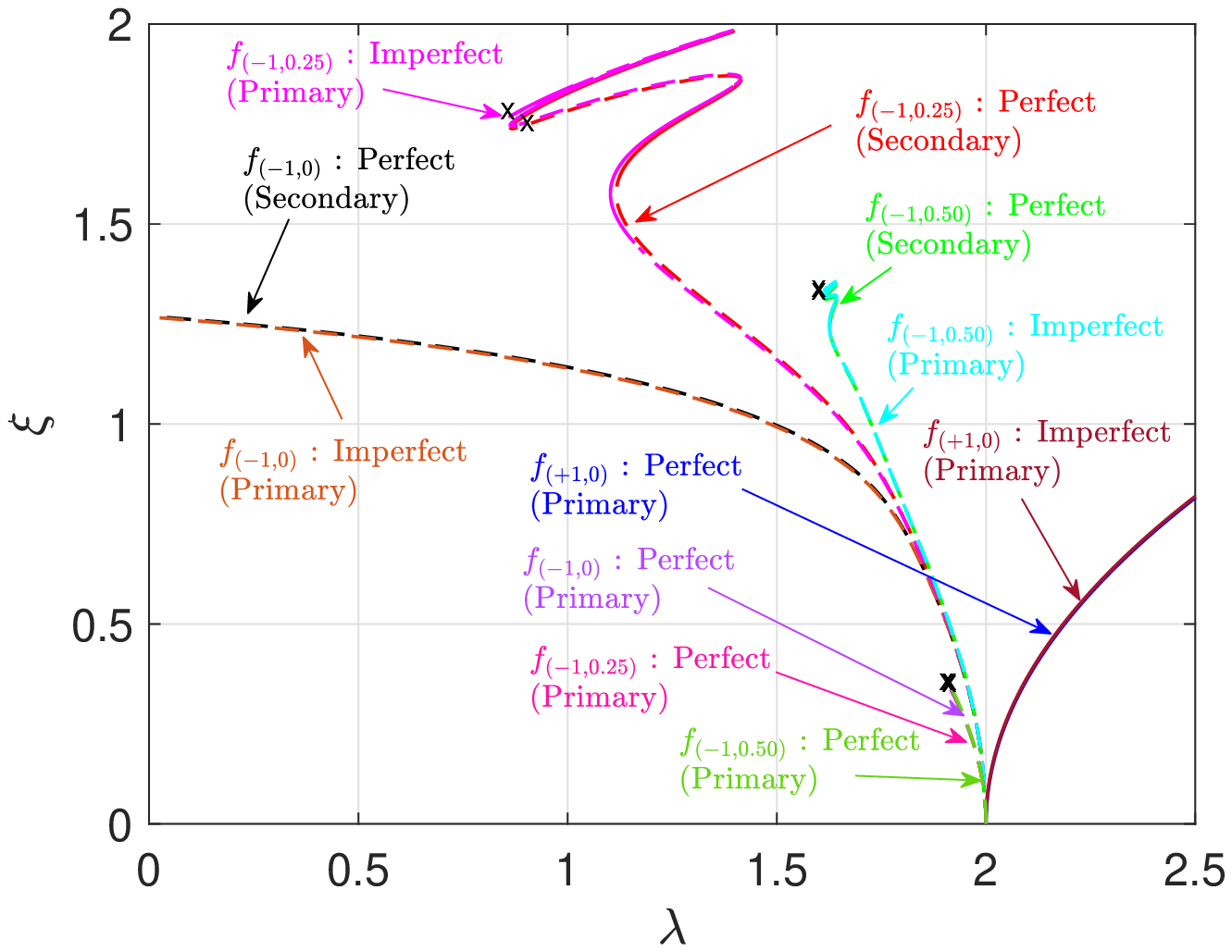}\\%
(b)\\%
\includegraphics[width=\textwidth]{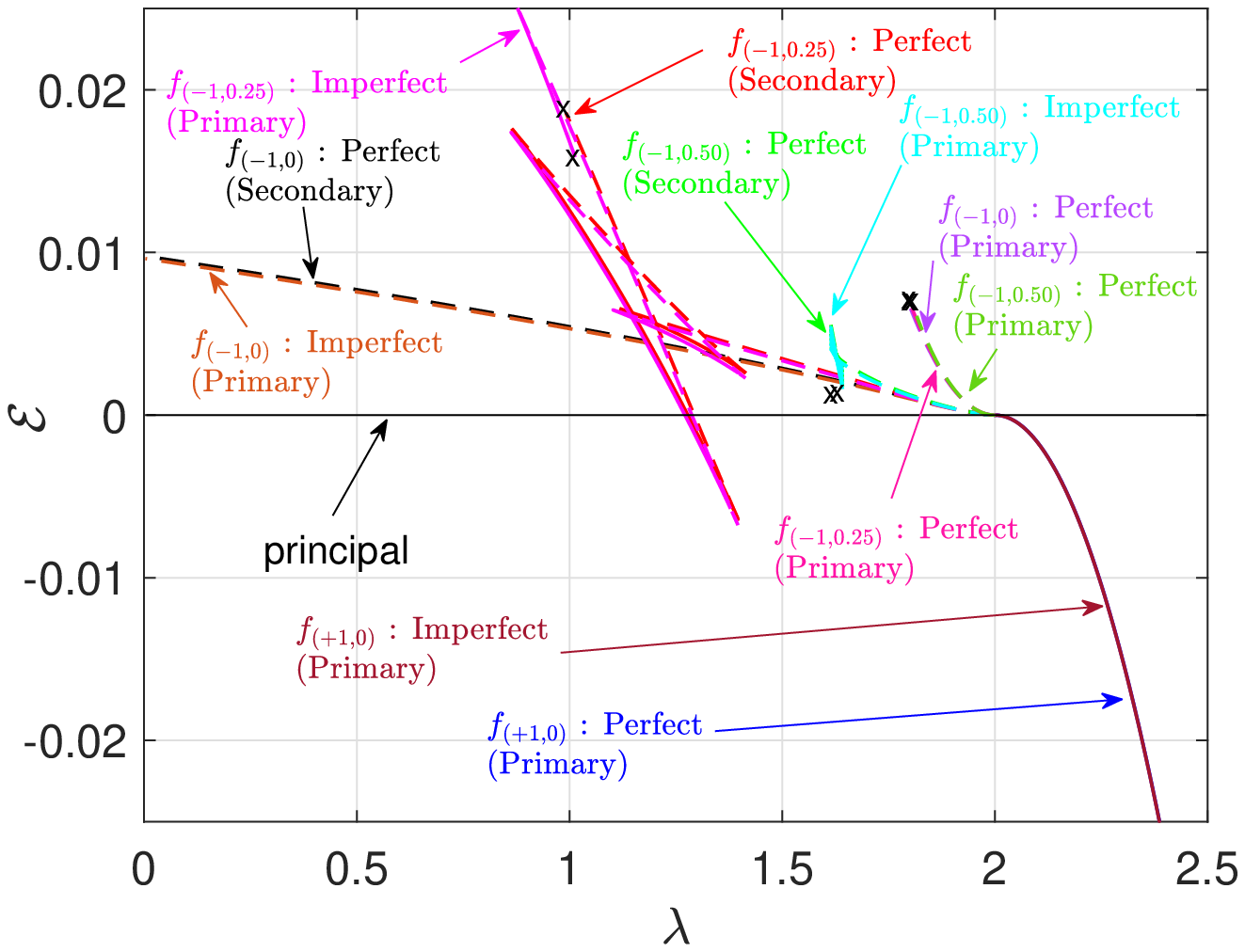}\\%
(d)%
\end{center}
\end{minipage}
\caption{Emergence of the primary bifurcation paths, termed {\it imperfect} solutions, from the principal solution ($w(x; \lambda) = 0$) for a periodically imperfect ($\zeta= -0.01$, $2x_0 = 0.01L_d$) infinite beam of period $L_d=40 \pi$ for the four different foundations considered in this work.  Results are compared to the corresponding secondary bifurcated equilibrium orbits of a perfect beam emerging from bifurcation points with an eigenmode wavenumber $k=1/20$, calculated for a $q=20$, $(L_d=40\pi$, same length) supercell (termed {\it perfect} solutions).  Solid and dashed lines correspond, respectively, to the stable and unstable parts of these different equilibrium paths, based on Bloch-wave analysis of a $q=20$ ($L_d=40\pi$) supercell.  Plots show the (a)~load $\lambda$ vs strain $\Delta$, (b)~amplitude $\xi$ vs load $\lambda$, (c)~amplitude $\xi$ vs strain $\Delta$, and (d)~energy density $\mathcal{E}$ vs load $\lambda$.}
\label{Fig:imperfect-dia}
\end{figure*}

Figure~\ref{Fig:imperfect-dia} shows a comparison of the imperfect and perfect beam solutions.  Results for each of the four different foundations considered in this work are included.  The primary bifurcation paths (termed {\it imperfect} solutions) of an imperfect ($\zeta= - 0.01$, $2x_0 = 0.01L_d$) periodic beam of period $L_d=40 \pi$ emerge from the bifurcation point with eigenmode $\stackrel{is}{w}(x)$ (see Eq.~\eqref{eq:symmetricmode}).  The corresponding secondary bifurcation orbits of the perfect periodic beam, (termed {\it perfect} solutions) are also shown.
The secondary bifurcated equilibrium orbits of a perfect infinite, periodic beam emerge from the bifurcation point with the lowest wavenumber $k=1/20$.  Plots in Fig.~\ref{Fig:imperfect-dia} show the (a) load $\lambda$ vs strain $\Delta$, (b) bifurcation amplitude $\xi$ vs load $\lambda$, (c) bifurcation amplitude $\xi$ vs strain $\Delta$, and (d) energy density $\mathcal{E}$ vs load $\lambda$.  Recall the convention that solid and dashed lines correspond, respectively, to the stable and unstable parts of the different equilibrium paths.

\begin{figure*}
\begin{minipage}{0.53\textwidth}%
\begin{center}%
\includegraphics[width=\textwidth]{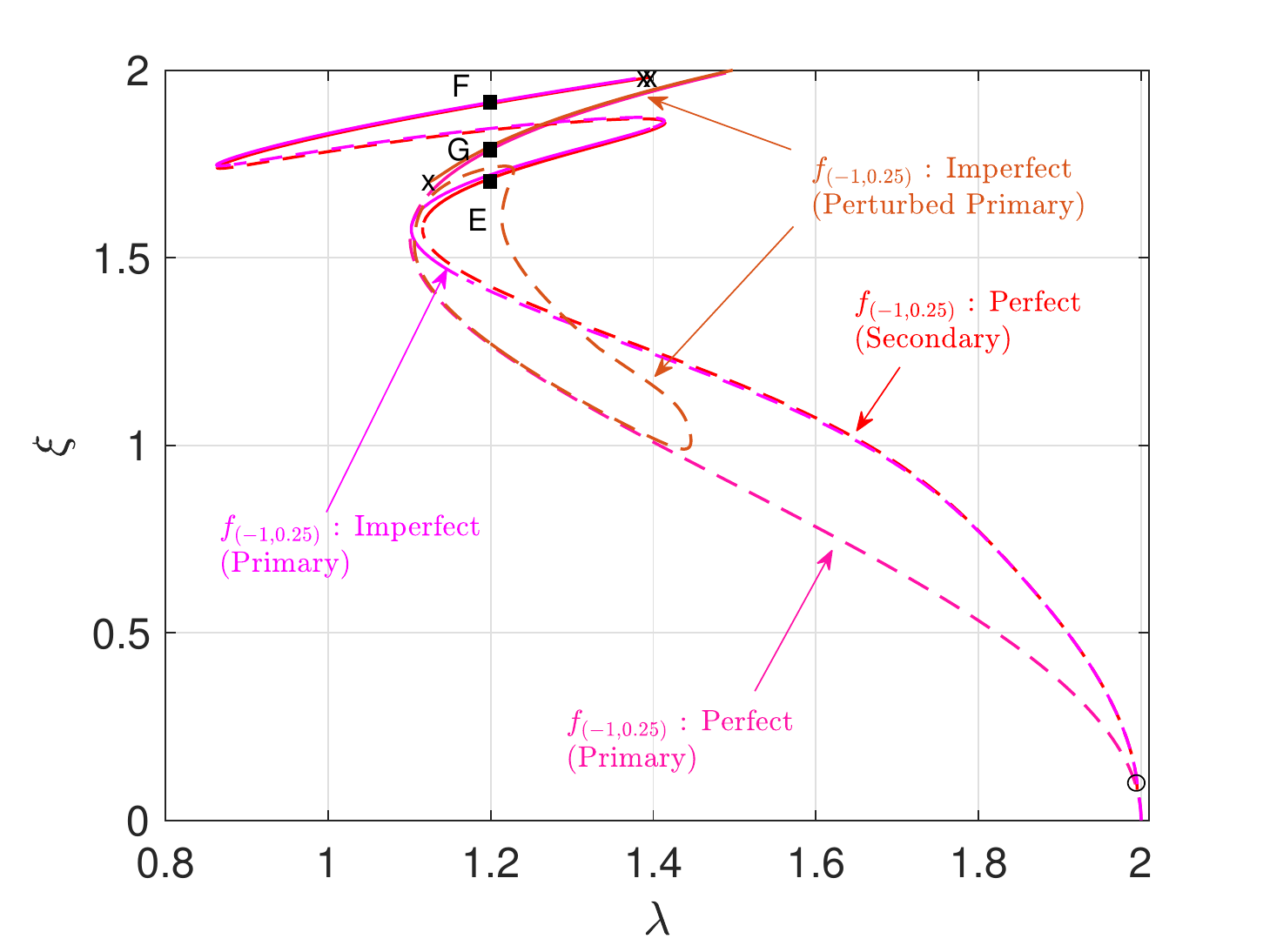}\\%
(a)\\
\includegraphics[width=\textwidth]{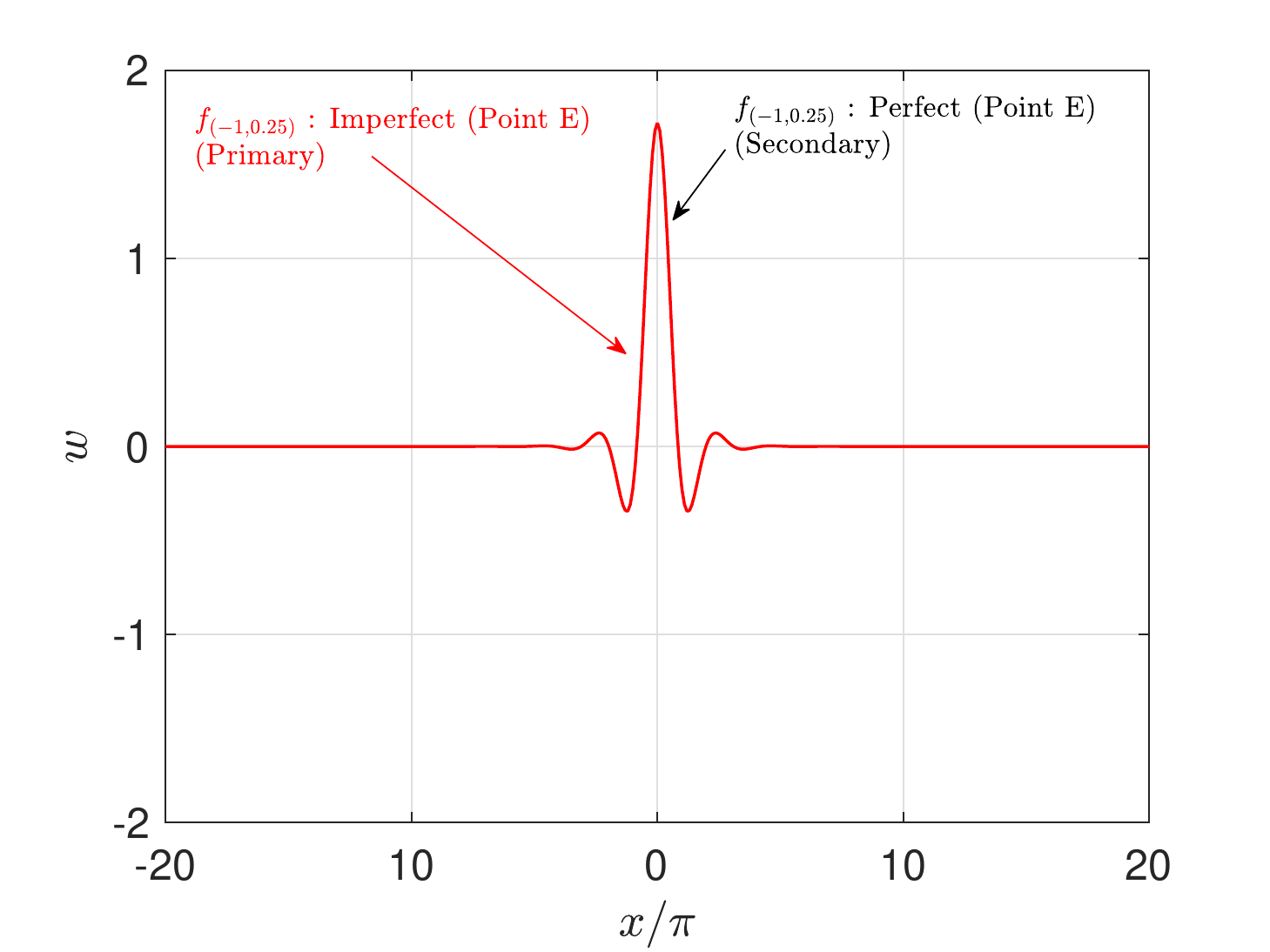}\\%
(c)%
\end{center}%
\end{minipage}%
\hspace*{-0.5cm}%
\begin{minipage}{0.53\textwidth}%
\begin{center}%
\includegraphics[width=\textwidth]{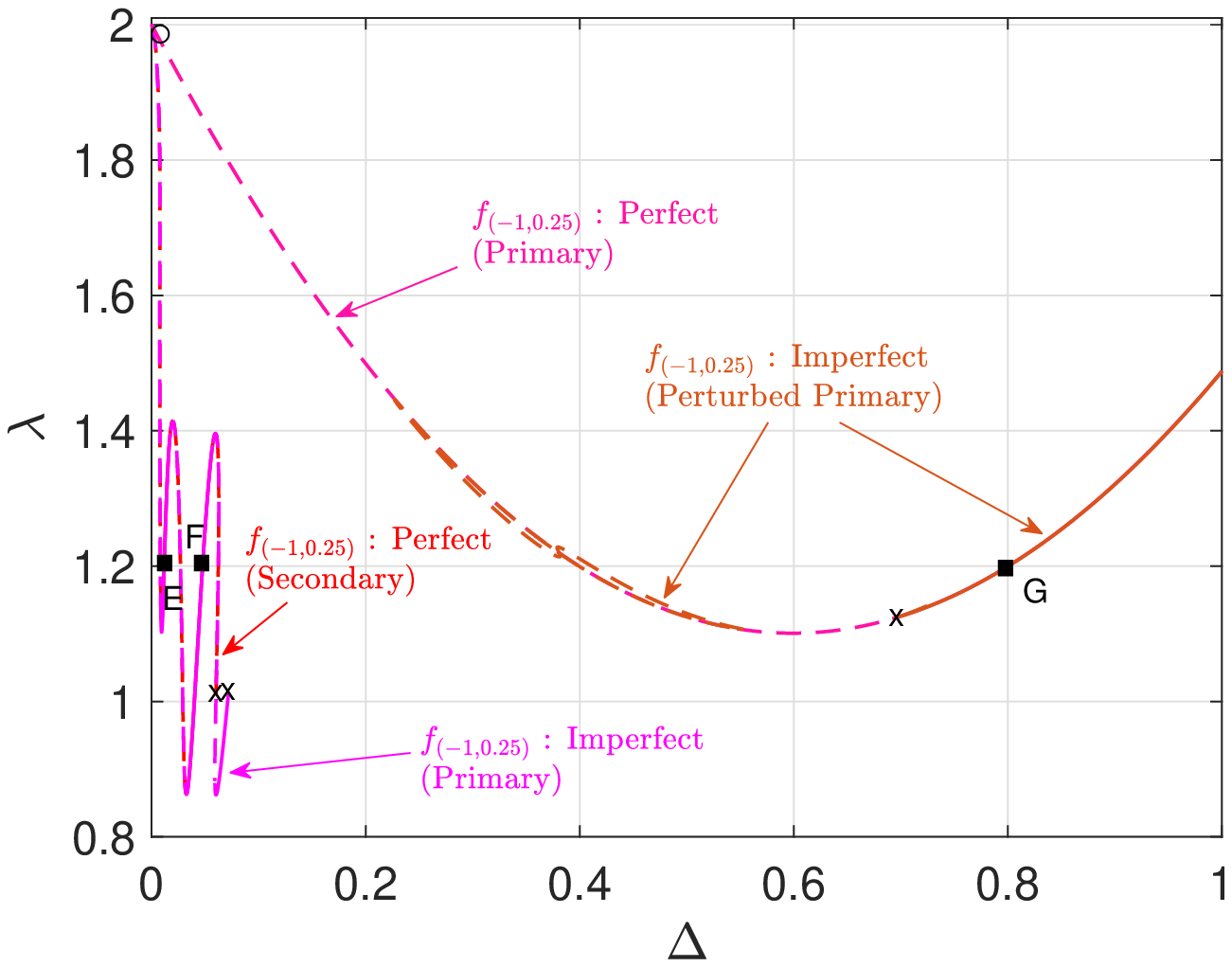}\\%
(b)\\%
\includegraphics[width=\textwidth]{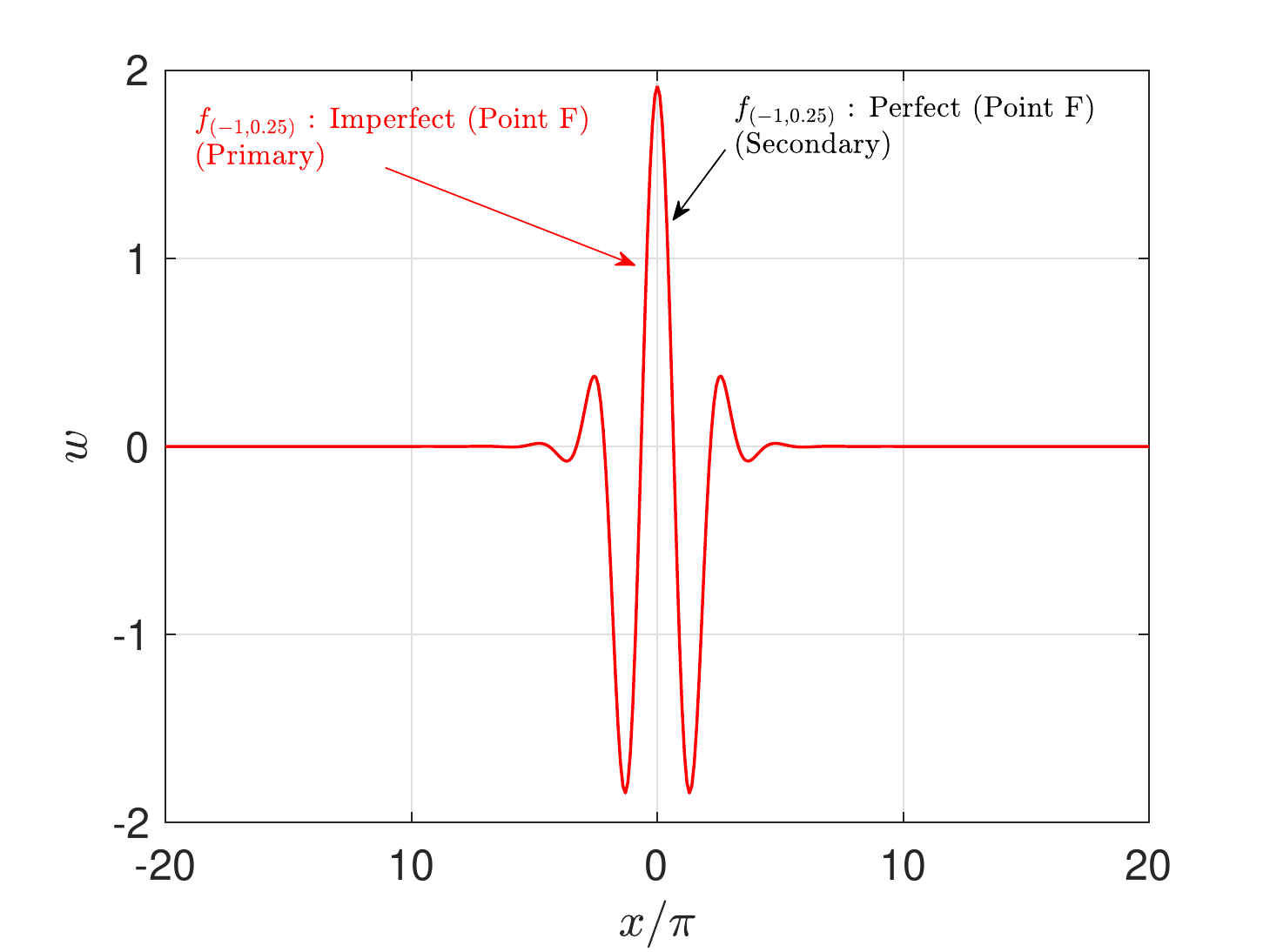}\\%
(d)%
\end{center}%
\end{minipage}\\%
\begin{center}%
\includegraphics[width=0.53\textwidth]{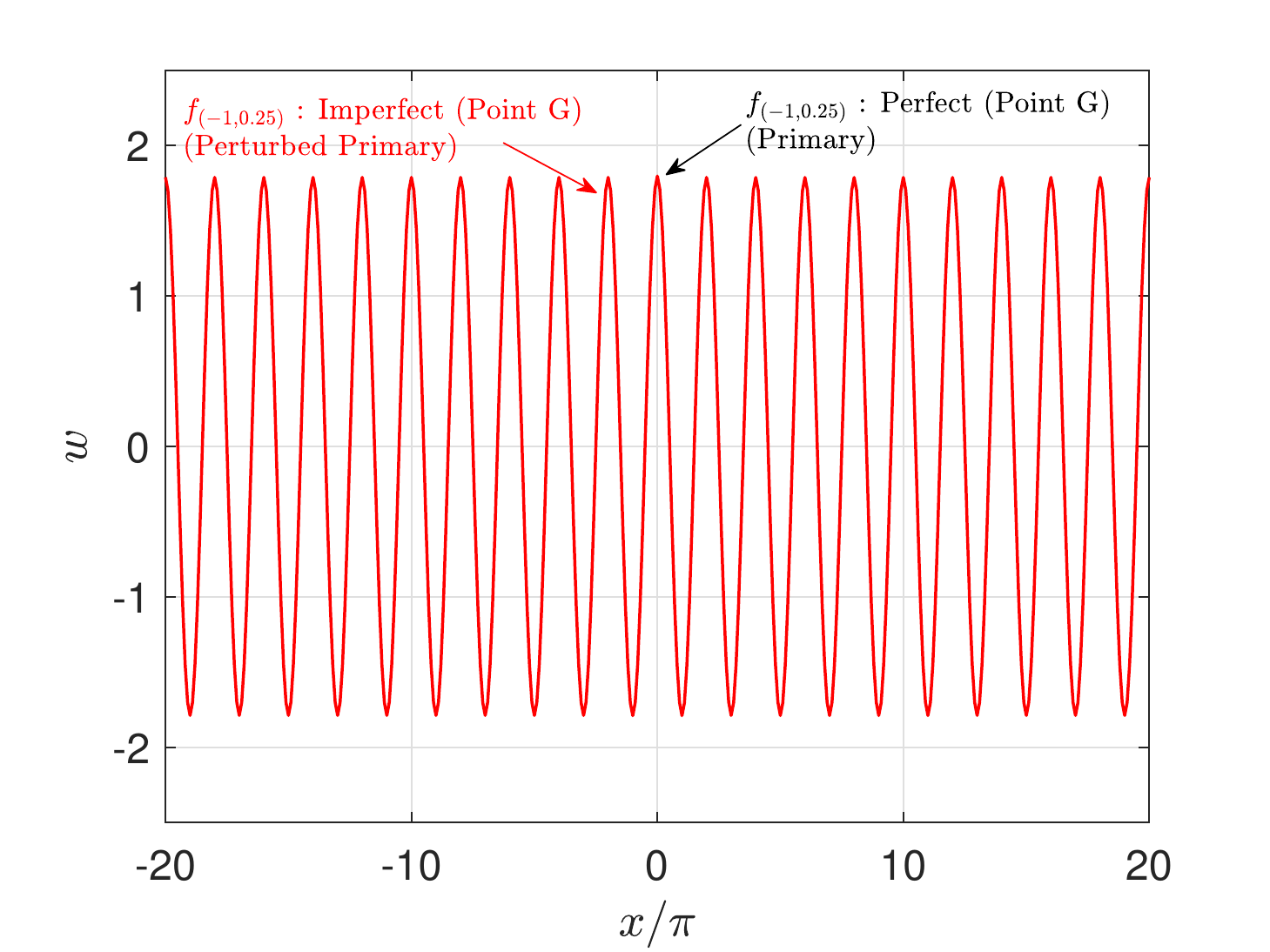}\\%
(e)%
\end{center}
\caption{Bifurcation diagram and stable deformed configurations emphasizing the coexistence of multiple stable solutions at the same value of load $\lambda \approx 1.2$, which is well below the critical load $\lambda_c = 2$ of the perfect, mild re-hardening ($\alpha = -1$, $\gamma = 0.25$) beam--foundation system.  (a) Amplitude $\xi$ vs load $\lambda$ bifurcation diagram showing only the perfect primary and $C_{1v} : k=1/20$ secondary bifurcated paths, and their imperfect counterparts, for the mild re-hardening foundation (cf.\ Fig.~\ref{Fig:soft-mild-dia}); (b) Load $\lambda$ vs strain $\Delta$ bifurcation diagram; (c) Symmetric fully localized deformation perfect and imperfect solutions (Point~E); (d) Symmetric localized packet deformation perfect and imperfect solutions (Point~F); and (e) Symmetric uniformly wrinkled deformation perfect and imperfect solutions (Point~G).}
\label{Fig:single-localized-dia}
\end{figure*}

For the case of the softening foundations ($\alpha = -1$), the main conclusion from Fig.~\ref{Fig:imperfect-dia} is that a small, localized imperfection results in a primary bifurcation orbit almost indistinguishable from the corresponding secondary bifurcation orbit with the longest wavelength emerging closest to $\lambda_c = 2$ of the perfect structure, as seen also in Figs.~\ref{Fig:soft-none-dia},~\ref{Fig:soft-mild-dia}, and~\ref{Fig:soft-strong-dia}.  Another noteworthy fact is that the stability of the two paths (imperfect and perfect) is the same.  Not only are the overall measures of these perfect and imperfect equilibrium orbits practically coincident (i.e.\ load vs strain, bifurcation amplitude vs strain, etc.), but their shapes (see Fig.~\ref{Fig:single-localized-dia}) are also practically indistinguishable.

For the case of the hardening foundation ($\alpha = +1$), one can see (by comparing the results in Fig.~\ref{Fig:imperfect-dia} to the corresponding ones for the perfect structure in Fig.~\ref{Fig:hard-dia}) that a small, localized imperfection results in a primary bifurcation orbit almost indistinguishable from the corresponding primary bifurcation orbit of the perfect structure.  Again, these two paths (imperfect and perfect) share nearly identical stability properties.

\begin{figure*}
\begin{minipage}{0.53\textwidth}
\begin{center}
\includegraphics[width=\textwidth]{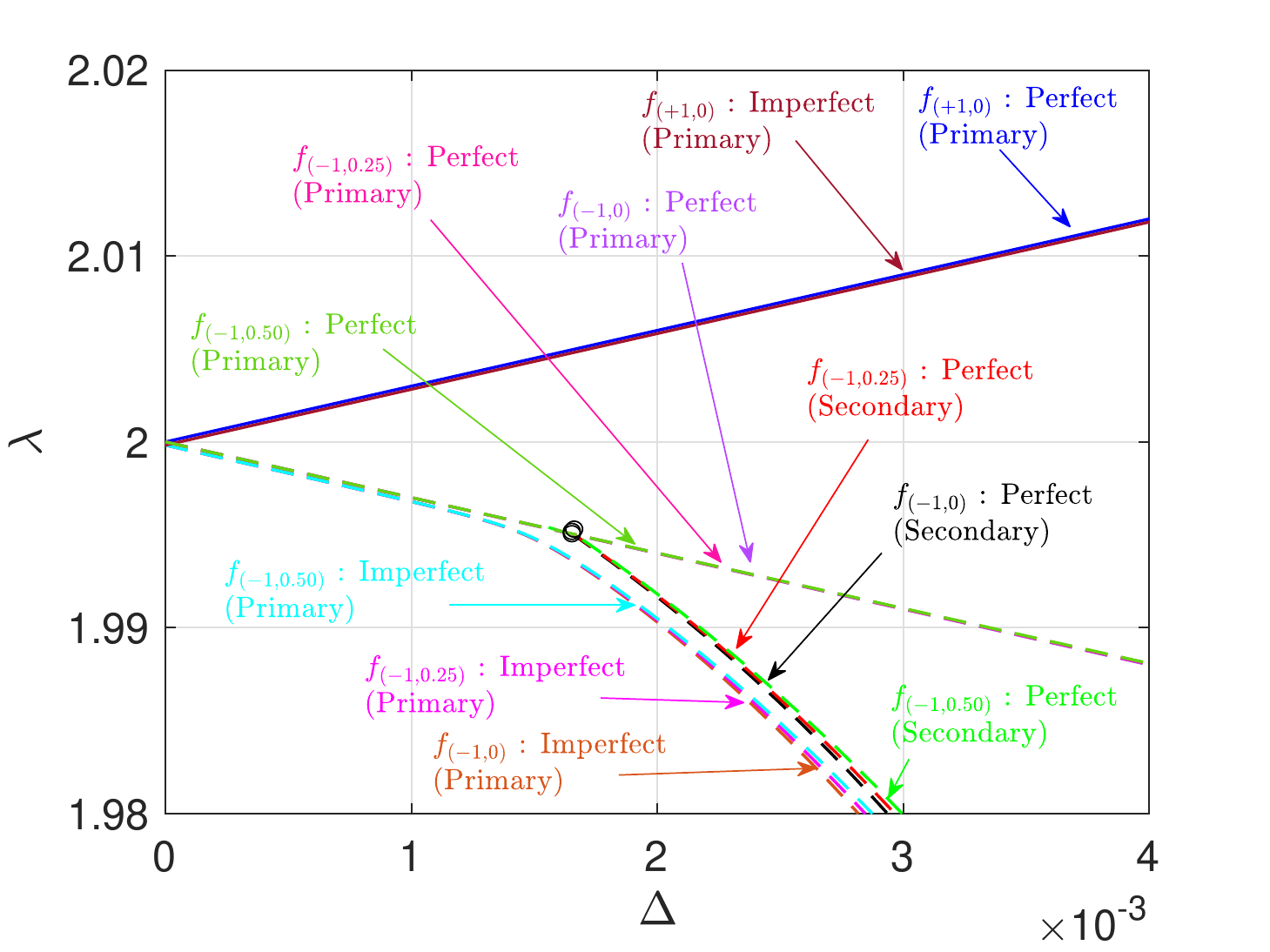}\\%
(a)\\%
\includegraphics[width=\textwidth]{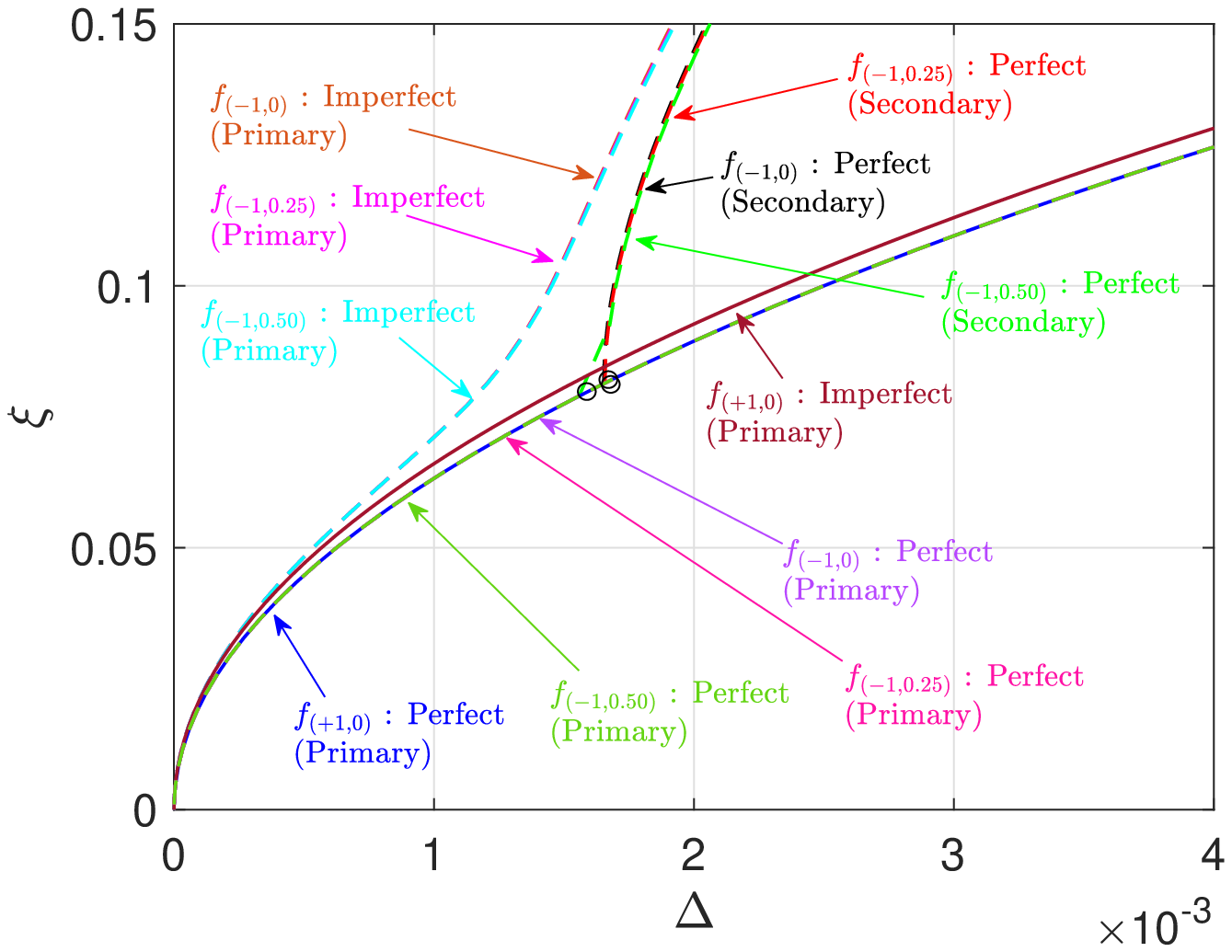}\\
(c)%
\end{center}
\end{minipage}%
\hspace*{-0.5cm}%
\begin{minipage}{0.53\textwidth}
\begin{center}
\includegraphics[width=\textwidth]{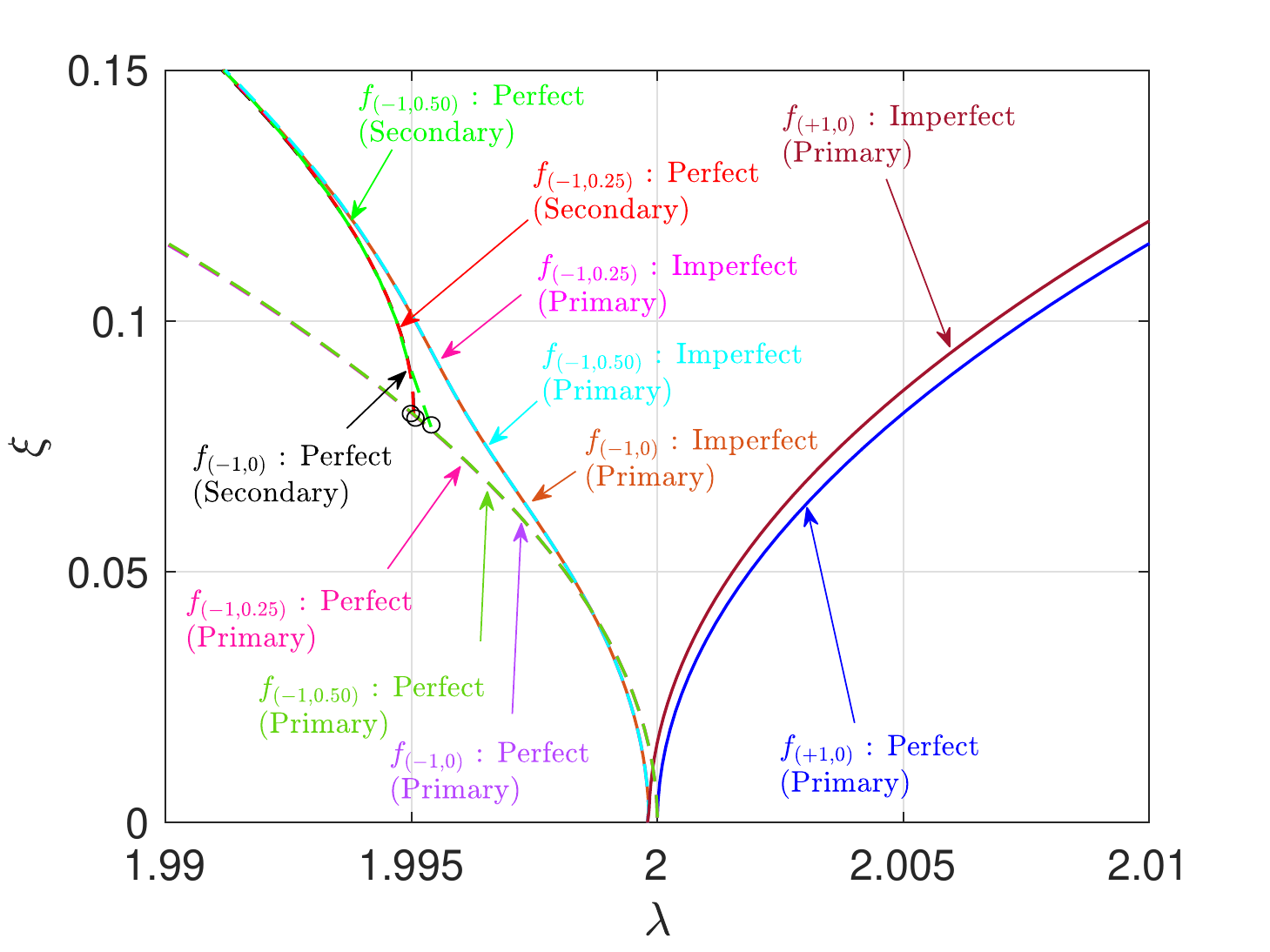}\\%
(b)\\%
\includegraphics[width=\textwidth]{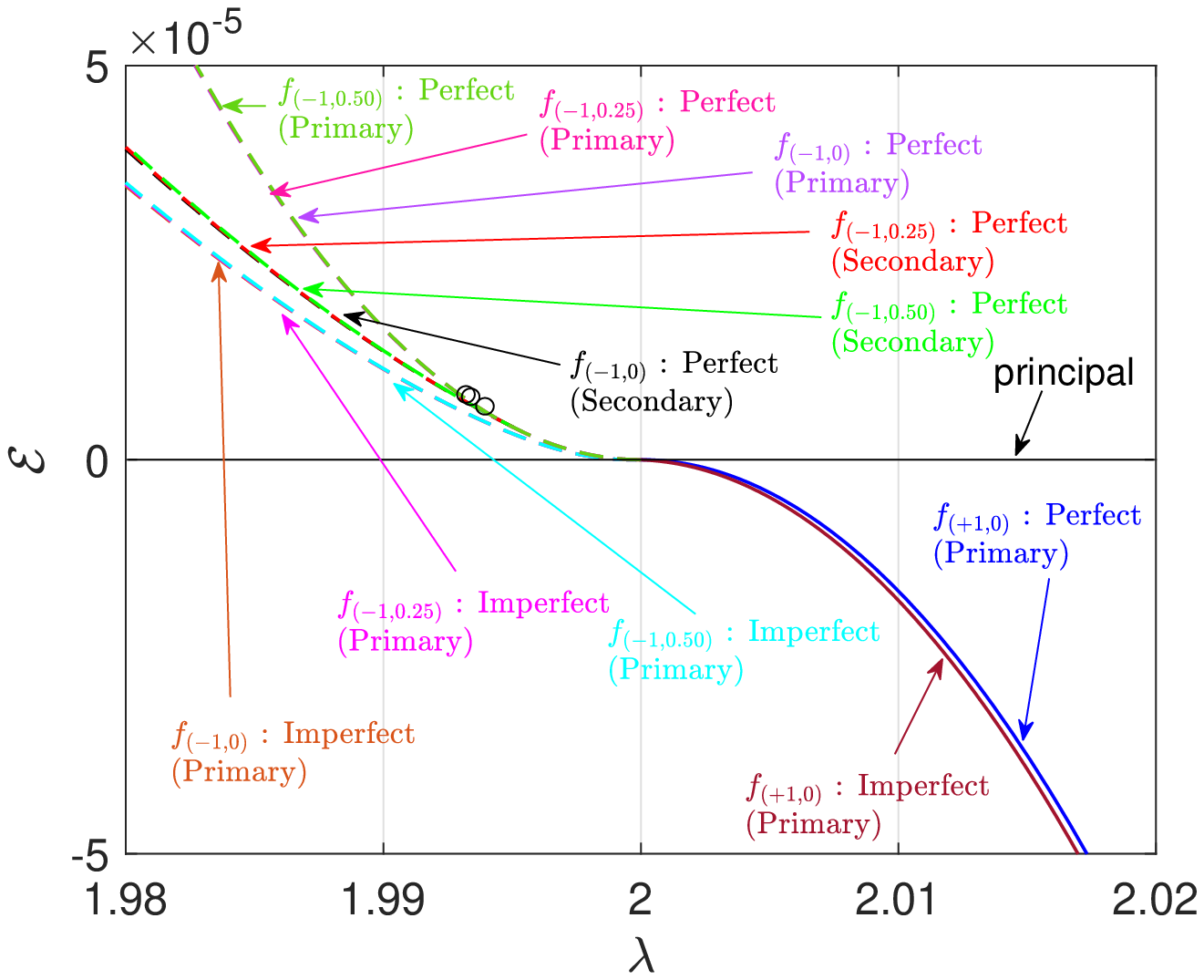}\\%
(d)%
\end{center}
\end{minipage}
\caption{Behavior of the perfect and imperfect beams in a region near the critical load $\lambda_c=2$.  Solid and dashed lines correspond, respectively, to the stable and unstable parts of the different equilibrium paths, based on Bloch-wave analysis of a $q=20\ (L_d=40\pi)$ supercell.  Plots show the (a)~load $\lambda$ vs strain $\Delta$, (b)~amplitude $\xi$ vs load $\lambda$, (c)~amplitude $\xi$ vs strain $\Delta$, and (d)~energy density $\mathcal{E}$ vs load $\lambda$.}
\label{Fig:imperfect-dia-zoom}
\end{figure*}

To understand the emergence of the imperfect equilibrium paths and their connection to the primary and secondary orbits of the perfect structure better, we zoom in to the region near the critical load $\lambda_c=2$ in Fig.~\ref{Fig:imperfect-dia-zoom}.
For the softening foundation case $\alpha=-1$, the graphs of Fig.~\ref{Fig:imperfect-dia-zoom} clearly show that the imperfect equilibrium path, starting from\footnote{From Appendix~\ref{A-imp-mode}, it is clear that  $2(1+\zeta)^{1/2} < \lambda_{cis} < 2$, for $\zeta < 0$.} $\lambda_{cis}$, skirts the primary bifurcated orbit of the perfect structure and subsequently follows the secondary bifurcated orbit with the smallest wavenumber $k=1/20$. If a longer domain had been used, the imperfect solution would, similarly, eventually follow the perfect structure's secondary bifurcation with the longest wavelength.  The imperfect primary path for the monotonically hardening foundation $\alpha=+1$, which also starts from $\lambda_{cis}$, is almost indistinguishable from the stable, perfect primary bifurcation branch emerging at $\lambda_c = 2$.

To conclude, in Fig.~\ref{Fig:single-localized-dia} we emphasize the main results of this investigation by presenting an uncluttered bifurcation diagram showing only the (perfect and imperfect) primary and localized $C_{1v} : k=1/20$ secondary bifurcated equilibrium paths corresponding to the mild re-hardening foundation (cf.\ Fig.~\ref{Fig:soft-mild-dia}).
The perfect and imperfect equilibrium paths and their corresponding deformed configurations are nearly indistinguishable; i.e.\ these solutions are insensitive to imperfections.  In Fig.~\ref{Fig:single-localized-dia}(a) we plot the amplitude $\xi$ vs load $\lambda$ for the (perfect and imperfect) stable spatially localized and stable uniformly wrinkled equilibrium solutions that exist simultaneously at a load value of $\lambda = 1.2$.  This load is significantly below the critical value $\lambda_c = 2$ of the perfect system.  Similarly, the load $\lambda$ vs strain $\Delta$ is plotted in Fig.~\ref{Fig:single-localized-dia}(b).  These are solutions for the mild re-hardening foundation.  Point~E corresponds to the fully localized deformation solutions (perfect and imperfect) shown in Fig.~\ref{Fig:single-localized-dia}(c); Point~F corresponds to the localized packet deformation solutions (perfect and imperfect) shown in Fig.~\ref{Fig:single-localized-dia}(d); and Point~G corresponds to the uniformly wrinkled deformation solutions (perfect and imperfect) shown in Fig.~\ref{Fig:single-localized-dia}(e).  Not shown is the stable principal solutions (perfect and imperfect), $w(x; \lambda) = 0$, at this load.  It is worth repeating that all of these solutions (and others not shown) are stable.  Thus, there exists a rich set of stable deformed equilibrium structures for the beam--foundation system at loads well-below the critical value predicted by the standard linearized buckling analysis of Section~\ref{sec:perfect:princ}.  Here, we have shown that a subset of those solutions, that correspond to highly localized deformations, occur as secondary bifurcations which emerge from the primary bifurcated path immediately after it emerges from the principal path.


\section{Conclusion}
\label{sec:Section_5}

The nonlinear beam--foundation model problem analyzed here, although it looks deceptively simple, has a very rich and complex set of solutions due to the high degree of its initial symmetry.  Unlike previously proposed methods that use asymptotic techniques (to obtain an equation for the localized amplitude of a rapidly-oscillating principal bifurcation eigenmode) near the critical load, we show that to obtain a stable fully localized deformation for the perfect structure, one has to follow the secondary bifurcating equilibrium path with the longest wavelength far away from the critical load.  The existence of stable---and hence observable---fully localized deformation solutions appears to require both initial-softening and later re-hardening of the nonlinear foundation.  Indeed, for monotonically softening foundations no \emph{stable} fully spatially localized solutions are found.  Similarly, for monotonically hardening foundations no stable \emph{fully localized} deformation solutions are found.

In addition to explaining how to find stable, fully localized deformation solutions in the problem at hand, the use of group-theory methods for this problem illustrates the general methodology for tackling nonlinear problems in mechanics with a high degree of symmetry for which the standard imperfection method is not a reliable option.

\begin{acknowledgements}
All authors would acknowledge support from the \'Ecole Polytechnique and its Laboratore de M\'ecanique des Solides (LMS).  The work was initiated in part by grants from \'Ecole Polytechnique and the C.N.R.S. (Centre National de Recherche Scientifique) during the AY 2017-2018, while TJH was a Distinguished Visiting Professor and SSP a visiting doctoral student.  RSE acknowledges several visits to LMS during this period supported by the LMS.  The work of SSP and TJH was also supported in part by the National Science Foundation (NSF) through grant DMS-1613753.  The work of RSE was also partially supported by the NSF grant CMMI-1462826.
\end{acknowledgements}
\bibliographystyle{spbasic}
\bibliography{citations}


\newpage
\appendix
\section{Group-Theoretic Considerations}
\label{appendix-A}

The fundamental concept used to study the bifurcated equilibrium paths and their stability in any conservative elastic system is the existence of a group $G$ of transformations that leave its energy $\mathcal{E}(w;\lambda)$---defined in Eq.~\eqref{eq:energy} for the problem at hand---unchanged, i.e.\ invariant under the action of all transformations $g \in G$.  More specifically, to each element $g \in G$ we associate an orthogonal transformation $T_g$ (termed ``{\it representation}" of $g$) acting on $w(x) \in H$ with image $T_g[w] \in H$ that satisfies
\begin{equation}
\mathcal{E}(T_g[w];\lambda) = \mathcal{E}(w ;\lambda) \, ;\quad \forall \lambda \ge 0  \, , \ \forall w \in H \, ,
\ \forall g \in G\; ,
\label{eq:invariance}
\end{equation}
where $\lambda$ is the scalar load parameter (assumed positive) and $H$ the space of admissible displacements.

It follows from Eq.~\eqref{eq:invariance} that the variation of $\mathcal{E}$ with respect to its argument $w$ (first order functional derivative $\mathcal{E}_{,w}$) possess the property of ``{\it equivariance}''
\begin{equation}
T_g[\mathcal{E}_{,w}(w;\lambda)] \delta w = \mathcal{E}_{,w}(T_g [w] ;\lambda) \delta w \, ; \quad  \forall \lambda \ge 0  \, ,
\ \forall w, \delta w \in H \, ,\  \forall g \in G\; .
\label{eq:equivariance}
\end{equation}

According to Eq.~\eqref{eq:equilibrium}, the system's equilibrium solutions $w(x ; \lambda)$ are found by extremizing its energy; consequently all solutions of the system $\mathcal{E}_{,w}(w;\lambda) \delta w = 0$ must satisfy Eq.~\eqref{eq:equivariance}.  It is more appropriate to talk about {\it orbits} of equilibrium paths since, in view of the equivariance described in \eqref{eq:equivariance}, applying to an equilibrium solution $w$ the transformation $T_g$ automatically generates another equilibrium solution $T_g[w]$.

A subset of these equilibrium solutions, termed ``{\it principal solutions}'' and denoted by ${\stackrel{0}{w}}(x; \lambda)$, are invariant under all transformations $T_g$.  These solutions belong to an invariant subspace of $H$, denoted ${\mathcal S}_G$ and called the ``{\it fixed-point space}''
\begin{equation}
\mathcal{E}_{,w}({\stackrel{0}{w}}(x; \lambda);\lambda) \delta w = 0 \; , \  \forall \lambda \ge 0 \; ; \quad {\stackrel{0}{w}} \in {\mathcal S}_G :=  \{ w \in H \;|\; \  T_g [w] = {w} \; , \ \forall g \in G \} \; .
\label{eq:fixedspace}
\end{equation}

To determine the stability of a principal solution, one has to check the positive definiteness of the self-adjoint bilinear operator ${\mathcal E}^0_{,ww}$, evaluated on the principal path ${\stackrel{0}{w}}(x; \lambda)$, by finding its eigenvalues $\beta(\lambda)$
\begin{equation}
({\mathcal E}^0_{,ww}\Delta w)\delta w = \beta(\lambda) <\Delta w, \delta w> \; ; \quad \forall \delta w \in H\;  ;
\ {\mathcal E}^0_{,ww} := {\mathcal E}_{,ww}({\stackrel{0}{w}}(x;\lambda);\lambda) \; ,
\label{eq:staboper}
\end{equation}
where $\Delta w$ is the corresponding eigenvector and $< \cdot \; ,\; \cdot >$ denotes an inner product in $H$.  A stable solution corresponds to a positive minimum eigenvalue\footnote{See Footnote~\ref{foot:zero:eigval}.} $\beta_{min} > 0$ (the number of eigenvalues depends on the dimension of $H$).  When ${\stackrel{0}{w}}(x;\lambda)$ is periodic, the Bloch-wave representation may be used, as described in Section~\ref{sec:stability}.  For a well-posed problem, its stress-free (unloaded) configuration at $\lambda=0$ is stable; as the load increases stability will be lost at the first bifurcation point encountered along the loading path at some $\lambda_b$.

It can be shown, e.g. \cite{GolStew88, mcweeny2002symmetry}, that the existence of the group $G$ implies the existence of a symmetry basis with respect to which (i) the operator ${\mathcal E}^0_{,ww}$ defined in \eqref{eq:staboper} can be block-diagonalized and (ii) the space of admissible functions $H=\oplus_{\mu=1}^h V^{\mu}$ can be uniquely decomposed into a direct sum of mutually orthogonal invariant subspaces $V^{\mu}$ (with $h$ being the number of equivalence, or conjugacy, classes for $G$).  Each subspace $V^{\mu}$ is associated with an $n_\mu$-dimensional irreducible representation $\tau^{\mu}$ of $G$, also termed ``{\it irrep}'' from which an appropriate projection operator can be constructed giving the $V^{\mu}$ component of any function in $H$.

Bifurcated equilibrium paths, termed {\it primary}, can emerge from the principal path at loads  $\lambda_b$ corresponding to zero eigenvalues of ${\mathcal E}^b_{,ww}$.   That is, $\beta(\lambda_b)=0$ so that
\begin{equation}
({\mathcal E}^b_{,ww} {\stackrel{i}{w}})\delta w =  0 \; ,
\ {\stackrel{i}{w}} \in {\mathcal N}_{\mu}\; ,\ <{\stackrel{i}{w}},{\stackrel{j}{w}}>=\delta_{ij}\; , \ i , j = 1,\dots,n_\mu \; ; \  \forall \delta w \in H\; ;\
{\mathcal E}^b_{,ww} := {\mathcal E}_{,ww}({\stackrel{0}{w}}(x;\lambda_b);\lambda_b) \; ,
\label{eq:bifurc}
\end{equation}
where the eigenmodes $\Delta w =\stackrel{i}{w}$ span ${\mathcal N}_{\mu}$, the $n_\mu$-dimensional null space of the operator ${\mathcal E}^b_{,ww}$. Some additional conditions, termed ``{\it transversality}'' conditions must also hold to ensure that $\lambda_b$ is a bifurcation and not a limit (i.e.\ turning) point\footnote{Compare Eq.~\eqref{eq:transversality2} to Eq.~\eqref{eq:transversality}.  In Eq.~\eqref{eq:transversality}, no assumptions about symmetry are made, and one must ensure that the entire operator is non-singular.  In Eq.~\eqref{eq:transversality2}, equivariance is assumed and this suffices to ensure that $\mathcal{E}^b_{,w\lambda} {\stackrel{i}{w}} = 0\;,\ i=1,2,\dots,n_{\mu}$ and that the operator $\mathcal{E}^0_{,ww}$ is a scalar multiple of the identity.  Thus, the two criteria are equivalent when the assumptions of equivariant bifurcation theory are satisfied.}:
\begin{equation}
([d{\mathcal E}^0_{,ww}/d\lambda]_b{\stackrel{1}{w}}){\stackrel{1}{w}} \neq 0\; .
\label{eq:transversality2}
\end{equation}
It can be shown that the null space ${\mathcal N}_{\mu}$ of the stability operator ${\mathcal E}^b_{,ww}$ is an $n_{\mu}$-dimensional subspace of the invariant subspace $V^{\mu}$ for some $\mu$, i.e.\ ${\mathcal N}_{\mu} \subseteq V^{\mu}$, associated to the irreducible representation $\tau^{\mu}$.  The (primary) bifurcated paths ${\stackrel{b}{w}}(x; \lambda)$ emerging from ${\stackrel{0}{w}}(x; \lambda)$ at $\lambda_b$, can be represented with the following asymptotic expressions, resulting from the projection of the equilibrium equations onto the null space ${\cal N}_\mu $ and termed ``{\it Lyapunov--Schmidt--Koiter}'' asymptotics:
\begin{equation}
\lambda(\xi)=\lambda_{b}+ \lambda_1\xi + {\frac{1}{2}} \lambda_2\xi^2 + O(\xi^3)\; , \qquad {\stackrel{b}{w}}(x; \xi) = {\stackrel{0}{w}}(x; \lambda(\xi)) + \xi \sum_{i=1}^{n_{\mu}} \alpha_i {\stackrel{i}{w}}(x) + O(\xi^2)\; ,
\label{eq:bifurcation}
\end{equation}
where $\xi$ is the path's ``{\it bifurcation amplitude}'' parameter and $\alpha_i$ the components of the bifurcated path's initial tangent vector.  From knowledge of the bifurcation irrep $\tau^\mu$, group theory results (such as the ``{\it lattice of isotropy subgroups}'' of $G$ \citep{chossat:lauterbach:2000}) allow the direct computation of the $\alpha_i$ tangents (one set for each bifurcating orbit) emerging at $\lambda_b$.  Once the asymptotic expression of the primary bifurcating equilibrium path ${\stackrel{b}{w}}(x; \xi)$ is established, the global solution path may be computed efficiently by using the path's isotropy group, i.e.\ the elements of the subgroup of $G$ satisfying $T_g [{\stackrel{b}{w}}]={\stackrel{b}{w}}$, to find its corresponding fixed-point space.  Along this path there may occur (secondary) bifurcation points.  In such cases, the above procedure begins once again with ${\stackrel{b}{w}}(x; \lambda)$ as the new principal path from which---secondary with respect to ${\stackrel{0}{w}}(x; \lambda)$---bifurcated orbits will emerge.

In order to determine the local (asymptotic) type---transverse or pitchfork---of the bifurcating paths (which may be different than the global type), we need to determine if $\lambda_1 = 0$ or $\lambda_1 \not= 0$ in Eq.~\eqref{eq:bifurcation}$_1$.  From an asymptotic analysis \citep[e.g.][]{triantafyllidis:peek:1992} $\lambda_1 = 0$ is guaranteed if
\begin{equation}
\mathcal{E}_{ijk} := (([\mathcal{E}_{,www}({\stackrel{0}{w}}(x;\lambda_b); \lambda_b)] {\stackrel{i}{w}}) {\stackrel{j}{w}}) {\stackrel{k}{w}} = 0, \qquad i,j,k = 1,\dots,n_{\mu}.
\label{eq:Eijk}
\end{equation}

To reiterate, the strategy followed in this work is to sequentially apply the above-described procedure to follow the bifurcating equilibrium orbits of the system by identifying, each time, their symmetry group and their corresponding fixed-point space.  As we proceed from the principal solution to the primary bifurcations emerging from it, then to the secondary bifurcations emerging from the primary ones, the corresponding symmetry groups and fixed-point spaces change accordingly.  Knowledge of the symmetries of a path allows for an efficient calculation of a unique solution in its own fixed-point space.  The method adopted here follows the procedures introduced by \cite{healey1988group,gatermann:hohmann:1991:ICSE}.  Moreover, following \cite{gatermann:hohmann:1991:ICSE,chossat:lauterbach:2000}, knowledge of the lattice of isotropy subgroups of the initial symmetry group guides the search for the bifurcated equilibrium paths in a systematic way and explains our findings.

\subsection{Principal Solution, Irreps, and Bifurcations---Group $G = D_{\infty h}$}

As described in Sec.~\ref{sec:symmetry}, the fixed-point space of $D_{\infty h}$ for the beam--foundation system consists of only the trivial principal solution $\overset{0}{w}(x; \lambda) = 0$.  According to group theory \citep[e.g.][]{ikeda:murota:2010} $D_{\infty h}$ has four 1-dimensional irreps (one being the trivial identity irrep).  These provide the possibility of simple bifurcations to paths with symmetry groups of $C_{\infty v}$, $C_{\infty h}$ or $D_{\infty}$.  There are also an infinity of 2-dimensional irreps, providing the possibility of double bifurcations.  These correspond to bifurcating paths with symmetry groups $D_{nh}$ or $D_{nd}$, where $n\in \mathbb N$, as shown in Table~\ref{tab:Dinftyirreps}.
\begin{table*}
\begin{center}
\def\matR{{\left[\begin{array}{cc} \displaystyle \cos(n\phi) & \displaystyle -\sin(n\phi) \\
\displaystyle \sin(n\phi) & \displaystyle \cos(n\phi) \end{array}\right]}}
\def\matP{{\left[\begin{array}{cc} 1 & 0 \\ 0 & -1 \end{array}\right]}}
\def\matIP{{\left[\begin{array}{cc} 1 & 0 \\ 0 & 1 \end{array}\right]}}
\def\matIM{{\left[\begin{array}{cc} -1 & 0 \\ 0 & -1 \end{array}\right]}}
\medskip
\begin{tabular}{|c|c|c|c|c|c|c|}
\hline
$n_{\mu}$ & Irrep $\mu$ & $\tau^{\mu}_{c(\phi)}$ &  $\tau^{\mu}_{\sigma_v}$ & $\tau^{\mu}_{\sigma_h}$ & $G^{\mu}$ & Bifurc.\ Orbit Sym.  \\
 \hline
 \hline
1 &  $A_1$ &   1 & 1 & 1 & $D_{\infty h}$ & No Bif.  \\
1 &  $A_2$ &   1 & 1 & -1 & $C_{\infty v}$ & $C_{\infty v}$  \\
1 &  $B_1$ &   1 & -1 & 1 & $C_{\infty h}$ & $C_{\infty h}$  \\
1 &  $B_2$ &   1 & -1 & -1 & $D_{\infty}$ & $D_{\infty}$  \\
 \hline
  & & & & & & \\
2 &  $E_{1n}$ &  $\matR$ & $\matP$ & $\matIP$ & $C_{nh}$ & $D_{nh}$  \\
 & & & & & & \\
2 &  $E_{2n}$ &   $\matR$ & $\matP$ & $\matIM$ & $S_{2n}$ & $D_{nd}$  \\
 & & $n= 1, 2, 3,\dots$ & & & & \\
\hline
\end{tabular}
\end{center}
\caption{Table of irreps and bifurcated orbit symmetries for $G = D_{\infty h}$.  The first column gives the dimension $n_\mu$ of the corresponding irrep; the second column gives a standard label/name for the irrep; the third through fifth columns provide the corresponding irrep matrix for the generators $\tau^{\mu}_{c(\phi)}$, $\tau^{\mu}_{\sigma_v}$, and $\tau^{\mu}_{\sigma_h}$, respectively; the sixth column gives the \emph{kernel of $\tau^{\mu}$} (i.e., $G^{\mu} = \{ g \in G \;|\; \tau^{\mu}_{g} = I\}$, where $I$ is the $n_\mu$-dimensional identity matrix.); the seventh column gives the symmetry group of the corresponding bifurcating equilibrium path(s).}
\label{tab:Dinftyirreps}
\end{table*}
In Section~\ref{sec:Stability} we find from Eq.~\eqref{eq:highercritical} only double bifurcations at $\lambda_n \geq 2$ with corresponding eigenmodes ${\stackrel{ni}{w}}(x), i=1,2$---as expected from the 2-dimensional irreps of $D_{\infty h}$ of Table~\ref{tab:Dinftyirreps}.  Notice that no simple bifurcations exist, in spite of the existence of 1-dimensional irreps of this group.  From the infinity of primary bifurcation paths, we follow next the bifurcation orbit emerging from the lowest load $\lambda_c = 2$ which corresponds to $n_c=q$ and $L_c = 2\pi$ (recall that the periodicity of the model is $L_d= L_c q$).

\subsection{Primary Bifurcation Orbit at $\lambda_c = 2$, Irreps, and Bifurcations---Group $G = D_{qd}$}

The bifurcated equilibrium paths emerging from the lowest critical load $\lambda_c$ correspond according to Eq.~\eqref{eq:critical} to a double bifurcation with eigenmodes ${\stackrel{c1}{w}}(x) = \sin(x)$ and ${\stackrel{c2}{w}}(x) = \cos(x)$.  Since every linear combination of these eigenmodes is left invariant by the elements of the group $S_{2q}$, the critical point corresponds to the $\mu = E_{2q}$ irrep, and according to the general theory (see Table~\ref{tab:Dinftyirreps}) the symmetry group of the bifurcating orbit is $D_{qd}$. This symmetry group is finite and has the following two generators: $\sigma_{h} c(\pi/q)$ and $\sigma_h\sigma_v$.

As indicated in Table~\ref{tab:Dqdirreps}, this group has four 1-dimensional irreps (one being the trivial identity irrep).  These provide the possibility of simple bifurcations to paths with symmetry groups of $S_{2q}$, $D_{q}$, or $C_{qv}$.  There are also $(q-1)$ 2-dimensional irreps, providing the possibility of double bifurcations.  These correspond to bifurcating paths with symmetry groups $D_{rd}$, $C_{rv}$ and $D_{r}$, where $r:={\rm gcd}(j,q)$, with $j=1,\dots,q-1$.
\begin{table*}
\begin{center}
\def\matQ{{\left[\begin{array}{cc} \displaystyle -\cos(\pi j/q) & \displaystyle \sin(\pi j/q) \\
\displaystyle -\sin(\pi j/q) & \displaystyle -\cos(\pi j/q) \end{array}\right]}}
\def\matPM{{\left[\begin{array}{cc} -1 & 0 \\ 0 & 1 \end{array}\right]}}
\def\matIP{{\left[\begin{array}{cc} 1 & 0 \\ 0 & 1 \end{array}\right]}}
\def\matIM{{\left[\begin{array}{cc} -1 & 0 \\ 0 & -1 \end{array}\right]}}
\medskip
\begin{tabular}{|c|c|c|c|c|c|}
\hline
$n_{\mu}$ & Irrep $\mu$ & $\tau^{\mu}_{\sigma_h c(\pi/q)}$ &  $\tau^{\mu}_{\sigma_h \sigma_v}$ & $G^{\mu}$ & Bifurc.\ Orbit(s) Sym.  \\
 \hline
 \hline
1 &  $A_1$ &   1 & 1 & $D_{qd}$ & No Bif.  \\
1 &  $A_2$ &   1 & -1 & $S_{2q}$ & $S_{2q}$  \\
1 &  $B_1$ &   - 1 & 1 & $D_q$ & $D_q$  \\
1 &  $B_2$ &   - 1 & -1 & $C_{qv}$ & $C_{qv}$  \\
 \hline
  & & & & & \\
2 &  $E^j$ &  $\matQ$ & $\matPM$ & $\begin{array}{c} S_{2r} :  {(q+j) / r} \, {\rm even} \\
C_r :  {(q+j) / r} \, {\rm  odd} \end{array}$ & $\begin{array}{c} D_{rd} \\ C_{rv} \ {\rm and}\ D_r \end{array}$  \\
 & & $1 \le j \le q -1$  & & $r:={\rm gcd}(j,q)$ & \\
\hline
\end{tabular}
\end{center}
\caption{Table of irreps and bifurcated orbit symmetries for $G = D_{qd}$.  The first column gives the dimension $n_\mu$ of the corresponding irrep; the second column gives a standard label/name for the irrep; the third and fourth columns provide the corresponding irrep matrix for the generators $\tau^{\mu}_{c(\pi/q)}$ and $\tau^{\mu}_{\sigma_h \sigma_v}$, respectively; the fifth column gives the \emph{kernel of $\tau^{\mu}$} (i.e., $G^{\mu} = \{ g \in G \;|\; \tau^{\mu}_{g} = I\}$, where $I$ is the $n_\mu$-dimensional identity matrix.); the sixth column gives the symmetry group of the corresponding bifurcating equilibrium path(s).  The function $\gcd(a,b)$ is the greatest common divisor of $a$ and $b$.}
\label{tab:Dqdirreps}
\end{table*}
The fixed-point space $\mathcal{S}_{ D_{qd}} := \{w(x) \in H \;|\; T_{g}[w(x)] = w(x), \; \forall g \in D_{qd}\}$ of the primary bifurcation solutions ${\stackrel{1}{w}}$ with $D_{qd}$ symmetry, as defined above, contains configurations $w(x)$ of the $L_d$-periodic beam that remain unaltered under the transformations of this group.  The primary bifurcation path ${\stackrel{1}{w}}(x; \lambda)$ is contained within $\mathcal{S}_{D_{qd}}$ and a representative configuration along this path parameterized with respect to the bifurcation amplitude $\xi$ (see Section~\ref{sec:Stability}) is plotted in Fig.~\ref{Fig:primary-w}.

\subsection{Secondary Bifurcation Orbits and their Symmetry}

Recall that for the numerical calculations reported here, we have selected $q=20$ which gives $L_d = L_c q = 40 \pi$.  Due to the (energy-induced) symmetry of the dispersion curve discussed in Section~\ref{sec:Stability}, all bifurcations are double.  Indeed, double bifurcation points corresponding to all of the 2-dimensional irreps $E^j$ of $D_{20d}$ are found along the primary bifurcation orbit ${\stackrel{1}{w}}(x; \lambda)$.  The correspondence between the 2-dimensional irrep index $j$ of Table~\ref{tab:Dqdirreps} and the Bloch-wave wavenumber $k$ is presented in Table~\ref{tab:Dqd-symmetry}, along with the symmetry group of each orbit bifurcating from a bifurcation point of this type.
\begin{table*}
\begin{center}
\medskip
\begin{tabular}{|c|c|c|c|c|c|}
\hline
$D_{20d}$ 2D Irrep no.\ $j$ & r={\rm gcd}(20,j) & (20+j)/r --- even/odd & orbit(s) sym. & global bif.\ type & wavenumber(s) $k$  \\
 \hline
 \hline
1 &  1 &    odd  & $C_{1v}$ {\rm and} $D_1$ & pitchfork & k = 3/20,\ 17/20 \\
2 &  2 &    odd &  $C_{2v}$ {\rm and} $D_2$ & pitchfork & k = 6/20,\ 14/20 \\
3 &  1 &    odd &  $C_{1v}$ {\rm and} $D_1$ & pitchfork & k = 9/20, \ 11/20 \\
4 &  4 &    even &  $D_{4d}$ & transverse & k = 8/20,\ 12/20 \\
5 &  5 &    odd &  $C_{5v}$ {\rm and} $D_5$ & pitchfork & k = 5/20,\ 15/20 \\
6 &  2 &    odd &  $C_{2v}$ {\rm and} $D_2$ & pitchfork & k = 2/20, \ 18/20 \\
7 &  1 &    odd &  $C_{1v}$ {\rm and} $D_1$ & pitchfork & k = 1/20, \ 19/20 \\
8 &  4 &    odd &  $C_{4v}$ {\rm and} $D_4$ & pitchfork & k = 4/20,\ 16/20 \\
9 &  1 &    odd &  $C_{1v}$ {\rm and} $D_1$ & pitchfork & k = 7/20, \ 13/20 \\
10 & 10 &    odd &  $C_{10v}$ {\rm and} $D_{10}$ & pitchfork & k = 10/20 \\
11 &  1 &    odd &  $C_{1v}$ {\rm and} $D_1$ & pitchfork & k = 7/20, \ 13/20 \\
12 & 4 &    even &  $D_{4d}$ & transverse & k = 4/20,\ 16/20 \\
13 & 1 &    odd &  $C_{1v}$ {\rm and} $D_1$ & pitchfork & k = 1/20, \ 19/20 \\
14 & 2 &    odd &  $C_{2v}$ {\rm and} $D_2$ & pitchfork & k = 2/20, \ 18/20 \\
15 & 5 &    odd &  $C_{5v}$ {\rm and} $D_5$ & pitchfork & k = 5/20,\ 15/20 \\
16 & 4 &    odd &  $C_{4v}$ {\rm and} $D_4$ & pitchfork & k =  8/20,\ 12/20 \\
17 & 1 &    odd &  $C_{1v}$ {\rm and} $D_1$ & pitchfork & k = 9/20, \ 11/20 \\
18 & 2 &    odd &  $C_{2v}$ {\rm and} $D_2$ & pitchfork & k = 6/20,\ 14/20 \\
19 & 1 &    odd &  $C_{1v}$ {\rm and} $D_1$ & pitchfork & k = 3/20,\ 17/20 \\
\hline
\end{tabular}
\end{center}
\caption{Correspondence between irreps of $D_{20d}$ and Bloch-wave wavenumber $k$.  Also shown are the symmetries and global bifurcation type (transverse or pitchfork) of the corresponding bifurcating orbits.}
\label{tab:Dqd-symmetry}
\end{table*}
The $j$--$k$ correspondence is obtained by using standard group representation theory results to decompose the irreps of $D_{20d}$ into direct sums of the irreps of its subgroup $C_{20}$.  In this way, we can find which wavenumbers $k$ (used in the Bloch-wave calculations of Sec.~\ref{sec:Section_3}) correspond to each 2-dimensional irrep of $D_{20d}$.  The global bifurcation type, either transverse or pitchfork, is obtained from a theorem in equivariant bifurcation theory \citep{gatermann:hohmann:1991:ICSE}, and indicates the nature of the global bifurcating equilibrium path.  In particular, a globally transcritical path $w(x; \xi)$ will have $w(x; -\xi) \not= w(x; \xi)$ for some value(s) of $\xi$, whereas a globally pitchfork path will have $w(x; -\xi) = w(x; \xi)$ for all $\xi$.  See Sec.~\ref{sec:Section_4} for some discussion of the different types of paths found in the present work.

In addition to the group-theory results in Table~\ref{tab:Dqd-symmetry}, we can also show by group-theoretic calculation that all the secondary bifurcating paths are locally pitchfork, i.e.\ $\lambda_1=0$ in Eq.~\eqref{eq:bifurcation}$_1$ since all the coefficients $\mathcal{E}_{ijk}$  of Eq.~\eqref{eq:Eijk} vanish.  This result is obtained by constructing an appropriate \emph{tensor-product rep} \citep{mcweeny2002symmetry} for the vector space of third-order tensors to which the ``vector'' $\mathcal{E}_{ijk}$ belongs.  Due to equivariance, $\mathcal{E}_{ijk}$ must be in the fixed-point space $\mathcal{S}_{D_{20d}}$ of this tensor-product space.  Further, it is easily found that $\mathcal{S}_{D_{20d}} = \{0\}$.  Thus, all paths bifurcating from the $D_{20d}$ primary orbit are locally pitchfork.  However, as indicated in Table~\ref{tab:Dqd-symmetry}, not all such bifurcating paths are globally pitchfork.  Again, see Sec.~\ref{sec:Section_4} for some discussion of this issue.

\subsection{Critical Load and Eigenmodes for the Imperfect Beam (Amplitude: $\zeta < 0$)}
\label{A-imp-mode}

We present here a calculation of the lowest load $\lambda_{cis}$ corresponding to a bifurcation with a symmetric (with respect to $x=0$) eigenmode ${\stackrel{is}{w}}(x)$ for an imperfect, periodic beam of length $L_d$.  The calculation of the lowest load corresponding to a bifurcation with an antisymmetric (with respect to $x=0$) eigenmode is similar and gives a slightly higher critical load $\lambda_{cia}$; the corresponding calculation is omitted here since we are interested in the bifurcated equilibrium paths emerging from the lowest load $\lambda_{cis}$.

By taking the functional derivative of the equilibrium Eq.~\eqref{eq:impequilibrium} and evaluating at $w(x)=0$ we obtain the equation for the symmetric with respect to $x=0$ eigenmode of the imperfect beam (${\stackrel{is}{w}}(x)={\stackrel{is}{w}}(-x)$).  The corresponding boundary conditions follow from Eq.~\eqref{eq:boundary} for $p=1, 3$ and hence
\newcommand\wis{{\stackrel{is}{w}}}
\begin{equation}
\frac{d^4 \wis}{d x^4} + \lambda \frac{d^2 \wis}{d x^2} + [1+z(x)]\wis = 0\; ,\ x \in (0,\frac{L_d}{2}) ;\qquad
\frac{d \wis}{d x} (0) =  \frac{d \wis}{d x} ({\frac{L_d}{2}}) = 0 \; ,\ \frac{d^3 \wis}{d x^3} (0) = \frac{d^3 \wis}{d x^3} ({\frac{L_d}{2}}) = 0\; .
\label{eq:impmode}
\end{equation}
Solving the problem in Eq.\eqref{eq:impmode} with the constant coefficient ordinary differential equation in the intervals $(0, x_0)$ and $(x_0, L_d/2)$, one has
\newcommand\az{a_{\zeta}}
\newcommand\bz{b_{\zeta}}
\begin{equation}
{\stackrel{is}{w}}(x) =
\left\{
\begin{array}{ll}
\left.
\begin{array}{l}
\eta \cos(\az x) + \theta  \cos(\bz x)  \\ \\
\az := \displaystyle \left[{\frac{\lambda+[\lambda^2-4(1+\zeta)]^{1/2}}{2}}\right]^{1/2},
\quad \bz := \left[{\frac{\lambda-[\lambda^2-4(1+\zeta)]^{1/2}}{2}}\right]^{1/2}
\end{array} \right\} &  x \in [0, x_0]   \\ \\
\left.
\begin{array}{l}
\displaystyle \delta \cosh[a(x-\frac{L_d}{2})] \cos[b(x-\frac{L_d}{2})] + \epsilon \sinh[a(x-\frac{L_d}{2})] \sin[b(x-\frac{L_d}{2})]    \\ \\
a  := \displaystyle {\frac{[2 -\lambda]^{1/2}}{2}}, \quad b := {\frac{[2 +\lambda]^{1/2}}{2}}
\end{array}  \quad \quad \quad \right\}  &  \displaystyle x \in [x_0 ,{\frac{L_d}{2}}]
\end{array} \right.
\label{eq:symmetricmode}
\end{equation}
where $\az,\; \bz,\; a,\; b$ are all positive constants. Notice from Eq.~\eqref{eq:symmetricmode} that the periodicity-imposed symmetry conditions at $L_d/2$ according to Eq.~\eqref{eq:impmode} are automatically satisfied.

Using the continuity condition at $x_0$ for ${\stackrel{is}{w}}(x)$ and its derivatives up to order three, one obtains a homogeneous linear system of four equations for the four constants $\eta, \theta, \delta, \epsilon$ appearing in Eq.~\eqref{eq:symmetricmode}.  The vanishing of the determinant of the corresponding $4 \times 4$ matrix (not recorded here in view of the cumbersome expressions) provides the equation for the critical load $2(1+\zeta)^{1/2} < \lambda_{cis} < 2$ (guaranteed by $\zeta <0$), which is obtained numerically.

\section*{Conflict of interest}
The authors declare that they have no conflict of interest.

\end{document}